\RequirePackage{fix-cm}
\documentclass[smallextended]{svjour3}
\smartqed  

\usepackage[utf8]{inputenc}
\usepackage[T1]{fontenc}
\usepackage{graphicx,psfrag}
\usepackage{mathrsfs}
\usepackage{amsmath,amsfonts,amssymb,pifont,gensymb, bm}
\usepackage{comment}
\usepackage{color}
\usepackage{acronym}
\usepackage{xspace}
\usepackage[normalem]{ulem}
\usepackage{mathtools}
\usepackage{subfigure}
\usepackage{makecell}
\usepackage{multirow,enumerate}
\usepackage[unicode=true, pdffitwindow=true,colorlinks=true, linkcolor=black, citecolor=black, urlcolor=black, hyperfootnotes=false]{hyperref}

\begin{document}
\sloppy

\title{Interpreting Binary Neutron Star Mergers}
\subtitle{Describing the Binary Neutron Star Dynamics, 
Modelling Gravitational Waveforms, and Analyzing Detections}

\author{Tim Dietrich \and
        Tanja Hinderer \and 
        Anuradha Samajdar 
}

\institute{T.~Dietrich 
           \at University of Potsdam, Institute for Physics and Astronomy, Karl-Liebknecht-Str. 24/25, 14476 Potsdam, Germany 
           \at Nikhef, Science Park 105, 1098 XG Amsterdam, The Netherlands, \\\           
           \email{tim.dietrich@uni-potsdam.de}\\
           \and
           T.~Hinderer 
           \at University of Amsterdam, Science Park 904, 1090 GL Amsterdam, The Netherlands, \\\
           \email{t.hinderer@uva.nl}\\
           \and              
           A.~Samajdar 
           \at Nikhef, Science Park 105, 1098 XG Amsterdam, The Netherlands, 
           \at Department of Physics, Utrecht University, Princetonplein 1, 3584 CC Utrecht, The Netherlands, \\\
           \email{a.samajdar@nikhef.nl}        
}

\date{Received: date / Accepted: date}

\maketitle

\begin{abstract}
Gravitational waves emitted from the coalescence of neutron star binaries
open a new window to probe matter and fundamental physics in unexplored, extreme regimes. 
To extract information about the supranuclear matter inside neutron stars 
and the properties of the compact binary systems, robust theoretical prescriptions
are required.
We give an overview about general features of the dynamics and the gravitational
wave signal during the binary neutron star coalescence. We briefly describe existing 
analytical and numerical approaches to investigate the highly dynamical, 
strong-field region during the merger. 
We review existing waveform approximants and discuss properties 
and possible advantages and shortcomings of individual waveform models, 
and their application for real gravitational-wave data analysis. 
\end{abstract}

\newpage

\setcounter{tocdepth}{3}
\tableofcontents

\newpage

\section{Introduction}
\label{intro}

The collision of two neutron stars (NSs) belongs to the most violent events in the
Universe and is connected to a variety of observables from gravitational wave (GW) and neutrino signatures
up to electromagnetic (EM) signals ranging through the full frequency band. 
This variety of possible `information channels' makes binary neutron star (BNS) mergers a 
perfect `tool' to study supranuclear dense material and strong gravity regions. 

The discovery of gravitational waves (GWs)~\cite{TheLIGOScientific:2017qsa} 
and electromagnetic (EM) signals~\cite{Monitor:2017mdv,Kasliwal:2017ngb,GBM:2017lvd} arising from a BNS coalescence 
detected in August 2017 marked a breakthrough in the field of multi-messenger astronomy. 
This multi-messenger observation represented a major leap forward in a number of research 
areas. 
It enabled a new and independent measurement of 
the Hubble constant~\cite{Abbott:2017xzu,Hotokezaka:2018dfi,Abbott:2019yzh,Dhawan:2019phb,Coughlin:2019vtv,Dietrich:2020lps}; 
it proved NS mergers to be a major cosmic source of heavy $r$-process elements, 
e.g.,~\cite{Cowperthwaite:2017dyu,Smartt:2017fuw,Kasliwal:2017ngb,Kasen:2017sxr};
it revealed that BNS mergers can act as the central engine for gamma-ray bursts (GRB)~\cite{GBM:2017lvd}; 
it enables an accurate measurement of the propagation speed of GWs~\cite{GBM:2017lvd}; 
it ruled out a number of alternative theories of gravity~\cite{Ezquiaga:2017ekz,Baker:2017hug,Creminelli:2017sry}; and
it allowed to place limits on the equation of state (EOS) of cold matter 
at supranuclear densities, e.g.,~\cite{TheLIGOScientific:2017qsa,Bauswein:2017vtn,Annala:2017llu,Most:2018hfd,Ruiz:2017due,Margalit:2017dij,Rezzolla:2017aly,Shibata:2017xdx,Abbott:2018exr,Abbott:2018wiz,De:2018uhw,Radice:2017lry,Coughlin:2018miv,Coughlin:2018fis,Dai:2018dca,Radice:2018ozg,Capano:2019eae,Dietrich:2020lps}. 

In addition, the increasing number of potential BNS
candidates and the second confirmed detection of a BNS merger, GW190425~\cite{Abbott:2020uma}, 
suggest that many more BNS systems will be detected in the near future. 

The extraction of information from the GW (and potentially EM) data relies on 
existing theoretical waveform models to perform a matched-filtering analysis, 
in which the data are cross-correlated with templates covering 
the possible physical parameter space, e.g.~\cite{Veitch:2014wba}. 
Cross-correlations have to be performed for a sufficient number of individual target 
waveforms (about a hundred millions of individual waveforms), which leads to large computational costs.
Therefore, the computation of each individual waveform needs
to be efficient and fast to ensure that the Bayesian parameter estimation
of signals, containing several thousand GW cycles, is at all manageable.
On the other hand, the construction of accurate template models requires a detailed
understanding of the late stages of the BNS coalescence, 
where matter effects on the GW signal become increasingly prominent.
In this regime, one has to solve Einstein's equations together with the 
equations of general relativistic hydrodynamics (GRHD). 
This challenging task is typically accomplished through numerical relativity (NR) 
simulations, see e.g.~\cite{Font:2007zz,Faber:2012rw,Baiotti:2016qnr,Dietrich:2018phi,Baiotti:2019sew,Kiuchi:2019kzt} 
and references therein. 

Unfortunately, NR simulations are computationally expensive and cannot be used 
to create template banks for BNS GW signals. Therefore, other approaches 
are needed to search for and interpret measured GW signals.
In the last years, there has been an increasing effort and success to construct
analytical and semi-analytical models to describe the GW signals emitted during the BNS coalescence. 

Despite the improvement, including the computation of higher-order tidal corrections, dynamical tidal effects,  and spin-tidal couplings,
e.g., Refs.~\cite{Damour:2012yf,Pani:2018inf,Banihashemi:2018xfb,Abdelsalhin:2018reg,Landry:2018bil,Jimenez-Forteza:2018buh}, 
the performance of Post-Newtonian (PN) waveform approximants becomes increasingly inaccurate towards the merger,
e.g.~\cite{Bernuzzi:2012ci,Favata:2013rwa,Wade:2014vqa,Hotokezaka:2016bzh,Dietrich:2018uni,Dudi:2018jzn,Samajdar:2018dcx}, 
which shows the need for possible alternatives. 
Therefore, most state-of-the-art time-domain tidal waveform
models~\cite{Baiotti:2010xh,Baiotti:2011am,Bernuzzi:2014owa,Hotokezaka:2015xka,Hinderer:2016eia,Steinhoff:2016rfi,Nagar:2018zoe,Nagar:2018gnk,Akcay:2018yyh,Nagar:2018plt}
are based on the EOB description of the general relativistic two-body problem~\cite{Buonanno:1998gg,Damour:2009wj}.
This approach has proven to be able to predict the BNS merger dynamics in large
regions of the BNS parameter space, but recent NR data
revealed configurations for which further improvements of the tidal EOB models might be
required~\cite{Hotokezaka:2015xka,Dietrich:2017feu,Akcay:2018yyh}.
While one can expect that over the next years,
these issues will be overcome due to further progress in the fields of NR,
gravitational self-force (GSF), PN theory, and applications of methods from high-energy physics,
the high computational cost for a single EOB waveform is yet another disadvantage.
One possibility to speed up the EOB computation is the use of
high-order post-adiabatic approximations of the EOB
description to allow an accurate and efficient evaluation of
the waveform up to a few orbits before merger~\cite{Nagar:2018gnk}.
The other possibility is the construction of a surrogate model~\cite{Lackey:2016krb,Lackey:2018zvw}.
Those models allow the fast computation of waveforms in the frequency
domain and are well suited for direct use in parameter estimation pipelines.

\begin{figure*}[t]
 \centering
 \includegraphics[width=\textwidth]{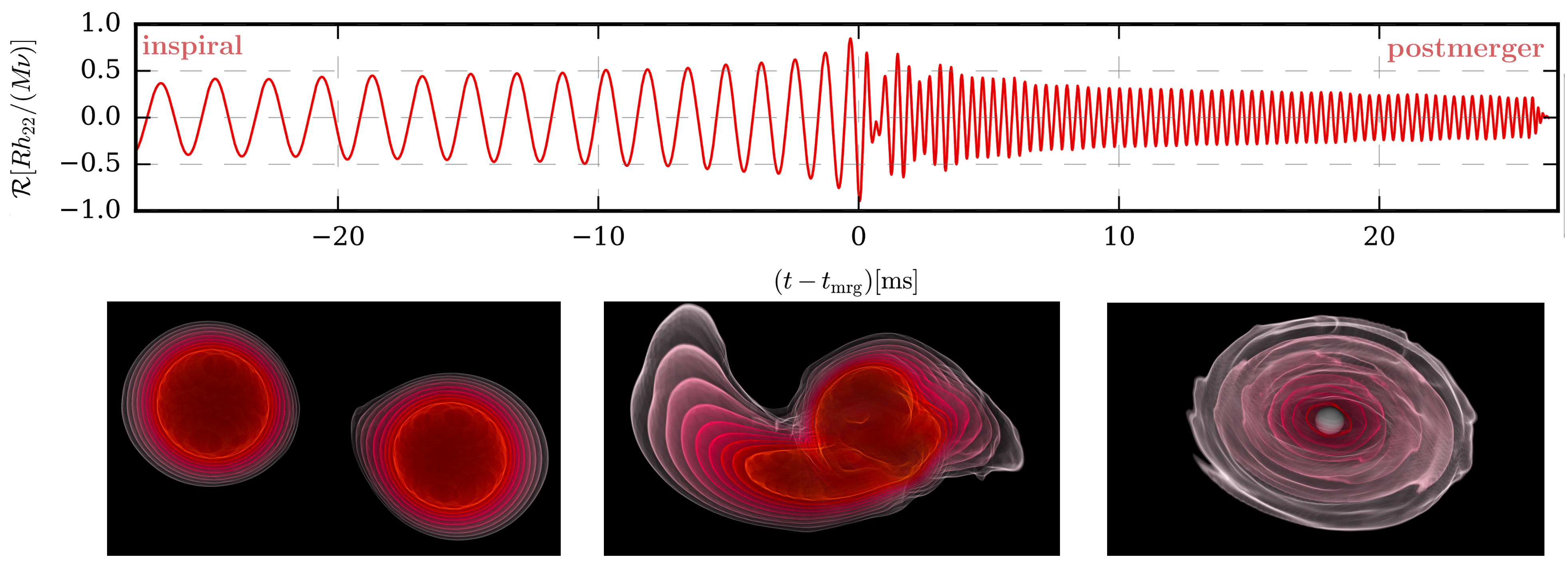}
 \caption{NR simulation of a BNS merger showing the GW signal and the matter evolution. 
          Top panel: GW signal emitted during the last orbits before the merger 
          (late-inspiral phase) and during the postmerger phase of the BNS coalescence. 
          Bottom panel: Rest-mass density evolution for the inspiral (first panel), 
          the merger (second panel) and the postmerger phase after the formation of the 
          black hole (third panel).}
 \label{fig:intro}
\end{figure*}

In addition to PN and EOB approximants, alternative ways to describe tidal GW signals have also been proposed.
Refs.~\cite{Lackey:2013axa,Pannarale:2013uoa,Thompson:2020nei}
develop phenomenological black hole-neutron star
(BHNS) approximants based on NR data.
Ref.~\cite{Barkett:2015wia} transforms NR
simulations of binary black hole (BBH) systems by adding PN tidal effects,
and Refs.~\cite{Lange:2017wki,Lange:2018pyp} develop a method to employ
NR waveforms or computationally expensive waveform approximants, e.g. tidal EOB approximants, 
directly for parameter estimation.

Yet another approach to describe BNS systems
was presented in Ref.~\cite{Dietrich:2017aum}, in which existing BBH models
have been augmented by an analytical closed-form expression correcting the
GW phase to include tidal effects~\cite{Dietrich:2017aum,Dietrich:2018uni,Kawaguchi:2018gvj,Dietrich:2019kaq}.\\

In this review article, we discuss the BNS and merger dynamics in Sec.~\ref{sec:coalescence}. 
In Sec.~\ref{sec:GWmodels_NR}, we briefly describe the use of 
NR simulations focusing on the imprint of the binary properties on the 
merger outcome and we review the current status in producing high-quality NR waveforms. 
In Sec.~\ref{sec:GWmodels_PN}, we give a brief overview about the existing PN knowledge 
for the description of the BNS coalescence. 
In Sec.~\ref{sec:GWmodels_EOB} and Sec.~\ref{sec:GWmodels_NRTidal}, we review EOB and 
phenomenological BNS models, respectively. 
Section~\ref{sec:PE} gives a short overview about the usage of BNS approximants for 
the parameter estimation of BNS signals. 

Throughout this work geometric units are used by setting $G=c=1$, and, additionally, 
$M_\odot=1$. In some places units are given explicitly to allow a better interpretation.
Further notations are $M=M_A + M_B$
for the total mass of the system, $\chi_A,\chi_B,\Lambda_A,\Lambda_B$
for the dimensionless spins and tidal deformabilities
of the individual stars. The mass ratio of the system is $q=M_A/M_B$ and
the symmetric mass ratio is $\nu=M_A M_B/(M_A+M_B)^2$.
We define the labeling of the individual
stars so their masses satisfy $M_A\geq M_B$.

\section{The BNS coalescence}
\label{sec:coalescence}

In contrast to BBH signals, BNS and BHNS mergers can produce a variety of 
EM counterparts such as sGRBs, a synchrotron afterglow, e.g., ~\cite{Eichler:1989ve,Paczynski:1986px,Narayan:1992iy,Lee:2007js,Nakar:2007yr,Sari:1997qe},
as well as a kilonova, see~\cite{Tanaka:2016sbx,Metzger:2019zeh} for reviews. 
The main difference between the GW signals emitted from the coalescence of BNS or BHNS systems and BBH 
systems are finite size effects that change the coalescence before the compact objects merge. 
Additionally, for BNS systems there is the possibility to have a postmerger GW signature. 
An illustration of the density evolution and the emitted GWs
of a BNS merger are presented in Fig.~\ref{fig:intro}, 
where we show approximately $60~\rm ms$ around the moment of merger. 
The figure includes the density profile obtained from a NR simulation 
during the late inspiral (bottom left panel), around the moment of merger (bottom middle panel), 
and the final BH surrounded by an accretion disk (bottom right panel).

Naturally, the full GW signal is composed of an inspiral and a postmerger evolution. 
These two regimes are separated by the moment of merger, i.e., 
the time when the amplitude of the GW signal reaches its maximum; cf.~Fig.~\ref{fig:intro}.

\subsection{Inspiral}

The inspiral of a BNS coalescence reaches frequencies up to about $1$-$2\rm kHz$, where the 
exact frequency depends on the binary properties. 
Current GW detectors are most sensitive at frequencies around $\sim100 \rm Hz$ and get  
less sensitive with increasing frequencies. 	
Because of this frequency-dependent sensitivity only the inspiral phase of BNS coalescences have been detected 
for GW170817~\cite{Abbott:2017dke,Abbott:2018wiz} and GW190425~\cite{Abbott:2020uma}.
While the detection of the postmerger signature is anticipated in the next years or decades, 
most information about the binary properties will come from the measurement of the inspiral. 

\begin{figure}[t]
 \centering
 \includegraphics[width=0.48\columnwidth]{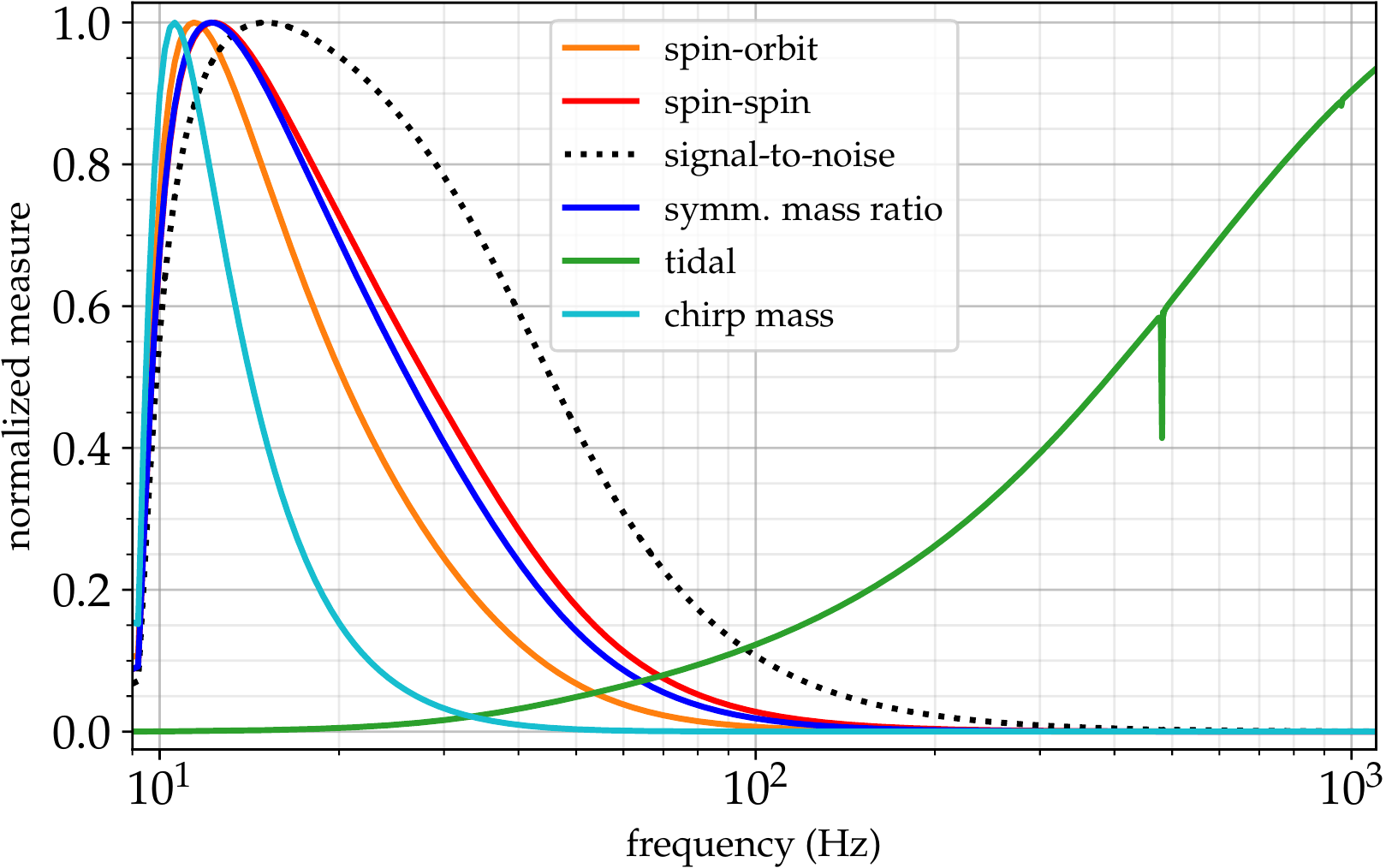}
  \includegraphics[width=0.48\columnwidth]{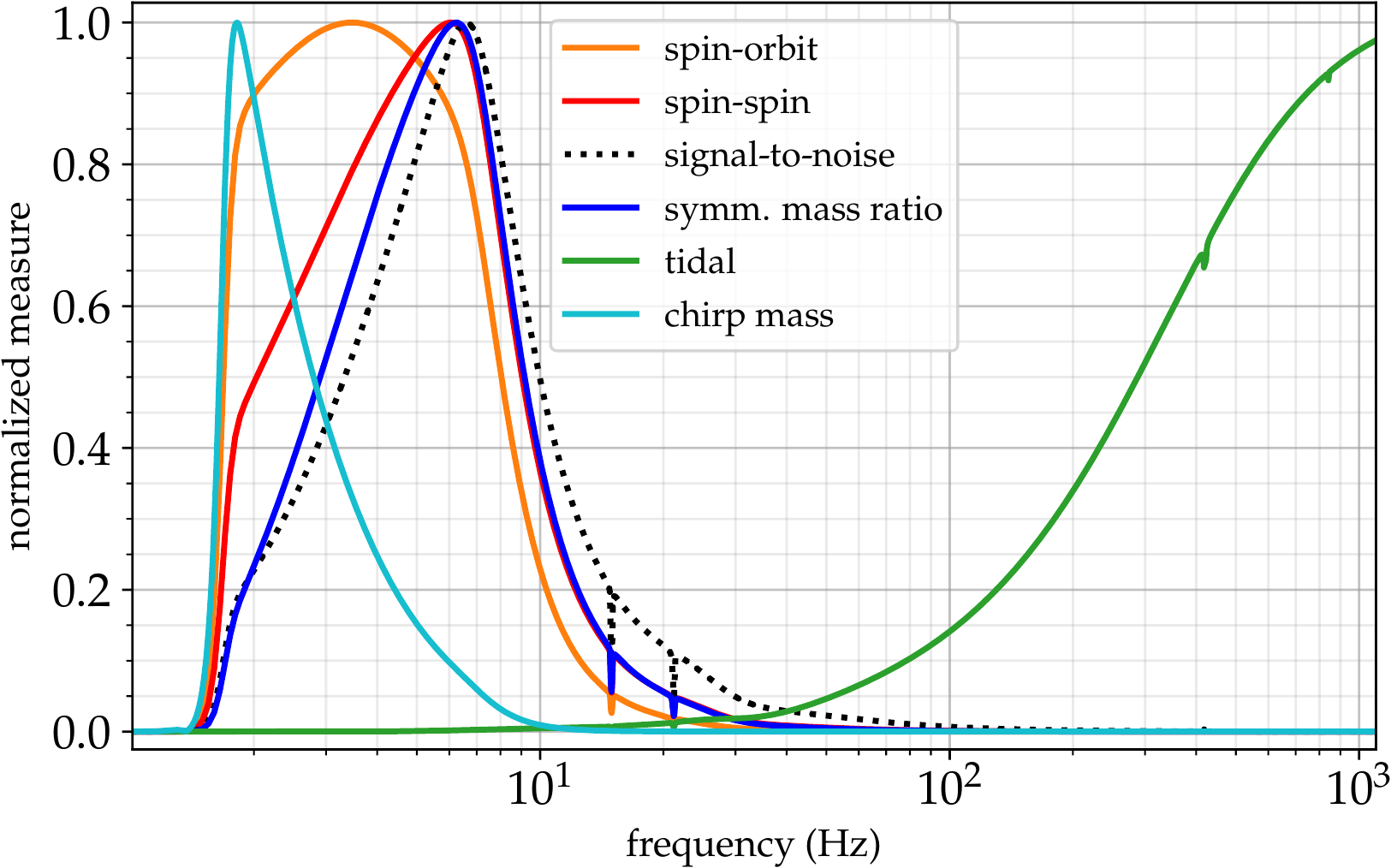}
 \caption{Accumulation of information about mass, spin, and tidal parameters per logarithmic frequency interval. \emph{Left panel}: aLIGO design sensitivity (zero detuned high power configuration)~\cite{TheLIGOScientific:2014jea,Sn:advLIGO}, \emph{right panel}: Einstein Telescope~\cite{Hild:2008ng,ET}. Here, 'normalized measure' is the leading-order contribution to the integrands appearing in the Fisher information matrix, with each curve normalized to its maximum value. To generate these plots, we used the PN TaylorF2 model with the nonspinning point-mass terms up to the highest order available (3.5PN)~\cite{Sathyaprakash:1991mt,Mishra:2016whh} combined with the leading-order tidal~\cite{Flanagan:2007ix} and spin effects~\cite{Kidder:1992fr,Damour:1990ji,Blanchet:1989ki}. Adapted from Ref.~\cite{Harry:2018hke}.}
 \label{fig:components_GWs}
\end{figure}

Therefore, the GW community has made significant efforts for a proper modeling of this regime. 
Here, we will focus mainly on the dominant tidal effects arising from the deformation of the NSs 
due to the external gravitational field of the companion, and matter effects introduced by 
the intrinsic rotation of the stars leading to a deformation related 
to the rotation frequency of the individual stars. 
These individual components affect the GW signal at different frequencies, 
cf.~Fig.~\ref{fig:components_GWs}. 
While the pure tidal effects dominate the late inspiral, 
the spin-induced quadrupole moment is mostly detectable from the GW signal around $20-30\rm Hz$ 
or at even lower frequencies for 3rd generation GW detectors.

\subsection{Coupling of NS matter to GWs}
During the early inspiral, the separation of scales in the binary system can be exploited for analytical approximations that enable tracing the information flow from matter properties to GW signals. In these approaches, the challenge of computing the dynamical spacetime is split into a coupled system of the binary dynamics, the asymptotic GWs, and the backreaction of GW losses on the dynamics. This requires using different approximation schemes in different regions of the spacetime and over different timescales. They are connected into a composite description through matched asymptotic expansions. For instance, the binary dynamics involve meshing a description adapted to the body zones close to each NS, where gravity is strong but the effects from the presence of a companion is small, with a description of the interaction zone where the NSs behave almost as point masses with small corrections due to their finite size~\cite{Flanagan:1997fn,Racine:2004hs}, see also~\cite{Dixon:1970zza,Thorne:1984mz,Itoh:1999bf,Kopeikin:2018zro,Kopeikin:2004ia,Vines:2010ca}. For weakly self-gravitating bodies described by PN gravity see also the seminal series of papers by Damour, Soffel, Xu~\cite{Damour:1991yw}. As will be discussed in detail in Sec.~\ref{sec:GWmodels_PN}, the multipole moments defined for the spacetime in the vicinity of the NSs play a key role for communicating information about NS matter between these descriptions. The multipole structure is affected by a variety of tidal effects, spins, and more complicated spin-tidal interactions. In addition to affecting the dynamics, the NS' multipole moments also give rise to additional imprints on the asymptotic gravitational radiation. The radiation can be described by double perturbation expansion around flat spacetime and an infinite series of radiative multipole moments, as explained in detail in the review article~\cite{Blanchet:2013haa}. The radiative moments are related in a complicated way, i.e., nonlinearly and non-locally in retarded time, to the total multipole moments of the binary system, which comprise contributions from the orbital motion and the NSs' multipoles. Problems such as the relativistic two-body problem that involve different scales can also efficiently be treated with effective-field-theory methods, see~\cite{Levi:2018nxp,Porto:2016pyg,Rothstein:2014sra,Foffa:2013qca} for comprehensive reviews and references.

\subsubsection{Dominant tidal effects}

\begin{figure}[t]
 \centering
 \includegraphics[width=.5\columnwidth]{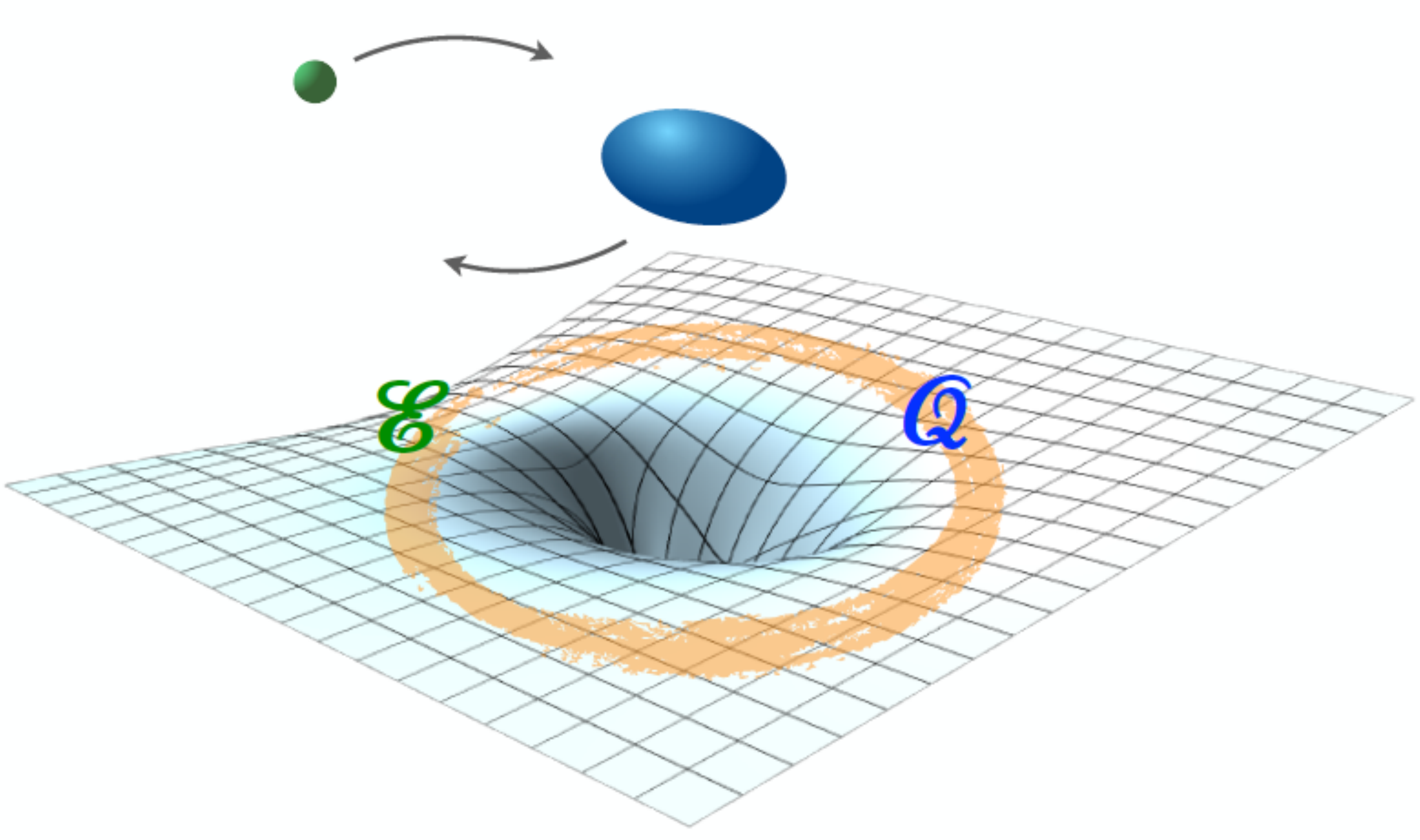}
  \caption{Cartoon depicting the definition of tidal deformability. 
  The tidal field ${\cal E}$ due to the spacetime curvature of the companion causes 
  the NS to deform as the matter adjusts to a new equilibrium configuration. 
  The relevant quantity influencing the GWs is the induced change in the multipole structure 
  of the NS's exterior spacetime ${\cal Q}$. The multipoles are also impacted by spin effects, and dynamical tidal effects. }
 \label{fig:tidalcartoon}
\end{figure}

In Newtonian gravity, tidal effects arise from the response of the NS to the gradient of the companion's gravitational field across its matter distribution. From the perspective of the NS, the companion is orbiting and produces a time-varying tidal field that slowly sweeps up in frequency. This quasi-periodic tidal forcing can excite characteristic oscillation modes in the NS that depend on the properties of matter in its interior. These concepts carry over to a General Relativistic description, where the modes are the NS's quasi-normal modes. A NS has a broad spectrum of modes~\cite{Kokkotas:1999bd}, several of which have sufficiently low frequencies to be relevant for the inspiral. The tidal excitation can either be a full resonant excitation of a mode or can be approximately adiabatic when the mode is driven well below its resonance frequency. The net effect of a mode on the GWs depends on (i) its frequency, which determines at what stage of the inspiral the mode gets excited and how long the effect accumulates, and (ii) on the tidal excitation factor, which characterizes how strongly the mode couples to the tidal field (analogous to the tidal deformability coefficients for the fundamental modes, as will be explained below). For circular binary inspirals, the fundamental modes have by far the largest tidal coupling~\cite{McDermott1985,Shibata:1993uk,Kokkotas:1995xe,Lai:1993di}, with the effect of the quadrupole $\ell=2$ being the largest. During most of the inspiral the fundamental modes generally have a higher frequency than the circular orbit tidal driving frequency and the effect is dominated by the adiabatic contributions characterized by the tidal deformability parameters~\cite{Flanagan:2007ix}. The tidal deformability $\lambda_\ell$ for each $\ell$-th multipole is defined in the limit that the timescale of variations in the tidal field are much faster than any internal timescales of the NS. It is the ratio of the induced multipole deformation in the NS's exterior spacetime $Q_\ell$ to the strength of the multipolar tidal field ${\cal E}_\ell$ associated with the curvature due to the companion as will be discussed in Sec.~\ref{sec:GWmodels_PN}:
\begin{equation}
Q_\ell =-\lambda_\ell \, {\cal E}_\ell
\end{equation}
The associated effect on GWs steeply increases with smaller separation or higher GW frequency, however, the information accumulates over all frequencies. 

\begin{figure}[t]
 \centering
 \includegraphics[width=.49\columnwidth]{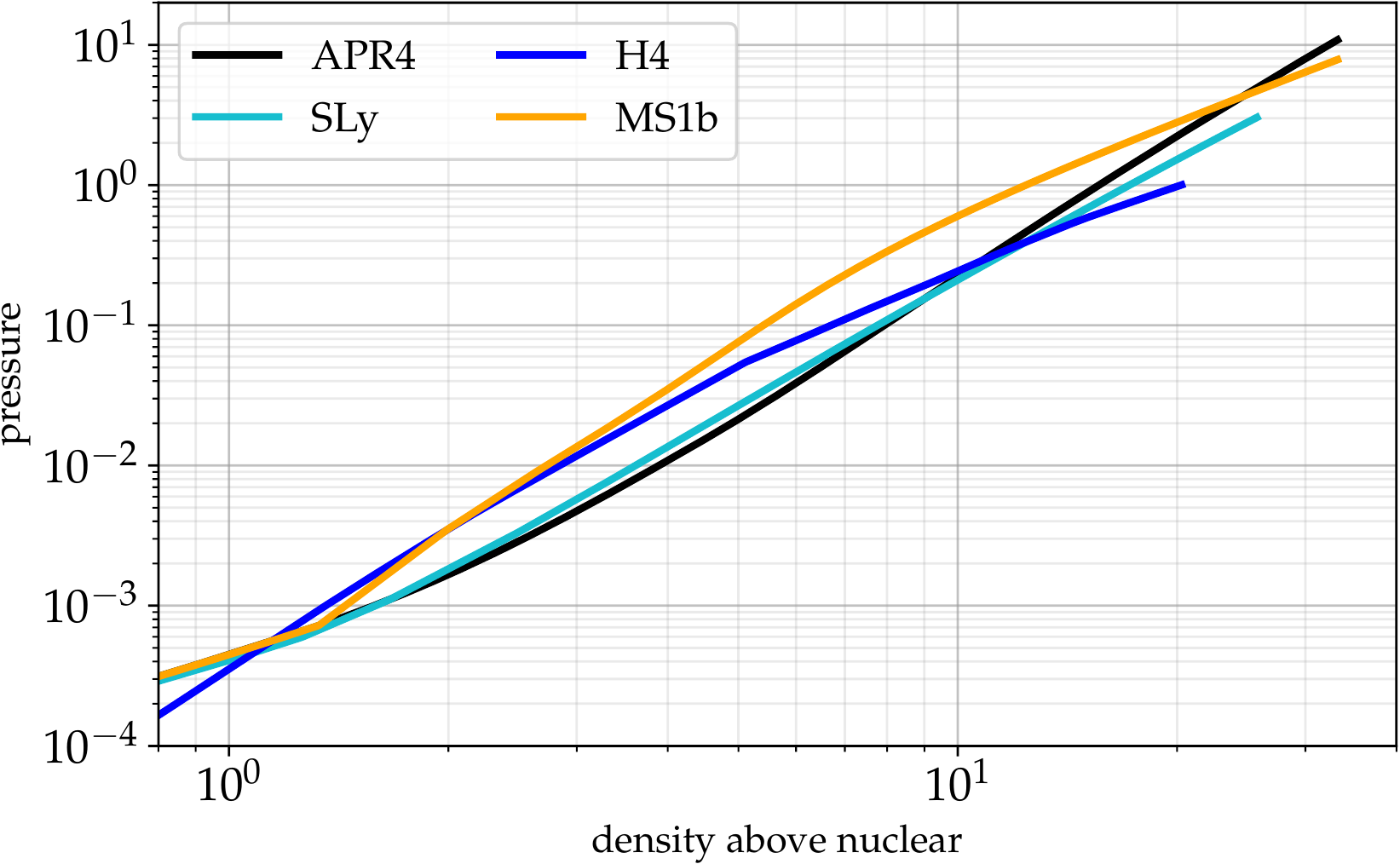}
  \includegraphics[width=.49\columnwidth]{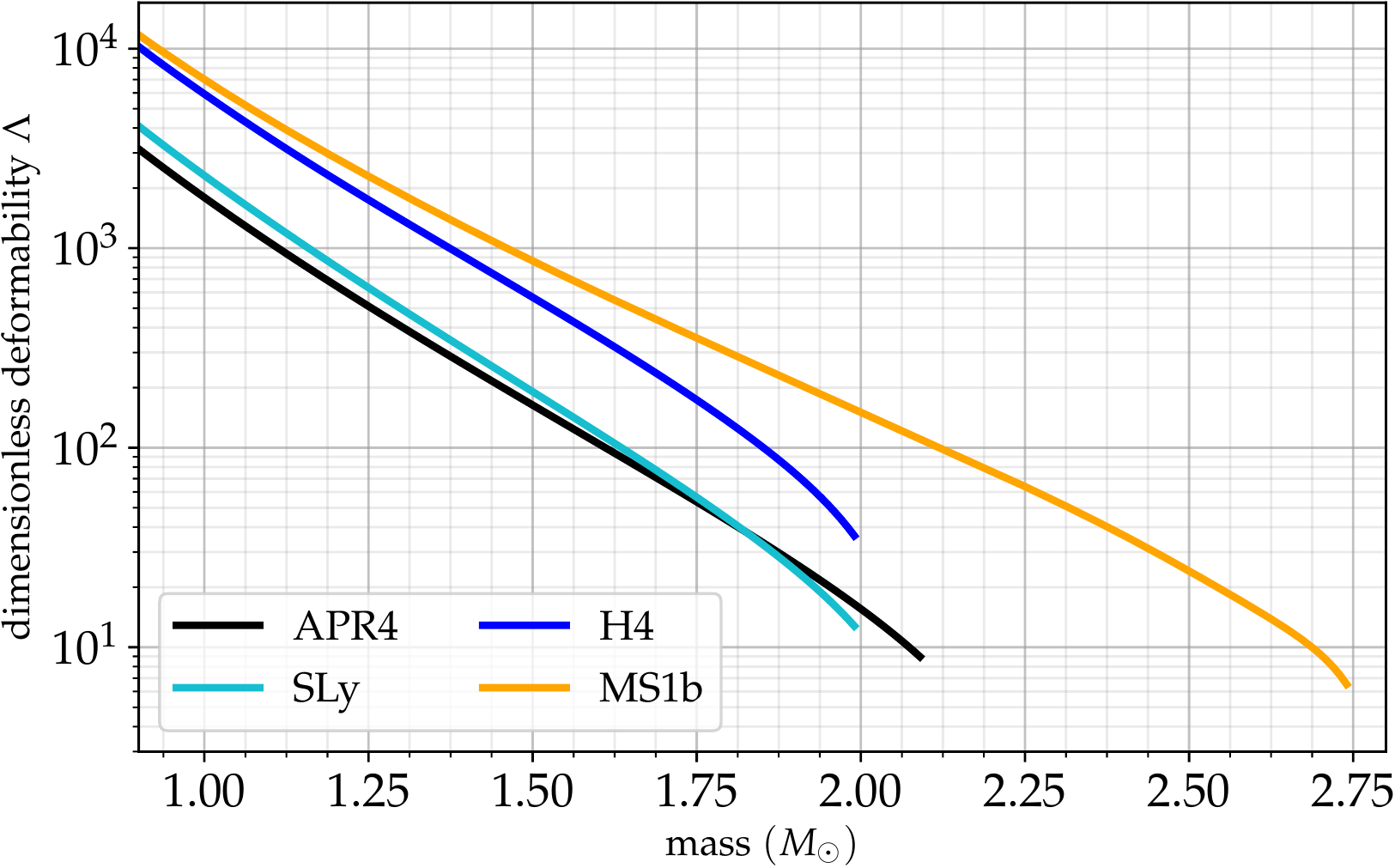}
  \caption{Examples of the link between the microphysics of cold, dense matter as encoded in the equation of state (EOS) and global NS properties such as tidal deformability as a function of mass. The connection occurs through the Einstein Field equations and energy-momentum conservation for linearized perturbations to an equilibrium NS configuration. The quantity $\Lambda=G\lambda_{\ell=2}(c^2/Gm)^5$ is a dimensionless tidal deformability coefficient characterizing the ratio of the induced quadrupole to the strength of the tidal field in the adiabatic limit. The colored curves correspond to a few illustrative EoS examples. The parameter that varies along the tidal deformability curves is the central density. For each EoS, there is a maximum central density and mass for which the material can resist gravitational collapse to a black hole. Adapted from Refs.~\cite{Hinderer:2009ca,GuerraChaves:2019foa}. }
 \label{fig:lambdavsmass}
\end{figure}

\subsubsection{Spin-induced multipole effects}
Multipole moments that couple to the companion's tidal field and lead to EOS signatures in GWs also arise from the NS's rotation. For rotating black holes, which are vacuum gravity in General Relativity, the multipole structure of the exterior spacetime is fixed by the no-hair property in terms of only their mass and spin~\cite{Carter:1971zc,Hawking:1971vc,Gurlebeck:2015xpa}:
\begin{equation}
\label{eq:nohair}
M^{\rm BH}_\ell+i S^{\rm BH}_\ell =m(i\chi m)^\ell,
\end{equation}
where $M^{\rm BH}_\ell$ and $S^{\rm BH}_\ell$ are the $\ell$-th mass and current multipole moments of the black hole, and $m$ and $\chi=cS/Gm^2$ are its mass and dimensionless spin.  
For material objects like NSs, the structure~\eqref{eq:nohair} no longer applies and instead the quadrupole and higher multipole moments also depend on the EOS of the matter. For instance, the spin-induced quadrupole is given by
\begin{equation}
M_{\ell=2}^{\rm NS}\approx -\kappa \chi^2 m^{3},
\end{equation} 
 The key characteristic matter-dependent coefficients $\kappa$ are the rotational Love numbers~\cite{Hartle:1967he,Laarakkers:1997hb,Poisson:1997ha,Mora04,Berti08}. Studies in Refs.~\cite{Yagi:2014bxa,Pappas:2013naa} have shown that the complicated multipole structure of a spinning NS can be approximated by three moments. The effects of spin-induced multipole moments in GWs are quadratic and higher order in the NS's spin (see~\cite{Krishnendu:2017shb} for the latest update on existing knowledge) but scale as a lower power of the frequency or orbital separation that fundamental-mode tidal effects. 

\subsection{Post-merger dynamics}

In contrast to BBH systems where the final state is always a BH, the postmerger of BNS allows for a variety of possible merger outcomes. In general the dynamics and the corresponding GW spectrum is complicated 
and influenced by several physical processes as thermal effects, turbulences, 
the magnetohydrodynamical instabilities, neutrino emissions, and possible phase transitions, 
e.g.,~\cite{Siegel:2013nrw,Alford:2017rxf,Radice:2017zta,Shibata:2017xht,Bauswein:2018bma,Most:2018eaw,DePietri:2018tpx,Ciolfi:2019fie,Weih:2019xvw}. 
Therefore, limited knowledge about this part of the BNS coalescence exists and  
NR studies are the only way of investigating this regime of the BNS coalescence. 
However, we note that even state-of-the-art simulations
contain just a subset of the important physical properties influencing the postmerger and 
lack a clean error budget during this part of the simulation due to shocks, turbulences, or 
discontinuities present at or after the merger; see e.g.~\cite{Most:2019kfe} for recent progress 
improving the accuracy of postmerger simulations. 

\begin{figure*}[t]
 \centering
 \includegraphics[width=1.\columnwidth]{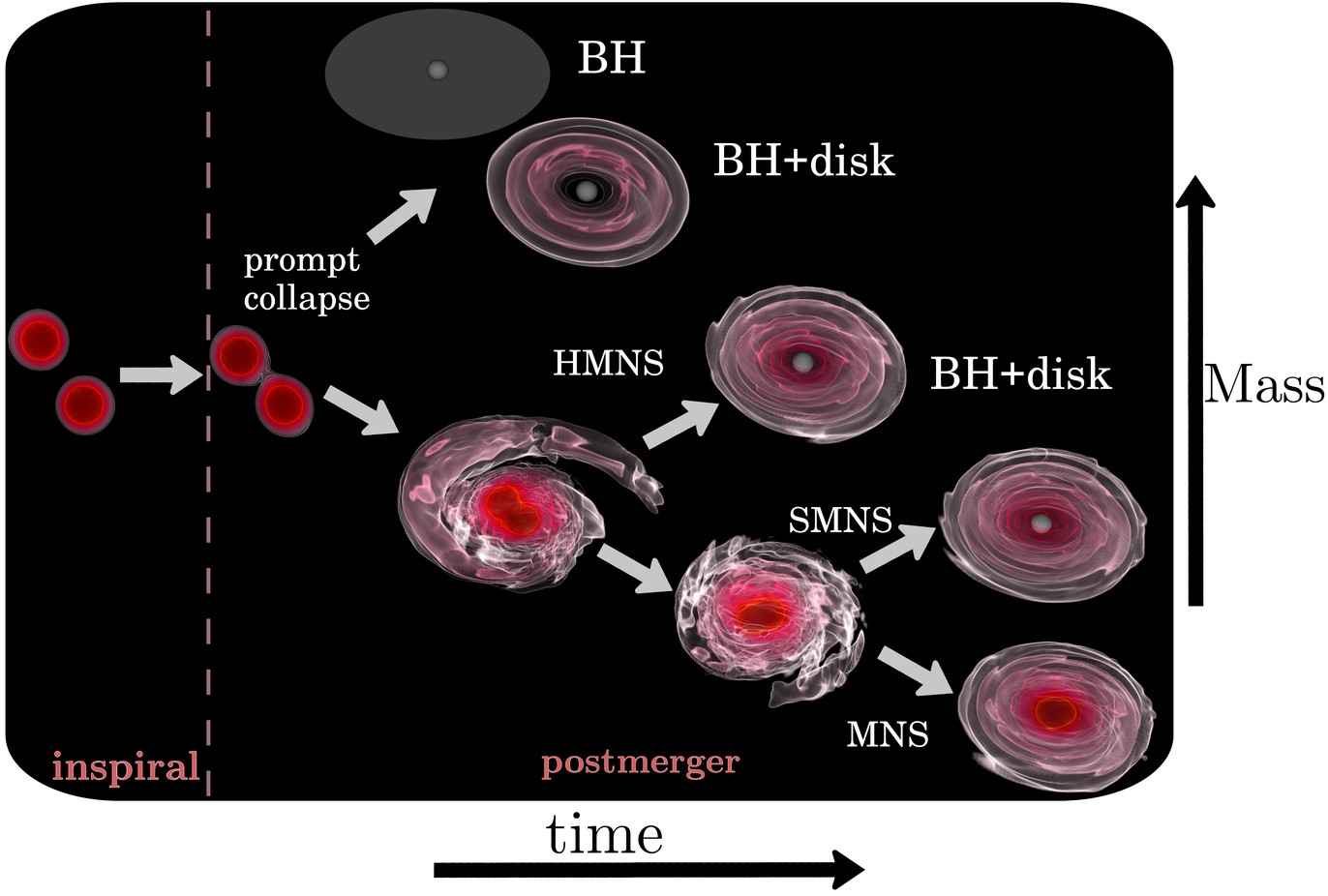}
 \caption{Overview about the possible postmerger dynamics and the possible postmerger outcome. 
 Mostly depending on the mass (but also the mass ratio and spin), the merger remnant 
 can promptly collapse to a BH, otherwise the 
 remnant can be a HMNS or SMNS collapsing on a longer timescale to a BH or, 
 if the mass is low enough, it can form a stable NS.}
 \label{fig:remnant_classification}
\end{figure*}

Independent of these caveats, the qualitative picture, 
as depicted in Fig.~\ref{fig:remnant_classification}, 
is robust and the possible merger outcomes are generally classified in four different categories~\cite{Baumgarte:1999cq}: 
prompt collapse, hypermassive NS (HMNS), supramassive NS (SMNS), and massive NS (MNS).

If the total mass of the binary is large enough, then the system will undergo a prompt collapse 
to a BH, i.e., the collapse happens within a few milliseconds after the merger. 
In this scenario, only a small amount of matter is ejected and the final BH 
is surrounded by a low mass accretion disk. 
Therefore, one does not expect bright EM counterparts, e.g,~\cite{Bauswein:2017vtn,Radice:2017lry,Margalit:2019dpi,Coughlin:2018miv,Dietrich:2020lps}, 
unless the system has a large mass ratio~\cite{Kiuchi:2019kzt}. 
The distinction between prompt collapse configurations and non-prompt collapse configuration is mainly determined 
by the ratio of the total mass $M$ and a given threshold mass $M_{\rm thr}$ as 
computed in~\cite{Bauswein:2013jpa,Koppel:2019pys,Agathos:2019sah}.

If the mass is smaller than the threshold mass and the baryonic mass 
is below the baryonic mass supported by a spherical, non-rotating star, 
the remnant is called a MNS. 
If the baryonic mass is below the maximum baryonic mass of a rigidly rotating star, 
it is called a SMNS, otherwise, if the remnant is supported by differential 
rotation, it is classified as a HMNS. 
It is important to emphasize that the latter three classifications were derived for 
cold-equilibrium configurations with zero temperature 
EOSs~\cite{Baumgarte:1999cq}, but has been applied to make qualitatively 
predictions, e.g.~\cite{Hotokezaka:2013iia,Zappa:2017xba}. 
A proper classification incorporating the temperature dependent effects are currently missing, 
but see~\cite{Kaplan:2013wra,Camelio:2019rsz} for first attempts. \\

While this review focuses on the inspiral part of the BNS coalescence, 
we will briefly present some characteristic features of the postmerger evolution, 
which for future generations of GW detectors might by important to place constraints 
on the EOS at densities and temperatures exceeding those which are probed during the inspiral~\cite{Radice:2016rys}. 

The postmerger waveform shows a non-monotonic amplitude and 
frequency evolution; cf.~Fig.~\ref{fig:intro}. 
The dominant and most robust feature of the postmerger GW signal
is the main emission frequency, called $f_2$ or $f_{\rm peak}$~\cite{Bauswein:2011tp,Clark:2014wua,Takami:2014zpa,Bernuzzi:2015rla,Bauswein:2015yca,Clark:2015zxa,Rezzolla:2016nxn,Tsang:2019esi}. 
In addition also a number of other, sub-dominant frequencies are present, e.g., Refs.~\cite{Clark:2015zxa,Bauswein:2015yca,Rezzolla:2016nxn}. 
In general, the frequency of the GW signal emitted during the postmerger phase is determined by 
the rotational state of the remnant and the EOS. Based on a large set of NR simulations, 
a number of groups have proposed different quasi-universal relations, 
which, in principle, can be used to determine the part of the EOS governing the 
remnant's density evolution~\cite{Radice:2016rys}. 
Recently, Refs.~\cite{Chatziioannou:2017ixj,Tsang:2019esi,Breschi:2019srl} employed a full Bayesian framework to 
investigate which SNR is required to extract information about the EOS from the 
postmerger GW signal and found that a minimum postmerger SNR of $3$ to $5$ is required. 
This corresponds to a total SNR of $100$ to $200$, which would allow very precise measurements 
about tidal effects from the inspiral GW signal. 

\section{Numerical Relativity Simulations}
\label{sec:GWmodels_NR}

NR simulations are a fundamental tool to study GWs from 
compact binaries in the late-inspiral and postmerger phase of the coalescence.
The main advantage of NR simulations is that the Einstein 
Equations can be solved directly by numerical 
discretization of the equations. 
Over the last years, the NR community has
made tremendous progress on several fronts:
(i) the exploration of the BNS parameter space by systematically varying
the EOSs, total mass, mass ratio, spin, and eccentricity, e.g.,\cite{Hotokezaka:2011dh,Bernuzzi:2014kca,Foucart:2015gaa,Bernuzzi:2015opx,Dietrich:2016hky,Lehner:2016lxy,Sekiguchi:2016bjd,Bernuzzi:2013rza,Tsatsin:2013jca,Kastaun:2013mv,Dietrich:2016lyp,East:2015vix,Kastaun:2016elu,Dietrich:2015pxa,Tacik:2015tja,Dietrich:2017xqb,Gold:2011df,East:2012ww,East:2016zvv,Radice:2016dwd,Papenfort:2018bjk,East:2019lbk,Tsokaros:2019anx,Most:2019pac,Ruiz:2019ezy,Kiuchi:2019kzt,Chaurasia:2020ntk};
(ii) the development of numerical schemes allowing many-orbits simulations with high-accuracy GWs, e.g.,
\cite{Bernuzzi:2011aq,Hotokezaka:2013mm,Radice:2013hxh,Bernuzzi:2016pie,Hotokezaka:2015xka,Dietrich:2017aum,Kiuchi:2017pte};
(iii) the increasing realism of microphysical schemes that incorporate 
magnetic effects, finite-temperature EOSs, composition effects, and neutrino
transport, e.g., \cite{Anderson:2008zp,Giacomazzo:2010bx,Rezzolla:2011da,Palenzuela:2013hu,Kiuchi:2014hja,Palenzuela:2015dqa,Ruiz:2017due,Ciolfi:2017uak,Kiuchi:2017zzg,Sekiguchi:2011zd,Galeazzi:2013mia,Neilsen:2014hha,Sekiguchi:2015dma,Palenzuela:2015dqa,Foucart:2016rxm,Foucart:2017mbt,Shibata:2012ty,Foucart:2015vpa,Foucart:2017mbt,Radice:2017zta,Shibata:2017jyf,Shibata:2017xht,Shibata:2017xdx,Fujibayashi:2017puw,Most:2019kfe};
(iv) the study of mass ejecta and EM counterparts, e.g., \cite{Hotokezaka:2012ze,Bauswein:2013yna,Wanajo:2014wha,Lehner:2016lxy,Radice:2016dwd,Dietrich:2016hky,Dietrich:2016lyp,Foucart:2016rxm,Fujibayashi:2017xsz,Perego:2017wtu,Hotokezaka:2018aui,Bovard:2017mvn}. 

In the following, we will focus on the computation of the GW signal and therefore, 
purely on GRHD simulations. We refer the interested reader to, e.g., \cite{Baiotti:2016qnr,Baiotti:2019sew,Rezzolla:2013gwa,Rezzolla:2013}, 
for discussions about effects of magnetic fields or neutrino radiation.

\subsection{Imprint of the binary properties and waveform catalogs}

\begin{figure}[t]
\includegraphics[width=\columnwidth]{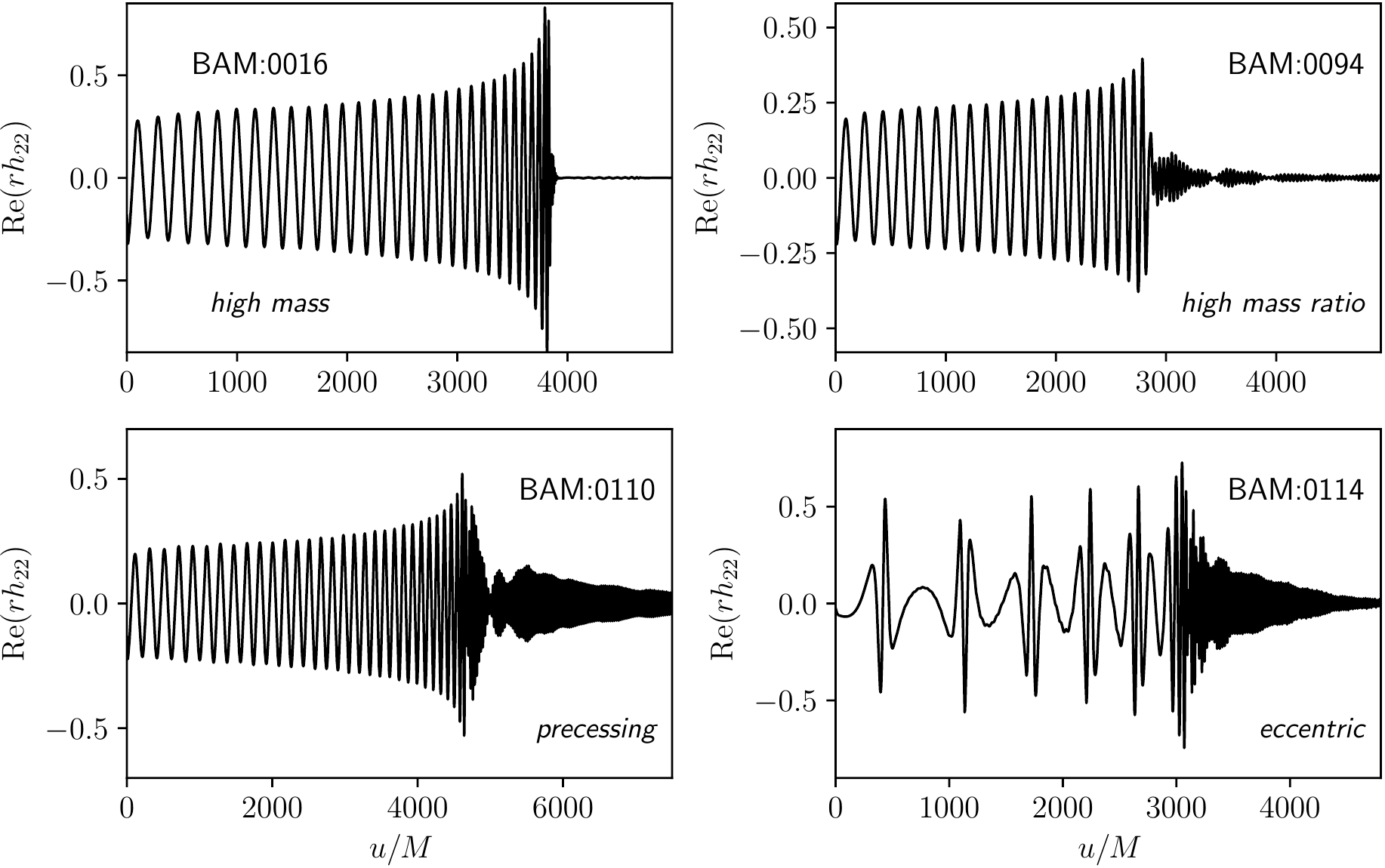}
\caption{GW signals for a variety of different binary parameters: a high total mass setup, 
a high mass-ratio configuration, a precessing simulation, 
and a highly eccentricity configuration.
The simulations are taken from the CoRe database and labeled accordingly. 
Adapted from~\cite{Dietrich:2018phi}.}
\label{fig:waveform_overview}
\end{figure}

A number of different groups have performed a large set of BNS simulations over the last years. 
The most ambitious simulation campaigns have led to the construction 
of the CoRe database\footnote{\texttt{www.computational-relativity.org}}~\cite{Dietrich:2018phi} 
and the SACRA Gravitational Waveform Data 
Bank\footnote{\texttt{https://www2.yukawa.kyoto-u.ac.jp$\sim$nr\_kyoto /SACRA\_PUB/catalog.html}}~\cite{Kiuchi:2019kzt}. 
Both catalogs combine several hundred simulations and are freely available. 
In addition large simulation campaigns have been performed by other groups, e.g., Refs.~\cite{Hotokezaka:2013iia,Takami:2014tva,Bauswein:2013yna}.
All these simulations allow the NR community to obtain a better understanding of the imprint 
of the individual binary properties on the merger process. 
In the following, we want to discuss shortly the imprint of the total mass, 
mass ratio, spin, and eccentricity. 

\subsubsection{Total mass}

In contrast to BBHs, BNS systems are not scale invariant and therefore can 
not be rescaled by the total mass since the mass enters in the computation 
of tidal interactions during the inspiral and determines the merger outcome; 
cf.~Fig.~\ref{fig:remnant_classification}. 

Based on the constraints derived from GW170817 one expects that 
the mass of an individual NS lies within 
$M \in [1.0,2.3 M_\odot]$~\cite{Lattimer:2012nd,ozel:2016oaf,Rezzolla:2017aly,Shibata:2017xdx,Ruiz:2017due,Margalit:2017dij,Shibata:2019ctb}. 

As discussed in Sec.~\ref{sec:coalescence}, 
systems with large total masses are typically characterized by a possible 
prompt collapse directly after the moment of merger. Such a system will be 
very similar to a BBH mergers, see e.g.~\cite{Haas:2016cop}. 

In addition to this prominent postmerger evolution, which is more similar to the BBH dynamics 
(top panel of Fig.~\ref{fig:waveform_overview})
than a typical BNS merger, massive NSs will have small tidal deformabilities
and consequently tidal effects are for massive BNS mergers less important than for 
lower-mass systems; cf.\ Fig.~\ref{fig:lambdavsmass}. 
Consequently a distinction between high mass BNS and low mass BBH 
systems will be problematic. This problem becomes even more pronounced since massive 
BNS mergers eject only a small amount of material and, thus, create no bright EM counterpart. 
Therefore, the correct astrophysical classification is complicated and 
is an active field of research, see e.g.~\cite{Chen:2020fzm}. 

\subsubsection{Mass ratio}

Based on population synthesis models 
a maximum mass ratio [$ q = \frac{M_A}{M_B} \geq 1$] of 
$q_{\max} \simeq 1.8 {-} 1.9$~\cite{Dominik:2012kk,Dietrich:2015pxa} 
might be possible for a BNS system. While observation of isolated NSs 
distribution would theoretically allow maximum mass ratios of $q_{\max} \simeq 2$, 
the largest observed mass ratio in BNSs is
$q\sim 1.3$~\cite{Martinez:2015mya,Lazarus:2016hfu}. 
Over the last years, the NR community managed to simulate systems with large $q$, 
at the boundary of physically realistic scenarios, see~\cite{Dietrich:2015pxa,Tichy:2019ouu} 
for simulations with $q\gtrsim 2$. 

With increasing $q$, one finds that the emitted GW energy decreases 
during both the inspiral and the postmerger phase; cf.\ Fig.~14 
of \cite{Dietrich:2016hky} and that the merger frequency decreases 
roughly as a function of $\propto q^{-1/2}$, cf.~\cite{Dietrich:2018uni}~\footnote{Alternatively, Ref.~\cite{Breschi:2019srl} finds a correlation dependent on the symmetric mass ratio $\nu$.}.

Considering the post-merger evolution, one finds that, while the EOS 
determines most of the evolution, systems with larger mass ratios tend to 
have merger remnants which seem to survive slightly longer before gravitational collapse and BH formation. 
In combination with the longer lifetime, the mass of the formed accretion disk increases with 
increasing $q$~\cite{Dietrich:2016hky,Kiuchi:2019kzt,Dietrich:2020lps}. 
Due to the large disk and ejecta mass of unequal mass systems 
one expects them to have brighter EM counterparts. 
In addition, one finds that systems in which the mass ratio is large enough to tidally disrupt 
the secondary star before the merger, the postmerger GW amplitude is significantly reduced, 
cf.~Fig.~\ref{fig:waveform_overview} (second panel) and~\cite{Tsang:2019esi}. 

\subsubsection{Spin}

The maximum NS spin is not precisely known, since it depends on
the unknown EOS, but EOS models predict breakup spins of up to 
$\chi \sim0.7$, which corresponds to spin periods of less than 1
ms~\cite{Lo:2010bj}. 

Observationally, a large number of highly-spinning NSs are known. 
The pulsar PSR J1748$-$2446ad~\cite{Hessels:2006ze} is, to date, the
fastest spinning NS with a frequency of $716$~Hz. 
PSR J1807-2500B~\cite{Lynch:2011aa} is the fastest spinning NS in a binary system with 
a rotation frequency of $239$~Hz and PSR J1946+2052~\cite{Stovall:2018ouw} 
is the fastest spinning NS in a BNS system ($59$~Hz). 
Due to magnetic dipole radiation, J1946+2052 will spin down and will  
presumably have a dimensionless spin of about $\chi \sim 0.05$ at merger. 

In the past, consistently evolved spinning BNS simulations have 
been presented in Refs.~\cite{Bernuzzi:2013rza,Dietrich:2015pxa,Dietrich:2016lyp,Dietrich:2018upm,Most:2019pac,Tsokaros:2019anx,East:2019lbk}
and precessing NR simulations are presented only in~\cite{Dietrich:2015pxa,Tacik:2015tja,Dietrich:2017xqb,Chaurasia:2020ntk}.

With respect of the dynamical evolution and energetics of spinning NS binaries during the inspiral one 
finds that even for astrophysical realistic spin magnitudes ($\chi \lesssim 0.05$)
spin-orbit interactions are larger than tidal effects up to or shortly before the contact 
of the two stars~\cite{Bernuzzi:2013rza}. Therefore, neglecting spin effects might lead to 
a wrong description of the binary evolution if one of the two binary constituents is a pulsar, 
which would introduce biases during the determination of tidal parameters from GW observations, 
e.g.~\cite{Favata:2013rwa,Agathos:2015uaa,Samajdar:2019ulq}.

Studying the energetics during the postmerger, one finds that spin aligned systems 
can create merger remnants with larger angular momentum than non-spinning systems. 
Consequently, collapse to a BH happens at later times~\cite{Chaurasia:2020ntk}. 
We note that simulations also indicate that characteristic frequencies in the postmerger, 
which generally also follow quasi-universal 
relations~\cite{Bauswein:2014qla,Bauswein:2015yca,Takami:2014tva,Takami:2014zpa,Bernuzzi:2015rla,Clark:2015zxa,Rezzolla:2016nxn}, 
might shift under the presence of spin. 

\subsubsection{Large Eccentricities}

While most NS binaries are expected to be formed out of an already existing 
binary system, there is a chance of producing BNS systems 
by dynamical capture in dense stellar environments such as globular clusters~\cite{Lee:2009ca}.
Those binaries can have a non-negligible eccentricity when they enter 
the LIGO frequency band. Unfortunately, estimates for the rates of eccentric BNS mergers
are highly uncertain. Ref.~\cite{Tsang:2013mca} gives a conservative 
estimate of a BNS merger rate of $0.5 {\rm yr}^{-1} {\rm Gpc}^{-3}$. 

While eccentric BNS mergers will occur infrequently, the detection of at least one of these events 
would allow for an intensive study of the strong-field regime and the properties of NS matter. 
Of particular interest for highly eccentric systems is the possibility to constrain the 
EOS during the inspiral from density oscillations induced by close encounters 
of the two stars. In fact, when the NSs approach each other, the induced tidal deformation 
of the star causes a mode excitation. 
Once excited, most energy is released through the $f$-mode. 
Consequently a measurement of this excitation frequency 
would allow to place additional constraints on the unknown supranuclear EOS, 
cf.~e.g.~\cite{Turner:1977b,Kokkotas:1995xe,Gold:2011df,Moldenhauer:2014yaa,Chirenti:2016xys,Parisi:2017kgx,Yang:2018bzx,Chaurasia:2018zhg}. 
 Furthermore, since information of the $f$-mode allows to place constraints 
on the binary properties, an accurate description also allows to understand possible 
resonance effects when the orbital motion approaches 
the $f$-mode frequency and due to a resonances~\cite{Steinhoff:2016rfi,Schmidt:2019wrl,Andersson:2019dwg}. 

\subsection{High-quality NR simulation}

\begin{figure}[t]
 \centering
 \includegraphics[width=0.7\columnwidth]{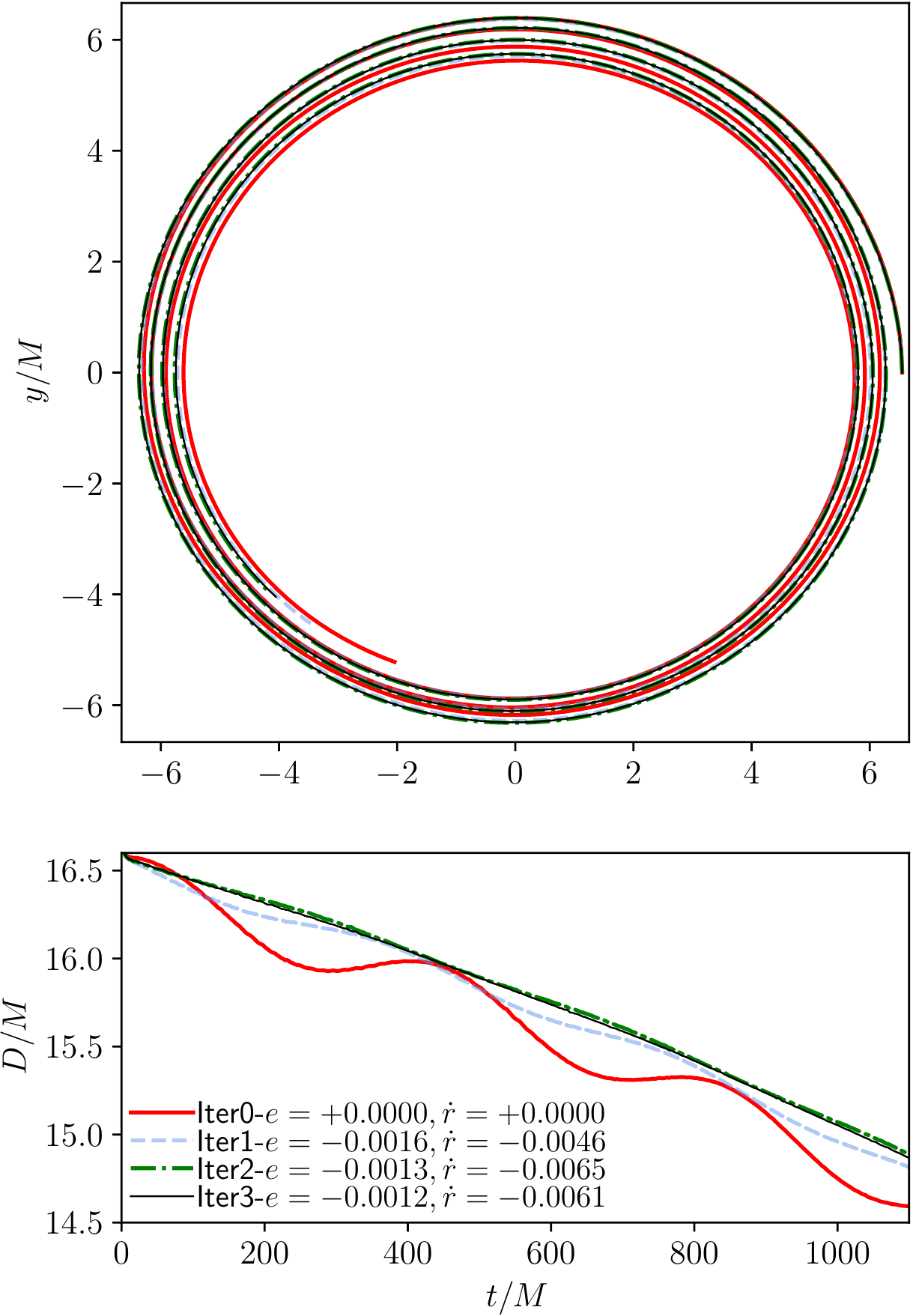}
 \caption{Eccentricity reduction procedure. 
 Top panel: NS track for one star for the different iteration steps during the 
 eccentricity reduction. The configurations need to be evolved for a few orbits 
 to estimate the parameters $(e,\dot{r})$ that are required for the next iteration step. 
 Bottom panel: Proper distance of the system as a function of time. 
   The artificial eccentricity is clearly visible as oscillations 
   in the proper distance. }
 \label{fig:ecc_red}
\end{figure}

As pointed out, there have been more than a thousand individual BNS simulations 
performed over the last decades, however, up to now, there is only a limited number of simulations 
which meet the accuracy standards necessary to develop and test existing GW models; see e.g.~\cite{Radice:2013xpa,Bernuzzi:2016pie,Hotokezaka:2015xka,Tsokaros:2016eik,Dietrich:2017aum,Kiuchi:2017pte,Dietrich:2019kaq,Most:2019kfe,Kiuchi:2019kzt} for further discussion. 
The main difficulty for the production of high-quality waveforms are: 
\begin{enumerate}[(i)]
 \item the reduction of artificial eccentricity; 
 \item the usage of high enough resolutions to reduce discretization errors; 
 \item the development of numerical schemes showing clear convergence across multiple resolutions and configurations. 
\end{enumerate}  

\subsubsection{Eccentricity reduction}

As an exemplary case addressing all listed items, 
we present a high-resolution simulation of~\cite{Dietrich:2019kaq}. 
The setup that we pick has been simulated for six different resolutions and  
describes an equal mass, non-spinning configuration with NS masses of 
$M_{A}=M_{B}= 1.35$ employing a piecewise-polytropic 
fit of the SLy EOS~\cite{Read:2008iy,Dietrich:2015pxa}. 
The eccentricity reduced initial configurations are obtained with the SGRID 
code~\cite{Tichy:2006qn,Tichy:2009yr,Tichy:2009zr,Dietrich:2015pxa,Tichy:2019ouu}. 
SGRID is one of the few codes able to produce data with 
arbitrary eccentricities. Up to our knowledge, other codes capable of 
computing low-eccentric data are SpEC's spectral solver 
SPELLs~\cite{Foucart:2008qt,Tacik:2015tja} and the
non-public extension of LORENE~\cite{lorene_web} discussed in Ref.~\cite{Kyutoku:2014yba}. 

The computation of low-eccentric data relies on an iterative procedure in which 
the binary's initial radial velocity and the eccentricity parameter (or the orbital frequency) 
are varied such that during an evolution the eccentricity reaches a certain threshold, 
see~\cite{Dietrich:2015pxa,Tichy:2019ouu}.

Figure~\ref{fig:ecc_red} shows the eccentricity reduction procedure 
for this configuration. 
In most cases it is sufficient to perform $2$ to $4$ iteration steps to 
achieve final eccentricities $\lesssim 10^{-3}$.

\subsubsection{Errors through numerical discretization}

The configurations which we want to discuss in the following 
have been simulated with the BAM 
code~\cite{Bruegmann:2006at,Thierfelder:2011yi,Dietrich:2015iva,Bernuzzi:2016pie} 
for six different resolutions covering the individual NSs with 
$64$, $96$, $128$, $192$, $256$, and $320$ points respectively. 

The highest resolution ($320$ points in the finest refinement level) 
has a spatial resolution of $0.047 M_\odot \approx  70 \ \rm m$ in 
the finest refinement boxes. 
The computational cost for all six resolutions is about $\sim 8$~million CPU-hours, 
where as for all 3+1-NR codes the computational expenses increase 
with the fourth power of the number of grid points, i.e., the $320$ point simulation requires 
more than $800$ times the resources compared to the lowest resolution setup.  

\begin{figure}[t]
 \centering
 \includegraphics[width=0.8\columnwidth]{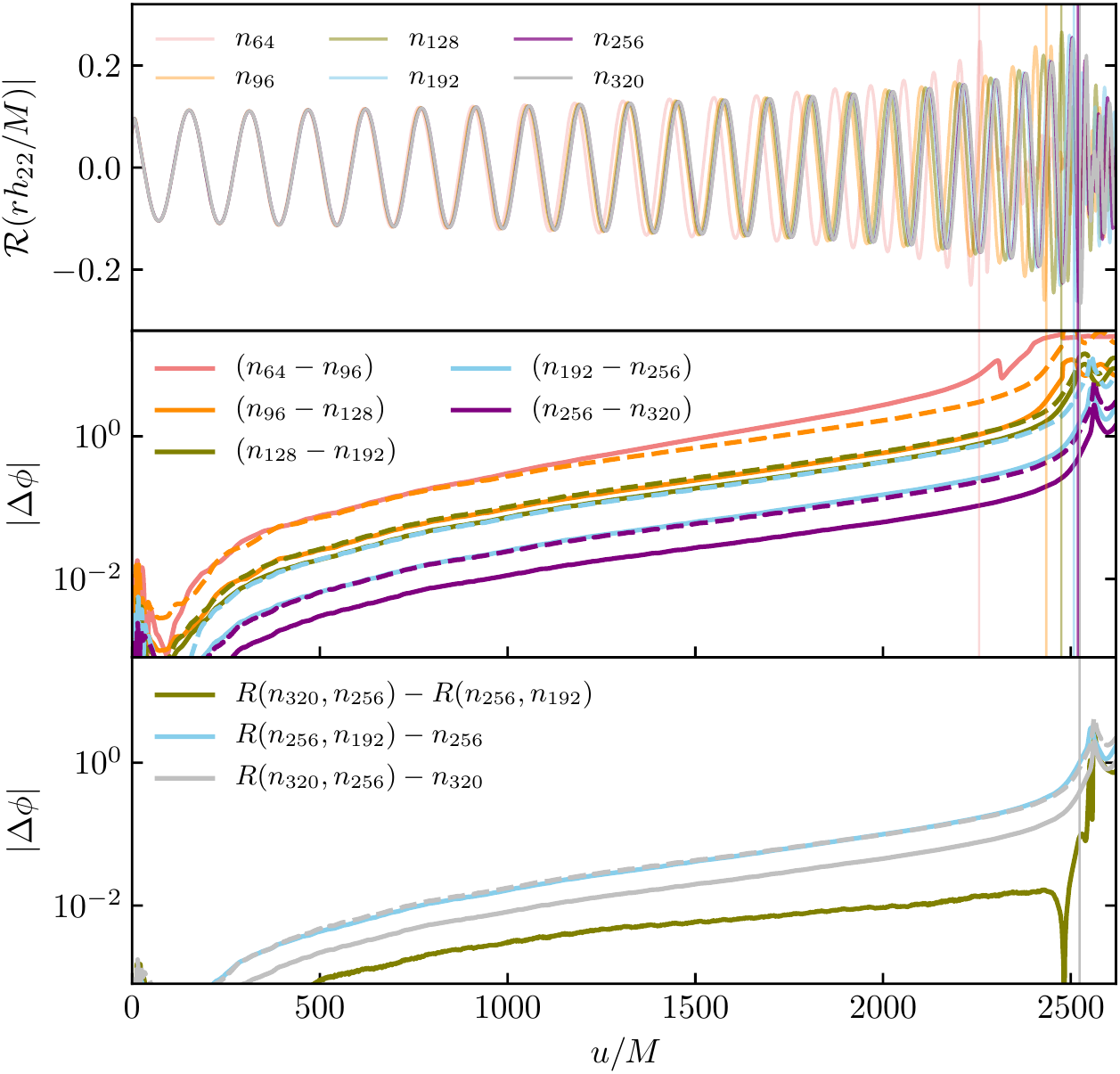}
 \caption{Top panel: Real part of the GW signal for the six different resolutions
employing $64$, $96$, $128$, $192$, $256$, and $320$ points in the refinement levels covering
the individual NSs as a function of the retarded time $u$.
Middle panel:
Phase difference between different resolutions.
Bottom panel:
Phase difference between different Richardson extrapolated
waveforms or between a Richardson extrapolated
waveforms and the waveform from an individual resolution.
The vertical lines in each panel refer to the time of merger, i.e.,
the peak time of the GW amplitude for the individual resolutions.
The dashed lines in the bottom two panels show the phase difference scaled to the next lowest pair of resolutions
assuming second order convergence. Adapted from~\cite{Dietrich:2019kaq}.}
 \label{fig:NR_highacc}
\end{figure}

The real part of the GW signal for all resolutions is shown in the top panel of Fig.~\ref{fig:NR_highacc}.
It is clearly visible that the merger time increases with an increasing resolution. 
This behavior is understandable since numerical dissipation, which decreases with an increasing 
resolution, adds dissipation and accelerates the inspiral. 
Therefore, the merger for an `infinite resolution' would be later than for the highest resolution 
shown in Fig.~\ref{fig:NR_highacc}. 
Because of the presence of a clean convergence order, as visible in the middle panel, one can use a Richardson 
extrapolation to obtain a better guess for such an `infinite resolution' scenario. 
In the bottom panel we show among others the phase difference between the Richardson extrapolation 
using the highest resolutions, $R(n_{320},n_{256})$, and the phase of the highest resolution $n_{320}$.

In addition, one can check the robustness of the extrapolation by computing 
the phase differences $R(n_{320},n_{256})-n_{320}$ and $R(n_{256},n_{192})-n_{256}$
in the bottom panel of Fig.~\ref{fig:NR_highacc}. Rescaling the phase difference
of $R(n_{320},n_{256})-n_{320}$ assuming second order convergence shows again
excellent agreement with $R(n_{256},n_{192})-n_{256}$. This proves that the 
leading error term scales quadratically with respect to the resolution. 

At merger, the phase difference $R(n_{320},n_{256})-n_{320}$ is only $0.37\ \rm rad$ and provides a good error 
estimate of our simulation. 
An alternative, but non-conservative error can be approximated by the difference between two Richardson 
extrapolated waveforms (green line in the bottom panel).
Throughout the whole inspiral, the difference between $R(n_{320},n_{256})$ and $R(n_{256},n_{192})$ 
is below $0.1\ \rm rad$.
However, due to the non-monotonicity of $R(n_{320},n_{256})-R(n_{256},n_{192})$ it is unclear to what extent higher-order error terms are properly modeled. \\

\subsubsection{Finite radius extraction}

In addition to the discretization errors, one has to consider for the full error budget of a simulation
that the GWs are extracted at a finite extraction radius, while in principle GWs should be extracted at 
future null infinity. We refer the interested reader to~\cite{Bishop:2016lgv} and references therein.

The best solution to obtain properly extracted results is the Cauchy Characteristic Extraction 
(CCE)~\cite{Bishop:1996gt,Reisswig:2009us}, see also~\cite{Babiuc:2005pg,Handmer:2016mls}. 

Another option and computationally cheaper approximation is the use of a polynomial extrapolation. 
For this purpose, the GWs are extracted at different radii 
$r_i$ with $i=0...N$ and then the phase and amplitude are extrapolated 
using a polynomial of the order $K$ with $K<N$: 
\begin{equation}
f(u; r_j) = f_0(u) + \sum^K_{k=1} f_k(u) r^{-k}_j \ .
\label{eq:poly_extrar}
\end{equation}
Using this extrapolation method and estimating the error as the difference between the extrapolated and non-extrapolated waveform, the finite radius extraction error is below $0.044\ \rm rad$ 
for our example shown in Fig.~\ref{fig:NR_highacc}. 

Another possibility is it to consider the next-to-leading-order (NLO)
falloff behavior of the $\Psi_4$ multipoles~\cite{Lousto:2010qx,Bernuzzi:2016pie}, 
\begin{equation}
\label{psilm_extrar}
r\,\psi_{\ell m} = r\,\ddot{h}_{\ell m} +
\frac{\ell(\ell+1)}{2r}\; r\,\dot{h}_{\ell m} + \O(r^{-2}) \ , 
\end{equation}
and obtains an extrapolation formula for $r\,\ddot{h}_{\ell m}$ which uses
$\psi_{\ell m}$ extracted at a given radius. 

It is interesting to note that, as shown explicitly in 
Fig.~12 of \cite{Bernuzzi:2016pie}, the finite extraction radius decreases 
for an increasing GW frequency. 
Thus, it is smaller at the moment of merger than during 
the early-part of the inspiral. 

\section{Post-Newtonian Waveform modelling}
\label{sec:GWmodels_PN}
The PN modeling of GWs from NS binary systems has significantly advanced in recent years. The description of spinning point-masses in PN theory, which applies for black holes up to the absorption of GWs by their horizons, have been iterated to high order. Progress has also been made on elucidating several subtleties that arise in higher-order calculations and on developing appropriate regularization schemes within different approaches for computing the dynamics. We refer the reader to the review article~\cite{Blanchet:2013haa} for a comprehensive list of references on this decades-long effort; for other reviews see ~\cite{Futamase:2007zz} and ~\cite{Schafer:2018kuf}. To date, the dynamics have been computed to 4PN order with three classes of PN methods: a Fokker-action approach, with the most recent new results obtained in~\cite{Marchand:2017pir,Bernard:2017bvn,Bernard:2017ktp}, a Hamiltonian approach~\cite{Damour:2016abl,Damour:2014jta}, and effective field theory methods~\cite{Foffa:2016rgu,Foffa:2019yfl,Levi:2018nxp,Levi:2017kzq,Blumlein:2020pog}. Other approaches including post-Minkowski expansions and further applications of methods from high-energy theory have garnered much recent interest and seen rapid developments that already enabled computing higher order results, e.g.~\cite{Bini:2019nra}, but will not be discussed here. For high-order computations of the gravitational radiation, however, only results from a single method, the multipolar post-Minkowski expansion~\cite{Blanchet:1986sp,Blanchet:2013haa,Thorne:1980ru}, are currently available. Other approaches have only been carried out up to second order, such as the direct integration of the relaxed Einstein equations~\cite{Blanchet:1996pi,Blanchet:1995ez,EpsteinWagoner} and effective field theory methods~\cite{Leibovich:2019cxo} based on tools developed in~\cite{Galley:2009px,Maia:2017yok,Galley:2015kus,Galley:2012qs,Birnholtz:2014fwa,Birnholtz:2013nta,Levi:2018nxp,Porto:2012as,Ross:2012fc,Foffa:2011np,Goldberger:2009qd,Goldberger:2007hy}. 

In this section, we will focus the discussion on modeling the additional contributions from the presence of matter on top of the point-mass descriptions. We will give details only for a few dominant effects that are included in current waveform models used in data analysis, specifically only the quadrupole and octupole tidal effects from the fundamental modes and their adiabatic limits.  As outlined in the Introduction, analytical modeling of the binary inspiral is based on a tapestry of approximation schemes for the dynamical spacetime describing the binary system. These perturbative results can then be re-summed and combined with additional information from test-particle limits and NR into EOB models as will be described in the next section. 

The starting point for approximations is to divide the problem into GW generation and propagation. In the distant wave zone, at distances from the binary much larger than the wavelength $\lambda$ of the emitted GWs as well as the binary's size and gravitational radius, and outside the regions of significant spacetime curvature produced by the binary, the perturbation expansion is a post-Minkowski multipolar scheme that yields a solution for the radiation field~\cite{Blanchet:2013haa}. The source of that radiation field considered here is a nonspinning  binary system with comparable masses on quasi-circular orbits in the regime where the orbital separation between the bodies $r$ is large compared to their characteristic size $R$. The orbital dynamics can be approximated by a PN expansion with dimensionless parameter $\epsilon \sim G M/rc^2\sim v^2/c^2 \ll O(1)$ and is dominated by point-mass contributions. Small corrections from finite-size effects are included through a tidal expansion in the parameter $\alpha=R/r$. This yields a double expansion where $\epsilon$ and $\alpha$ are treated as distinct parameters that correspond to different physical effects, although they scale in the same way with the orbital separation~\cite{Flanagan:1997fn}. As we will discuss below, the finite size effects involve characteristic matter parameters such as tidal deformabilities that enter as coupling constants in the effective action for the dynamics and influence the radiative multipole moments of the binary and hence are imprinted in the GWs~\cite{Flanagan:2007ix,Hinderer:2007mb}. The link between these constants and the microphysics of matter are determined by considering the physics in the body zone, close to each NS, where the self-gravity of the NS is strong but the influence from the distant companion is a small perturbation. Solving the linearized Einstein field equations for perturbations to the NS's equilibrium configuration yields the perturbed NS matter distribution and spacetime geometry, and enables extracting the knowledge of the characteristic matter coefficients.

In this framework, Newtonian quadrupolar tidal effects in the dynamics first appear at $O(\alpha^5)$. Relativistic corrections have been computed up to $O(\alpha^5\epsilon^2)$ for the quadrupole and $O(\alpha^7\epsilon^2)$ for the octupole~\cite{Henry:2019xhg,Bini:2012gu}. The knowledge of effects in the GW signals will be described in more detail below. We also note that the scalings discussed above apply only for~\emph{adiabatic} tidal effects, and in particular, the resonant excitation of a generic mode does not fit into that counting scheme.

For simplicity we first consider a NS with a point-mass companion. To linear order in the finite-size effects, the case of two extended objects can be recovered by adding the same contribution with the body labels interchanged. We will assume that the bodies are nonspinning and spherically symmetric in isolation. To motivate the relativistic calculations of multipole moments and their impact on the dynamics and GWs, we will first discuss the Newtonian description expressed in a form that carries over to General Relativity (GR) with appropriate modifications. We will specialize to the same assumptions made in models that are currently well-developed for data analysis, and limit the details to the dominant mass-quadrupolar tidal effects for simplicity; extensions to higher mass multipole moments are straightforward. At the end of this subsection we will discuss various other effects that have been studied and are expected to give rise to interesting subdominant contributions.

\subsection{Multipole expansion: NS's mass moments and tidal moments }

Below, we will first introduce the multipole moments generated by the NS itself, then consider the multipolar tidal moments due to the companion that are felt by the NS.

\subsubsection{ NS's mass multipole moments in Newtonian gravity}
In Newtonian gravity the gravitational potential $U$ generated by a mass density $\rho$ at a field point $\bm{x}$ is a solution to Poisson's equation $\nabla^2 U=-4\pi\rho$. The Green function solution is:
\begin{equation}
U(t,\bm{x})=\int d^3 x^\prime \frac{\rho(t, {\bm x}^\prime)}{|{\bm x}-{\bm x}^\prime|}. \label{eq:Uintegral}
\end{equation}
For points $\bm{x}>\bm{x}^\prime$ outside the mass distribution we can perform a Taylor series expansion around a moving reference point $\bm{z}(t)$ to write Eq.~\eqref{eq:Uintegral} as
\begin{equation}
\label{eq:Uselfexpansion}
U=\int d^3 x^\prime \rho(t, {\bm x}^\prime) \sum_{\ell=0}^\infty \frac{(-1)^\ell}{\ell !} (x^\prime-z)^L\partial_L\frac{1}{|{\bm x}-{\bm z}|}. 
\end{equation}
Here, $L = a_1 a_2 \cdots a_\ell$ denotes a string of $\ell$ indices,
$x^L=x^{a_1} \cdots x^{a_\ell}$, and the summation over repeated indices is implied. Likewise, the notation for $\ell$ derivatives is $\partial_L=\partial_{ x^{a_1}} \ldots \partial_{ x^{ a_\ell}}$ with $\partial_x=\partial/\partial x$. 

The expansion of the potential~\eqref{eq:Uselfexpansion} can be written in a more suggestive form by defining the NS's Newtonian mass multipole moments by the following integrals
\begin{equation}
M_{\rm Newt}=\int d^3x\;\rho (t, \bm{x}), \ \ \ \ \ \ \ \ 
 Q^{L}_{\rm Newt}=\int d^3x\;\rho(t, \bm{x}) (x-z)^{<L>}, \label{eq:Newtmultis}
\end{equation}
where $M_{\rm Newt}$ is the mass monopole and $Q^L$ the $\ell$-th multipole moment. The angular brackets around indices denote the symmetric and trace-free projection of a tensor. For example, $x^{<ij>}=x^ix^j-\delta^{ij}|\bm{x}|^2$, see Refs.~\cite{Thorne:1980ru,Hartmann94} for more details. We have defined the moments as their trace-free parts because the contribution to Eq.~\eqref{eq:Uselfexpansion} involves the derivative $\partial_L |{\bm x}-{\bm x}^\prime|^{-1}$ which projects out the trace-free piece $Q^L$. Because our reference point $z^i(t)$ was chosen as the NS's center of mass the dipole term ($\ell=1$) is absent from the expansion~\eqref{eq:Newtmultis}. With these definitions the exterior potential takes the form
\begin{equation}
\label{Uself}
U(t,\bm x)=\frac{m}{|{\bm x}-{\bm z}|}+\sum_{\ell=2}^\infty \frac{(-1)^\ell}{\ell!}Q^L\partial_L\frac{1}{|{\bm x}-{\bm z}|}.
\end{equation}
This expression in terms of Cartesian tensors is useful for computing the dynamics of a binary system. When calculating the deformation of the NS, it is more convenient to use spherical coordinates $(x-z)^i/|{\bm x}-{\bm z}|=( \sin\theta\cos\phi,\sin\theta\sin\phi,\cos\theta)$ and express \eqref{Uself} as an equivalent spherical-harmonic expansion
\begin{equation}
U(t,\bm x)=\frac{m}{|{\bm x}-{\bm z}|}+\sum_{\ell=2}^{\infty}\sum_{m=-\ell}^\ell Q_{\ell m}\frac{Y_{\ell m}(\theta, \phi)}{|{\bm x}-{\bm z}|^{\ell+1}}, 
\end{equation}
where the relation between spherical and Cartesian multipole moments is \cite{Thorne:1980ru}:
 \begin{equation}
 Q_{\ell m}=\frac{4\pi}{2\ell+1}{\cal Y}_{L}^{*\, \ell m}Q_L . \label{eq:Qcarttospher}
 \end{equation}
The tensors ${\cal Y}_{L}^{\ell m}$ are defined by $Y_{\ell m}={\cal Y}_{L}^{\ell m}n^{<L>}$ and involve complex coefficients that convert between the unit vector $\bm{n}$ and spherical harmonic representations~\cite{Thorne:1980ru}. 
\subsubsection{Multipole moments of the NS's spacetime in GR}
The Newtonian expansion around the center-of-mass position can be generalized to an expansion around a reference center-of-mass worldline $z^\mu(\sigma)$ where $\sigma$ is an evolution parameter along the worldline. This is known as the worldline-skeleton approach~\cite{Dixon:1970zza}, and also arises naturally in frameworks based on an effective action~\cite{Levi:2018nxp}. The NS's multipole moments are defined in a region of spacetime at large distance compared to the size of the NS. In this region, and when expressing the asymptotic metric in a local asymptotic frame~\cite{Thorne:1980ru,Thorne:1997kt}, the time-time component of the metric involves a potential $U_{\rm eff}$ that is analogous to the Newtonian gravitational potential $U$:
\begin{equation}
g_{tt}=-\left(1-2 U_{\rm eff}\right), 
\end{equation}
For a spherical object $U_{\rm eff}=M/r$ where $r$ is the distance from the body, while for a nonspherical, nonspinning body it has the asymptotic form
\begin{equation} \lim_{r\to \infty} U_{\rm eff}=\frac{M}{r}+ \frac{3n^{<ij>}Q_{ij}}{2r^{3}}+{\cal O}(r^{-4}). \label{eq:Qdef}
\end{equation}
In this setting, the $\ell$-th mass multipole moment is associated with the coefficient of the term that falls off as $r^{-(\ell+1)}$. This definition is equivalent to the Geroch-Hansen multipole moments for stationary spacetimes ~\cite{Guersel83}. We note that the multipole moments can be defined by considering the energetics of test particle orbits~\cite{Ryan:1995wh}, and that in a binary system a small ambiguity remains in the definition of the multipole moments~\cite{Thorne:1984mz,Gralla:2017djj}. 

\subsubsection{Tidal moments in Newtonian gravity}
Next, we consider the gravitational potential due to the binary companion that is felt by the NSs. We will denote it by $U^{\rm{ext}}$ to indicate that the source of this potential is external to the NS. A Taylor expansion around the NS's center of mass leads to:
\begin{subequations}
\label{eq:Uext}
\begin{equation}
U^{\rm{ext}}(t,\bm{x})=U^{\rm{ext}}(t,\bm{z})+(x-z)^j \left[\partial_jU^{\rm{ext}}(t,\bm{x})\right]_{\bm{x}=\bm{z}}-\sum_{l=2}^\infty\frac{1}{\ell!}(x-z)^{L}{\cal E}^L_{\rm Newt}.
\end{equation}
The coefficients ${\cal E}^L$ in \eqref{eq:Uext} are the NS's tidal moments~\cite{Thorne:1984mz}. We are interested in specializing the general expansion~\eqref{eq:Uext} to the case that $U_{\rm ext}$ is sourced by a binary companion whose gravitational potential we denote by $U_{\rm c}$. The NS's tidal moments are then given by
\begin{equation}
{\cal E}^L_{\rm Newt}=- {\partial_{L}}U_{\rm c}(t,\bm{x})\mid_{\bm{x}=\bm{z}(t)}. \label{eq:calEdef}\end{equation}
\end{subequations}
\subsubsection{Tidal moments in GR} 
The relativistic tidal moments are projections of the curvature tensor due to the companion. The analogue to the Newtonian ${\cal E}_L$ is the `electric part' of the companion's Weyl  tensor $C_{\mu\alpha\nu\beta} $ as~\cite{Thorne:1984mz}
\begin{equation}
{\cal E}_{\mu \nu}=z^{-2}C_{\mu\alpha\nu\beta} u^\alpha u^\beta \; \label{eq:relatE}
\end{equation}
where $u^\mu=dz^\mu/d\sigma$ is the tangent to the NS's worldline and we have defined 
\begin{equation}
z=\sqrt{-u_\alpha u^\alpha}\label{eq:zdef}
\end{equation}
The tidal tensor ${\cal E}_{\mu\nu}$ is symmetric and trace free. In the NS's rest frame $u^\mu=(-1,0,0,0)$ it is a purely spatial tensor ${\cal E}_{\mu\nu} u^\nu = 0$ and given by ${\cal E}_{ij}=C_{titj}$. Higher multipole tidal moments are defined by
\begin{equation}
{\cal E}_L=\frac{1}{(\ell-2)!} C_{\langle 0 a_10 a_2;a_3 \cdots a_\ell\rangle},
\end{equation}
where the semicolon denotes covariant derivatives.
\subsection{Equations of motion and action principle}
We next consider the effects in the interaction zone, where the dynamics are dominated by the point-mass motions with small corrections from (i) the coupling of the NS's multipole moments to the tidal field, and (ii) the energetics associated with the internal dynamics of the multipoles. 

\subsubsection{Tidal effects on Newtonian binary dynamics}
 The center-of-mass position of a NS in a binary system moves according to Newton's second law:
\begin{equation}
\label{eq:eomNewt}
m\ddot{z}^j=m\frac{\partial}{\partial x^j }U^{\rm{ext}}(t,\bm{x})\mid_{\bm{x}=\bm{z}}-\sum_{\ell=2}^\infty \frac{1}{\ell !}Q^{L}{\cal E}_{jL},
\end{equation}
Note that only the potential sourced by the companion contributes to the body's motion, as can be verified by direct calculation.

The dynamics can be concisely summarized by an action principle. The action is constructed from the Lagrangian ${\cal L}=T-V$, where $T$ and $V$ are the kinetic and potential energies for the system. These can be separated into contributions from the NS's center-of-mass motion and from its internal dynamics~\cite{Vines:2010ca}:
\begin{equation}
T=\frac{1}{2}\int d^3 x \rho \dot{{\bm{z}}}^2+T^{\rm int}, \qquad \qquad V= \frac{1}{2}\int d^3 x \rho U_{\rm c}+V_{\rm int}.
\end{equation}
Expanding around the NS's center of mass, adding the contributions from the companion, and transforming to the barycentric frame of the binary system leads to 
\begin{equation}
T=\frac{1}{2}\mu v^2+T_{\rm int}, \qquad
V=-\frac{ \mu M}{r}+\sum_{\ell\geq 2}\frac{1}{\ell!}Q_{L}{\cal E}_{L}+V_{\rm int}.
\end{equation}
Here, $M$ and $\mu$ are the total and reduced mass respectively. The relative separation is $\bm{r}=\bm{z}-\bm{z}_{\rm c}$ with magnitude $r=|\bm{r}|$, and the relative velocity is $v^2=\dot{\bm{r}}\cdot \dot{\bm{r}}$. The action for the binary dynamics is then given by~\cite{Flanagan:2007ix,Rathore:2002si,Lai:1993di} 
\begin{equation}
S=S_{\rm pm}+\int dt \sum_{\ell\geq 2}\bigg[-\frac{1}{\ell!}Q_{L}{\cal E}_{L} +{\cal L}^{\rm int}\bigg], \label{eq:Newtactiongen}
\end{equation}
where $S_{\rm pm}=\int dt \, \left[(\mu/2) v^2+\mu M/r\right]$ describes the orbital motion of point-masses and ${\cal L}^{\rm int}$ the internal dynamics of the multipole moments. We will assume that the multipoles arise only from the response to the companion's tidal field \cite{Bildsten:1992a,1994ApJ...426..688R,Lai:1993ve,Lai:1993di,1994ApJ...426..688R,1977A&A....57..383Z,1970A&A.....4..452Z,Kopal78,Kochanek:1992wk,Hansen:2005qv,Mora04,Kokkotas:1995xe,Flanagan:2007ix,Ferrari:2011as,Damour:1991yw,Shibata94}, which excites the oscillation modes of the NS. We will consider here only the fundamental ($f$-) modes. They behave as harmonic oscillators described by a Lagrangian~\cite{Flanagan:2007ix,Rathore:2002si,Lai:1993di,Kokkotas:1995xe,Chakrabarti:2013xza} 
\begin{equation}
{\cal L}^{\rm int}=\frac{1}{ 2\ell! \lambda_\ell \omega_{0\ell}^2} \left[ {\dot Q}_{L} {\dot
    Q}^{L} - \omega_{0\ell}^2 Q_{L} Q^{L} \right]. \label{eq:finalactionNewt}
\end{equation}
Here, the quantities $\omega_{0\ell}$ denote the $f$-mode frequencies and we have neglected contributions from other quadrupolar $\ell=2$ modes. The parameters $\lambda_\ell$ are the tidal deformability coefficients. They are defined by considering the \emph{adiabatic} limit, where the NS's internal time scales $\tau^{\rm int}\sim \omega_{0\ell}^{-1}\sim \sqrt{R^3/m}$ are fast compared to the time scale of variations in the tidal field $\tau_{\rm orb}\sim \sqrt{r^3/M}$ as
\begin{equation}
Q_{L}^{\rm adiab}=-\lambda_{\ell} \, {\cal E}_L ,\label{eq:lamdef}
\end{equation}
The tidal parameters $\lambda_\ell$ are related to the NS's dimensionless tidal Love numbers $k_\ell$  ~\cite{AEHLove} by $\lambda_\ell=\frac{2}{(2\ell-1)!!} k_\ell R^{2\ell+1} $ and depend on the NS's interior structure. For data analysis applications, it is useful to define the dimensionless tidal deformabilities
\begin{equation}
\Lambda_\ell=\frac{\lambda_\ell}{m^{2\ell+1}}.
\end{equation}
For adiabatically induced multipoles, $dQ_L/dt=0$, the internal Lagrangian is only the elastic potential energy associated with the deformation ${\cal L}^{\rm int}_{\rm adiab}=- Q_{L} Q^{L}/(2\ell! \lambda_\ell )$.
In this limit the mass quadrupole finite size effects on the dynamics depend only on the orbital variables and $\lambda_\ell$:
\begin{equation}
S_{\rm adiab}=S_{\rm pm}+\int dt \bigg[\frac{\lambda_\ell}{2\ell!}{\cal E}_{L}{\cal E}^{L} \bigg], \label{eq:Newtactionadiab}
\end{equation}
with ${\cal E}_L=-m_{\rm c} \partial_L r^{-1}$.

\subsubsection{Tidal effects on relativistic dynamics}
The action describing tidal interactions in a \emph{relativistic} binary system can be obtained by expressing the Newtonian result from Eq.~\eqref{eq:Newtactiongen} in a covariant form, inserting the appropriate redshift factors to ensure invariance under re-parametrizations, and replacing all time derivatives by covariant derivatives along the center-of-mass worldline~\cite{Steinhoff:2016rfi}:
\begin{equation}
\label{eq:generalaction}
S = S_{\rm pm}+\int d\sigma \, \left\{- \frac{z}{2} {\cal E}_{\mu\nu} Q^{\mu\nu}+ {z \over 4 \lambda z^2 \omega_{02}^2} \left[ \frac{D{Q}_{\mu \nu}}{d\sigma}\frac{D {Q}^{\mu \nu}}{d\sigma} - z^2\omega_{02}^2 Q_{\mu\nu} Q^{\mu\nu} \right]\right\}.
\end{equation}
Here, total derivatives are denoted by $\frac{D}{d\sigma}=u^\beta\nabla_\beta$, where $\nabla_\alpha$ is the covariant derivative. Contributions from quadrupolar modes with higher radial nodes and other terms due to the incompleteness of quasi-normal modes have been omitted. The interaction terms in the action \eqref{eq:generalaction} also follow from an effective-field-theoretical approach~\cite{Goldberger:2004jt,Endlich:2015mke,Steinhoff:2016rfi} when considering all possible terms consistent with the symmetries (general covariance, parity, and time reversal) and re-defining variables to eliminate accelerations. 
 After decomposing Eq.~\eqref{eq:generalaction} into the time and space components and imposing the constraints to eliminate all unphysical gauge degrees of freedom we obtain
\begin{equation}
\label{eq:action3plus1}
S = S_{\rm pm}+\int d\sigma \left[ - \frac{z}{2} {\cal E}_{ij} Q^{ij} +  \frac{z}{4 \lambda z^2 \omega_{02}^2}  \left(\dot{Q}_{ij} \dot{Q}^{ij}
 -z^2\omega_{02}^2Q_{ij} Q^{ij} \right)+ L_\text{FD}   \right].
\end{equation}
Here, the term $L_{\rm FD}$ describes relativistic frame-dragging effects that arise kinematically when expressing $(D Q^{\mu\nu}/d \sigma)^2$ in terms of $ (d{Q}_{ij}/dt)^2 $. The term $L_\text{FD}$ describes the coupling of the orbital angular momentum to the angular momentum (or spin) associated with the quadrupole
\begin{equation}
\label{eq:tidalspin}
S_Q^{i}=\frac{1}{\lambda \omega_{02}^2} \epsilon_{ijm}\left[ Q^{kj} \dot{Q}^{m}{}_k-Q^{km} \dot{Q}^{j}{}_k\right].
\end{equation}
To proceed with computing the dynamics from~\eqref{eq:action3plus1} and the GWs requires explicit expressions for the various quantities such as ${\cal E}_{ij}$ and $z$ and the frame-dragging terms appearing in the action. These have been computed in post-Newtonian (PN) theory~\cite{Vines:2010ca,Bini:2012gu,Steinhoff:2016rfi}, see also ~\cite{Levi:2018nxp} for the effective field theory description, in the test-particle limit~\cite{Bini:2015kja} or the gravitational self-force formalism~\cite{Dolan:2014pja,Bini:2014zxa,Nolan:2015vpa,Shah:2015nva}. An alternative approach based on an affine model has also been studied~\cite{Ferrari:2011as,Maselli2012}. 


\subsection{Computation of tidal deformability}


We have seen that the action summarizing the dynamics involves the characteristic information about matter properties through parameters tidal deformability and, in the non-adiabatic case, the mode frequencies. The mode frequencies are routinely computed in calculations of stellar oscillations in Newtonian gravity and of quasi-normal modes in GR~\cite{Kokkotas:1999bd,Ferrari:2007dd}. Thus we will focus here on the tidal deformability parameters defined in terms of spacetime multipole moments and their relation to the tidally perturbed interior structure of the NS. 

\subsubsection{Newtonian calculation of tidal deformability}
We start from an equilibrium configuration for the NS in isolation. It is a solution to the Poisson equation for the gravitational potential together with conservation of mass and momentum:
\begin{equation}
\label{eq:flu}
\nabla^2 U=-4\pi  \rho, \qquad 
\frac{\partial \rho}{\partial t}+\nabla \cdot (\rho {\bm v})=0, \qquad
\frac{\partial v^i}{\partial t}+({\bm v}\cdot \nabla)v^i=-\frac{\partial^i p}{\rho}+\partial_i U +a^i_{\rm ext}. \qquad\end{equation}
Here, $a^i_{\rm ext}$ is the acceleration due to external forces which vanishes for a star in isolation. Solving the system of equations, \eqref{eq:flu}, for a given EOS model yields a background configuration of an isolated NS with pressure $p_0$, density $\rho_0$, and gravitational potential $U_0$. 
Next, we consider the effect of external static tidal field ${\cal E}_L$. This causes the NS matter to adjust to a new static configuration with an exterior gravitational potential given by
\begin{equation}
U_{\rm total}=U+U_{\rm ext}=\frac{m}{r}+\sum_{\ell=2}^\infty \sum_{m=-\ell}^\ell Y_{\ell m} \left[\frac{Q_{\ell m}}{r^{\ell+1}} -\frac{1}{(2\ell-1)!!} {\cal E}_{\ell m} r^\ell \right] \label{eq:fullUexpan}
\end{equation}
The linear response defining $\lambda$ in \eqref{eq:lamdef} can equivalently be written as
\begin{equation}
Q_{\ell m}=-\lambda_\ell {\cal E}_{\ell m} \label{eq:lamm}
\end{equation}
for each $\ell$-th multipole and azimuthal quantum number $m$. To obtain $\lambda_\ell$, we need to consider only a single value of $m$, and it is easiest to choose $m=0$. We will need to compute the perturbed interior solution for the gravitational potential and matching it to the exterior description \eqref{eq:fullUexpan} at the NS surface. 

The equations \eqref{eq:flu} also describe the perturbed interior, with $p=p_0+\delta p$, $\rho=\rho_0+\delta \rho$, $U=U_0+\delta U$, and an external acceleration ${\bm a}_{\rm ext}=\bm{\nabla} U_{\rm tidal}$. The fluid perturbation can be represented by a Lagrangian displacement $\xi(x,t)$, which maps a fluid element at position $x$ in the unperturbed NS to $x+\xi(x,t)$ in the perturbed star. Expanding the Euler equation to linear order in the perturbations leads to the differential equation
\begin{equation}
\frac{d^2 \xi^i}{dt^2}=-\frac{\partial_i \delta p}{\rho_0}+\frac{\partial_i p_0}{\rho_0^2}\delta \rho+\partial_i \delta U+a^i_{\rm ext}.\label{eq:pertEuler}
\end{equation}
This expression is used in general for calculating stellar oscillations. Here we are interested in specializing to static perturbations with $(\dot \xi=0)$. For a barotropic EOS relation $p=p(\rho)$ we can eliminate $\delta p$ from \eqref{eq:pertEuler} in favor of $\delta \rho$, combine all terms that involve $\delta \rho$ into a total derivative and integrate. We then decompose the perturbations as $\delta \rho=f(r)Y_{\ell m}(\theta,\phi)$ and $\delta U_{\rm tot}=H(r)Y_{\ell m}(\theta,\phi)$, substitute into the linearized Poisson's equation, and after manipulations obtain the differential equation for $H(r)$ in the region $r\leq R$:
\begin{equation}
H^{\prime\prime}+\frac{2}{r}H^\prime-\frac{\ell(\ell+1)}{r^2}H=-{4\pi}\left(\frac{1}{\rho_0}\frac{dp_0}{d\rho_0}\right)^{-1}H.\label{eq:singleODE}
\end{equation}
Except for special choices of the EOS, solving \eqref{eq:singleODE} requires numerically integrating in the NS interior, with the boundary condition that ensures regularity at the center of the star, $H\propto r^\ell$ for $r\to 0$. For $r>R$ the exterior solution is 
\begin{equation}
H^{\rm ext}=\frac{Q_{\ell m}}{r^{\ell+1}} -\frac{1}{(2\ell-1)!!} {\cal E}_{\ell m} r^\ell =-{\cal E}_{\ell m}\left[\frac{\lambda_\ell}{r^{\ell+1}} +\frac{1}{(2\ell-1)!!} r^\ell\right]. \label{eq:Hext}
\end{equation}
To extract the Love numbers we eliminate ${\cal E}_{\ell m}$ by considering the logarithmic derivative
\begin{equation}
y(r)=\frac{r H'(r)}{H(r)}. \label{eq:yNewt}
\end{equation}
We solve for $k_\ell$ by using \eqref{eq:Hext} in \eqref{eq:yNewt} and matching the result for $y(R)$ obtained from the interior and exterior solutions at the star's surface:
\begin{equation}
k_\ell=\frac{\ell-y(R)}{2\left[\ell+1+y(R)\right]|} 
\label{eq:k2result}
\end{equation}
For an incompressible $n=0$ polytrope the exact solution is
$k_\ell=3/(4\ell-4)$. 

\subsubsection{Calculation of tidal deformability in GR} 
The definition of $\lambda_\ell$ from Eq.~\eqref{eq:lamdef} is general and holds with the relativistic definitions of ${\cal E}_L$ and $Q_L$. The computation of $\lambda_\ell$ in GR follows a similar method as the Newtonian case but requires replacing the gravitational potential by the spacetime metric, the Poisson equation by the Einstein field equations, the density by the stress-energy tensor, and the continuity and Euler equations by the relativistic energy-momentum conservation. This complicates the equations so we will only outline the process here. 

For the equilibrium solution we use the metric of a spherically symmetric, isolated nonspinning NS solution expressed in the form~\cite{Hartle67}
\begin{equation}
ds_0^2 =g_{\mu\nu}^{(0)}dx^\mu dx^\nu= -e^{\nu(r)} dt^2 + e^{\gamma (r)} dr^2 + r^2 (d\theta^2 + \sin^2 \theta d\varphi^2)\,. 
\end{equation}
The NS matter is described by the stress-energy tensor $T_{\mu \nu} = (\rho + p) u_{\mu} u_{\nu} + p g_{\mu\nu}$,  
where $u^\mu$ is the fluid's four-velocity and $\rho$ the energy density. In the NS's rest frame, the normalization condition $u_\mu u^\mu = -1$ implies that $u^{\mu} = (e^{-\nu/2}, 0, 0,0)$. The Einstein field equations  and energy-momentum conservation $\nabla_\nu T^{\mu\nu}=0$ determine the NS structure
\begin{equation}
\label{eq:GReqs}
G_{\mu\nu}=8\pi T_{\mu\nu} , \qquad\qquad  \nabla_\nu T^{\mu\nu}=0.
\end{equation}
This system reduces to the Tolman-Oppenheimer-Volkoff equations that in general must be solved numerically except in special cases. 

When considering the tidal perturbations of interest here, the metric is the background plus the linear perturbations $g_{\mu\nu}=g_{\mu\nu}^{(0)}+h_{\mu\nu}dx^\mu dx^\nu$. For our purposes we can decompose the perturbation as~\cite{Regge57,Campolattaro1,IpserPrice1992,Detweiler1985}
\begin{equation}
h_{\mu\nu}dx^\mu dx^\nu=\sum_{\ell,m}\bigg[-e^{\nu}H_0^{\ell m}dt^2+2H_1^{\ell m}dtdr+e^\gamma H_2^{\ell m}dr^2+r^2K^{\ell m}d\Omega^2\bigg]Y^{\ell m}
\end{equation}
To compute tidal Love numbers we consider static perturbations where $H_0$, $H_2$, and $K$ depend only on $r$. The perturbations to the stress-energy tensor are likewise decomposed into spherical harmonics. Substituting these decompositions into Eqs.~\eqref{eq:GReqs}, and extracting only the pieces that are linear in the perturbations in all the components leads to various relations between the different perturbation functions. Ultimately this yields a master equation for $H\equiv H_0^{\ell m}$~\cite{Hinderer:2007mb,Damour:2009vw,Binnington:2009bb}
\begin{eqnarray}
\label{eq:H0}
0&=&\frac{d^2 H}{dr^2} + \left\{ \frac{2}{r} + e^{\gamma} \left[ \frac{2M}{r^2} + 4 \pi r (p-\rho) \right] \right\} \frac{dH}{dr} \nonumber\\
&&+ \left\{ e^\gamma \left[ - \frac{\ell (\ell +1)}{r^2} + 4 \pi (\rho + p) \frac{d\rho}{dp} + 4 \pi (5 \rho + 9 p) \right]  - \left( \frac{d\nu}{dr} \right)^2 \right\} H \,.
\end{eqnarray}
The initial condition at the center is that $H \propto r^\ell$ to ensure regularity of the solution.

Outside the star, the metric perturbation has the general form
$H_\ell =  a_{\ell}^{Q} {Q}_{\ell 2}(x)+a_{\ell}^{P} {P}_{\ell 2}(x) $,
where $x \equiv r/M -1$ and ${P}_{\ell 2}(x)$ and ${Q}_{\ell 2}(x)$ are the normalized associated Legendre functions of the first and second kinds respectively. They are normalized such that for $x \rightarrow \infty$, ${P}_{\ell 2}(x) \sim x^{\ell}$ and ${Q}_{\ell 2}(x) \sim x^{-(\ell + 1)}$.  The constants $a_{\ell}^{P}$ and $a_{\ell}^{Q}$ are determined by matching the logarithmic derivative of the interior and exterior solutions, 
\begin{equation}
\label{eq:yl}
y_\ell\equiv \frac{r}{H_\ell} \frac{dH_\ell}{dr}\,,
\end{equation}
at the NS surface~\cite{Hinderer:2007mb}. Comparing with the definition of $Q_\ell$ and ${\cal E}_\ell$ in the asymptotic metric \eqref{eq:Qdef} and the definition of the Love numbers shows that 
\begin{equation}
\label{eq:biglambdal}
(2\ell-1)!!\Lambda_\ell = -\left.\frac{{P}_{\ell 2}'(x)-C y_{\ell} {P}_{\ell 2}(x)}
{{Q}_{\ell 2}'(x)- C y_{\ell} {Q}_{\ell 2}(x)}\right\vert_{x=1/C-1}, 
\end{equation}
For the dominant quadrupolar effect the explicit expression is
\begin{align}
\Lambda\mid_{\ell=2}&=  \frac{16}{15} (1-2 C)^2[2+2C(y-1)-y] \nonumber \\
&\times  \left\{ 2C[6-3y+3C(5y-8)]  +4C^3[13-11y+C(3y-2)+2C^2(1+y)]  \right.\nonumber\\
&\quad \left.+3(1-2C)^2[2-y+2C(y-1)] \ln(1-2C)  \right\}^{-1}\,,
\end{align}
where $y$ is evaluated at the surface $r=R$. 
For an incompressible star with $\rho=\text{const}$ or $p=K \rho^{1+1/n}$ with $n=0$, the density profile is a step function and the matching of the interior and exterior solutions must be modified as follows~\cite{Damour:2009vw}. 
From Eq.~\eqref{eq:yl}
in the interior we obtain $y^{\rm in}(R)$. The density discontinuity introduces a correction to $y$ just outside the star $y^{\rm out}$ given by
\begin{equation}
\label{correcty}
y^{\rm out}_{\rm incompressible} = y^{\rm in}- 3.
\end{equation}
This is valid for any $\ell$ and  \eqref{correcty} is to be substituted for $y$ when evaluating $\Lambda$ from \eqref{eq:biglambdal}.

A number of studies have computed examples of $\Lambda$ for a wide range of EOS models and delineated interpretations of its physics content such as symmetry energy, phase transitions, and multi-body interactions~\cite{Hinderer:2009ca,Postnikov:2010yn,Fattoyev:2012uu,Fattoyev:2013rga,Steiner:2014pda,Lattimer:2012xj,VanOeveren:2017xkv,Penner:2011pd,Lindblom:2013kra,Kumar:2016dks,Annala:2017llu,Raithel:2018ncd,Lim:2018bkq,Zhu:2018ona,Zhang:2018vrx,Krastev:2018wtx,Nandi:2017rhy,Abdelsalhin:2017cih,Paschalidis:2017qmb,Fattoyev:2017jql,Annala:2017tqz,Zhou:2017pha,Bhat:2018erd,Lindblom:2018ntw,Char:2018grw,Wolter:2018qqt,Gomes:2018eiv,Malik:2018zcf,Marczenko:2018jui,Alvarez-Castillo:2018pve,Kim:2018aoi,De:2018uhw,Han:2018mtj,Lalit:2018dps,Montana:2018bkb,Wei:2018dyy,Christian:2018jyd,Lau:2018mae,Zhang:2018vbw,Hornick:2018kfi,Li:2018ayl,Zhao:2018nyf,Zhang:2018bwq,Carson:2018xri,Biswas:2019ifs,Raithel:2019uzi,Forbes:2019xaz,Li:2019tjx,Mena-Fernandez:2019njf,Baym:2019iky,Hirayama:2019vod,Lim:2019som,Fasano:2019zwm,Gamba:2019kwu,Li:2019akk,Lourenco:2019zng,Malik:2019whk,Yamazaki:2019tuo,Lourenco:2018dvh,Piekarewicz:2018sgy,Lourenco:2020qft,Perot:2020gux,Ferreira:2019bgy,Zhang:2019xhu,Ji:2019hxe,Zacchi:2019ayh,Perot:2019gwl,Tsang:2019ymt,Datta:2019ueq,Liu:2019ntc,Llanes-Estrada:2019wmz,Carson:2019xxz}. Tidal parameters can also distinguish exotic objects, matter halos, and modified gravity~\cite{Cardoso:2017cfl,Sennett:2017etc,Mendes:2016vdr,Uchikata2015,Pani2015,Maselli:2017vfi,Emparan:2017qxd,Nelson:2018xtr,Baumann:2018vus,Ellis:2018bkr,Chakravarti:2018vlt,Maselli:2018fay,Lai:2018ugk,Das:2018frc,Giddings:2019ujs,Quddus:2019ghy,Cardoso:2019upw,Sham:2013cya,Pani:2014jra,Doneva:2017jop,Gupta:2017vsl,Yazadjiev:2018xxk,Chagoya:2018lmv,Cardoso:2019vof,Bernard:2019yfz}. Substantial recent interest has quasi-universal relations~\cite{Yagi2013a,Yagi2013b} that link dimensionless parameters characterizing various global properties of the NS in an approximately EOS-independent way~\cite{Lattimer:2012xj,Pappas:2013naa,Yagi2014,Pappas:2016sye,Pappas:2015mba,Haskell2014,Chakrabarti2014,Maselli2013,AlGendy:2014eua,Chirenti:2015dda,Pannarale:2015jia,2016EPJA...52...18S,Haskell2014,Breu2016,Silva:2016myw,Yagi:2013sva,Reina:2017mbi,Chan:2014tva,Chan:2015iou,Chatziioannou:2018vzf,Kumar:2019xgp,Samajdar:2020xrd}.

\subsection{Other matter effects}

In the discussion above, we specialized to tidal effects for a nonspinning NS from its fundamental modes. A NS has a rich spectrum of modes. Those with sufficiently low frequencies could be tidally excited during an inspiral and yield spectroscopic information about matter in NS interiors; several examples have been studied to date~\cite{Pratten:2019sed,Andersson:2017iav,Ho:1998hq,Flanagan:2006sb,Shibata94,Yu:2016ltf,Lai:1993di,Kokkotas:1995xe,Tsang:2013mca,Tsang2012,Pnigouras:2019wmt}. Effects from nonlinear mode coupling have also been studied~\cite{Xu:2017hqo,Essick:2016tkn,Landry:2015snx}. In GR, besides the tidal couplings to the gravito-electric fields ${\cal E}_L$ that induce mass multipole moments, there are also gravitomagnetic tidal fields that induce current multipoles. These effects have no Newtonian analogue but their mathematical  description is similar to the gravitoelectric case, although the interpretation and adiabatic limits are more subtle. In the local frame of the NS, the gravitomagnetic part of the curvature due to the companion is given by 
$ {\cal B}_L=3 \epsilon_{<a_1 jk}C^{jk\,}_{\; \; a_20; a_3 \cdots a_\ell>}/2(\ell+1)(\ell-2)!$
where $\epsilon_{ijk}$ is the completely antisymmetric permutation tensor. The induced current moments  ${\cal S}_L$  appear in the time-space part of the asymptotic metric in a local asymptotic frame, see Refs. ~\cite{Favata:2005da,Landry:2015cva,Pani:2018inf,Landry:2015snx,Flanagan:2006sb} for more details. 
Similar to the tidal deformability coefficients $\lambda_\ell$, the relation between ${\cal S}_L$ and ${\cal B}_L$ is characterized by a set of gravitomagnetic Love numbers $\sigma_\ell=-
 {\cal S}_L / {\cal B}_L$.
 
 When considering rotating NSs, the additional spin degrees of freedom introduce a yet richer phenomenology. Spins give rise to new spin-tidal couplings~\cite{Jimenez-Forteza:2018buh,Abdelsalhin:2018reg,Pani:2015nua,Pani:2015hfa,Gagnon-Bischoff:2017tnz,Landry:2017piv,Landry:2015zfa,Dietrich:2019kaq} and can significantly impact the tidal excitation of quasi-normal modes~\cite{Doneva:2013zqa,Ho:1998hq,Foucart:2018lhe,Ma:2020rak}. As discussed in the previous section, eccentricity also greatly enriches matter phenomena and mode excitations~\cite{Gold:2011df,Chirenti:2016xys,Yang:2019kmf,Vick:2019cun}. While this review focuses on double neutron star systems, we mention here for completeness that for NS-BH binaries there exist parameter regimes of mass ratios, spins, and NS compactness where the NS is tidally disrupted. This leads to a sudden shutoff of the GW signal, with the disruption frequency depending on the parameters including the EOS~\cite{Vallisneri00,Shibata:2011jka,Pannarale:2011pk,Ferrari:2009bw,Maselli:2013rza,Foucart2014,Kawaguchi:2017twr,Lackey:2013axa}, see Ref.~\cite{Shibata:2011jka} for a review. An additional distinction between NSs and black holes is that the neutron star has a surface while a black hole possesses an event horizon that absorbs all incoming GWs; this effect is also imprinted in the GWs~\cite{Poisson:1994yf,Teukolsky:1974yv,Tagoshi:1997jy,Maselli:2017cmm,Poisson:2004cw,Fang:2005qq,Porto:2007qi,Nagar:2011aa,Bernuzzi:2012ku,Taracchini:2013wfa,Isoyama:2017tbp}.

\subsection{ Matter signatures in GW signals}
The asymptotic GWs are computed by combining an expansion using powers of $G$ as the bookkeeping parameters around flat space with a multipolar decomposition~\cite{Blanchet:2013haa}. The radiative post-Minkowski solution can be formally written in a radiative coordinate system $(u,{\bm X})$ where $u=T-|X|/c$ is the retarded time, as
\begin{eqnarray}
\label{eq:radiativeh}
h_{ij}^{\rm rad}(u, X)&=&\frac{4G}{D c^2} {\cal P}_{ijab}(N) \sum_{\ell=2}^\infty \frac{1}{c^\ell \ell!}\bigg[N_{L-2}U_{abL-2}(u)\\
&& \qquad \qquad \qquad \; \; -\frac{2\ell}{c(\ell+1)}N_{dL-2}\epsilon_{de(a}V_{b)eL-2}(u)\bigg] + O(D^{-2}) \nonumber 
\end{eqnarray}
Here, ${\cal P}_{ijab}(N)$ is a transverse-traceless projection operator, $N$ is the vector pointing between the source and the observer, $D$ is the distance to the source, and $U_L$ and $V_L$ are \emph{radiative} mass and current multipole moments. Their relation to the multipole moments of the binary system is nonlinear and non-local in time, and complicated in general. At the leading Newtonian order, there is a simple relation to the source quadrupole $U_{ij}^{\rm Newt}=d^2 Q^{\rm T}_{ij}/du^2$, where the total quadrupole moment of the system is the sum of the orbital and NS contributions $Q_{ij}^{\rm T}=\mu r^2 n^{<ij>}+Q_{ij}$. In Eq.~\eqref{eq:radiativeh} we have displayed explicit factors of $c$ because they indicate the PN order, with each factors of $c^{-2}$ being considered one PN order. Using \eqref{eq:radiativeh}, the energy and angular momentum fluxes in GWs can also be expressed in terms of the moments $U_L$ and $V_L$ ~\cite{Blanchet:2013haa}.

An approximation for the GW phase evolution can be computed by imposing that the power radiated by a binary system is balanced by a change in the energy of the binary. We will briefly review this approach to obtain the lowest-order matter signatures in GWs. The energy of the binary is $E=E_{\rm pm}+E_{\rm tidal}$, where the subscript ``pm'' denotes point-masses. Specializing to circular orbits $\ddot{r}=\dot{r}=0$ and $\dot \phi=\Omega$, $\ddot\phi=0$, and to perturbatively solving the dynamics described by the action~\eqref{eq:Newtactionadiab}, and eliminating $r$ in terms of the frequency variable $x=(M\Omega)^{2/3}$ we find the following tidal contribution in the limit of adiabatic tides~\cite{Vines:2010ca}
\begin{equation}
E_{\rm tidal}(x)= -\frac{1}{2}\mu x\left[-9\frac{m_B}{m_A}  \frac{\lambda_{\ell=2,A}x^5}{M^5}-65\frac{m_B}{m_A}  \frac{\lambda_{\ell=3,A}x^7}{M^7}\right] +(A\leftrightarrow B).
\end{equation}
Here, the subscripts $A,B$ label the two objects, and the factor outside the brackets is the result for Newtonian point masses. The leading order adiabatic tidal corrections contributions to the energy flux from the mass quadrupole radiation are \cite{Vines:2011ud}
\begin{equation}
P_{\rm GW}^{\rm tidal}=\frac{32\mu^2}{5M^2} x^{5/2}\left[\left(\frac{18M}{m_A}-12\right)\frac{\lambda_{\ell=2,A }x^5}{M^5}+\frac{80m_B}{m_A}\frac{\lambda_{\ell=3,A}x^7}{M^7}\right].
\end{equation}
By requiring that $P_{\rm GW}$ be balanced by a change in the energy $E$ of the binary one can derive the evolution equations
\begin{equation} \label{orbitalODEs}
\frac{d\phi}{dt}=\frac{x^{3/2}}{M}\ \qquad \qquad
\frac{dx}{dt}=\frac{-{P}_{\rm GW}}{dE/dx}
\end{equation}
There are several ways to solve for $\phi$ in a PN approximation. For example, one can numerically solve Eqs.~\eqref{orbitalODEs} for $\phi(t)$ and $x(t)$ after first expanding the ratio ${P}_{\rm GW}/({dE/dx})$ about $x=0$ to the consistent PN order. These waveforms are known as \texttt{TaylorT4} approximants as reviewed in Ref.~\cite{Buonanno:2009zt}, where the point-mass terms are also given explicitly. The adiabatic quadrupolar tidal corrections that add linearly to the point-mass contributions:
\begin{equation}
\frac{d{x}}{dt}\mid_{\rm tidal}=\frac{32}{5}\frac{m_B}{M^7}\lambda_{\ell=2,A} \, x^{10}\left[ 12\left(1+11\frac{m_B}{M}\right) \right]+(A\leftrightarrow B).
\end{equation}
Other possibilities to perturbatively solve~\eqref{orbitalODEs} and obtain the tidal contributions to different approximants for the gravitational waveform are detailed in the Appendix of Ref.~\cite{Wade2014}. Another widely utilized class of template waveforms for data analysis are TaylorF2 waveforms which provide a fully analytic frequency-domain model and are thus very fast to generate. The derivation is explained e.g. in Ref.~\cite{Cutler:1994ys} and leads to a signal of the form
\begin{equation}
\label{eq:tildeh}
\tilde{h}(f)={\cal A}f^{-7/6}\exp\left[i\left(\psi_{\rm pm}+\psi_{\rm tidal}\right)\right],
\end{equation}
where $f$ is the GW frequency, ${\cal A}\propto{\cal M}^{5/6}/D$, where ${\cal M}$ is the chirp mass ${\cal M}=\eta^{3/5}M$, and $D$ is the distance between the GW detector and the binary. Extrinsic parameters of the source such as the location on the sky are also contained in ${\cal A}$, where higher PN order corrections to the amplitude enter. The point-mass phase $\psi_{\rm pm}$ to the current best knowledge for nonspinning binaries is given e.g. in Eq.(3.18) of~\cite{Buonanno:2009zt}. The leading order tidal contributions to $\psi_{\rm tidal}$ can be expressed as:
\begin{subequations}
\begin{equation}\label{eq:deltapsi}
\delta\psi_{\rm tidal}=\frac{3}{128\nu x^{5/2}}\left[ -\frac{39}{2}\tilde{\Lambda}x^5-\frac{4000 }{9}\frac{x^7}{M^7}\left(\Lambda_{\ell=3,A} m_A^6 m_B+\Lambda_{\ell=3, B} m_B^6 m_A\right) \right],
\end{equation}
where
\begin{equation}
\label{eq:lambdatilde}
\tilde{\Lambda}=\frac{16}{13}\left[\left(1+\frac{12 m_B}{m_A}\right)\frac{ \Lambda_{\ell=2,A}m_A^5}{M^5}+\left(1+\frac{12 m_A}{m_B}\right)\frac{ \Lambda_{\ell=2,B}m_B^5}{M^5}\right]
\end{equation}
\end{subequations}
The parameter $\tilde \Lambda$ plays an analogous role as the chirp mass. For equal-mass BNSs $\tilde{\Lambda}$ reduces to $\Lambda$ of the individual neutron stars. Other combinations of the two parameters $\Lambda_A$ and $\Lambda_B$ are also in use and have advantages in different contexts, e.g. to characterize the dominant effect in the conservative dynamics~\cite{Damour:2009wj}, or to improve the measurability~\cite{Yagi:2015pkc,Yagi:2016qmr}. This shows that for a double neutron-star system GW measurements are most sensitive to a weighted average of the deformability parameters of the two objects, which complicates the interpretation of measurements. 

The effects of the fundamental mode excitations can be approximately included in this waveform model by adding to \eqref{eq:deltapsi} the following mode-frequency-dependent and regularized contributions~\cite{Schmidt:2019wrl}:
\begin{eqnarray}
\delta\psi_{\rm tidal}^{\rm f-modes}&\approx&-\frac{\left[10\sqrt{3}\pi-27-30\log(2)\right]}{96\nu}\frac{\Lambda_{\ell=2,A}m_A^4}{M^6 \omega_{02,A}^2}\left(155-147\frac{m_A}{M}\right)x^{11/2}\nonumber\\
&&-\frac{1875\left[5-6\log(2)\right]}{16}\frac{\Lambda_{\ell=3,A}m_A^5}{M^7 \omega_{03,A}^2}x^{15/2}
\end{eqnarray}
Finally, we give the full analytical knowledge to the highest complete PN order $O(\alpha^5 \epsilon^{3/2})$, i.e., 1.5PN corrections to tidal effects for the quadrupolar adiabatic phasing~\cite{Damour:2012yf}:
\begin{equation}
\delta\psi_{\rm tidal} = - \Lambda_{\ell=2,A} c^A_{\rm Newt} x^{5/2}
\left(1 + c^A_1 x + c^A_{3/2} x^{3/2} \right) +  A \leftrightarrow B , \label{eq:PhiT_DNV}
\end{equation}
with the dimensionless EOB tidal parameter $\kappa_A$ (defined below) and
$x(M \omega)=(M \omega/2)^{2/3}$.
The individual coefficients $c_i^A$ are
\begin{subequations}
\begin{eqnarray}
 c^A_{\rm Newt}  &  = &  -\frac{3m_A^3}{8M^3}  \left(\frac{m_A}{m_B}+12\right), \\
 c^A_1 & = & -\frac{5 \left(260 \frac{m_A^3}{M^3}-2286 \frac{m_A^2}{M^2}-919 \frac{m_A}{M}+3179\right)}{336 \left(11 \frac{m_A}{M}-12\right)} , \\
 c^A_{3/2} & = & -\frac{5}{2} \pi  , \label{eq:PN_coeff}
\end{eqnarray}
\end{subequations}
and similarly with $A \leftrightarrow B$.  Partial knowledge is available at higher orders. The $O(\epsilon^{5/2})$ terms are known, however, the lower order $O(\epsilon^2)$ contributions have only been computed for the dynamics but not the fluxes. As all pieces of known contributions are used in the phenomenological models discussed below, we also give the two additional terms that contribute with a corresponding power of $x$ to \eqref{eq:PhiT_DNV}: 
\begin{eqnarray}
c_2^A & = & \left[ 5 \left(67702048 \frac{m_A^5}{M^5}-223216640\frac{m_A^4}{M^4}+337457524 \frac{m_A^3}{M^3} \right . \right.  \nonumber \\
            &&  \left. \left. -141992280 \frac{m_A^2}{M^2}+96008669 \frac{m_A}{M}-143740242\right) \right]/ \nonumber \\
            &&  \left[ 3048192 \left(11 \frac{m_A}{M}-12\right) \right] , \\
 c_{5/2}^A & = & -\frac{\pi  \left(10232 \frac{m_A^3}{M^3}-7022 \frac{m_A^2}{M^2}+22127 \frac{m_A}{M}-27719\right)}{192 \left(11 \frac{m_A}{M}-12\right)}
\end{eqnarray}

The effects of spin-multipole moments also couple to the companion's tidal tensors and influence the GWs in an analogous way as tidally-induced multipole moments except the scalings with parameters are very different. For the most recent developments on modeling the effects of spin-induced multipole moments on the GW signals see Refs.~\cite{Krishnendu:2017shb,Bohe:2015ana,Marsat:2014xea,Levi:2015msa,Porto:2012as,Porto:2010zg,Levi:2014gsa,Buonanno:2012rv}. The leading-order contribution to the GW phase from the spin-induced quadrupole is given by~\cite{Krishnendu:2017shb}:
\begin{equation}
\delta\psi^{\rm spin-quad}=\frac{30}{128\nu^{4/5}(\pi {\cal M}f)^{1/3}} \left[-5\kappa_A\chi_A^2\frac{m_A^2}{M^2}+(A\leftrightarrow B)\right]
\end{equation}

\begin{figure}[t]
 \centering
 \includegraphics[width=\columnwidth]{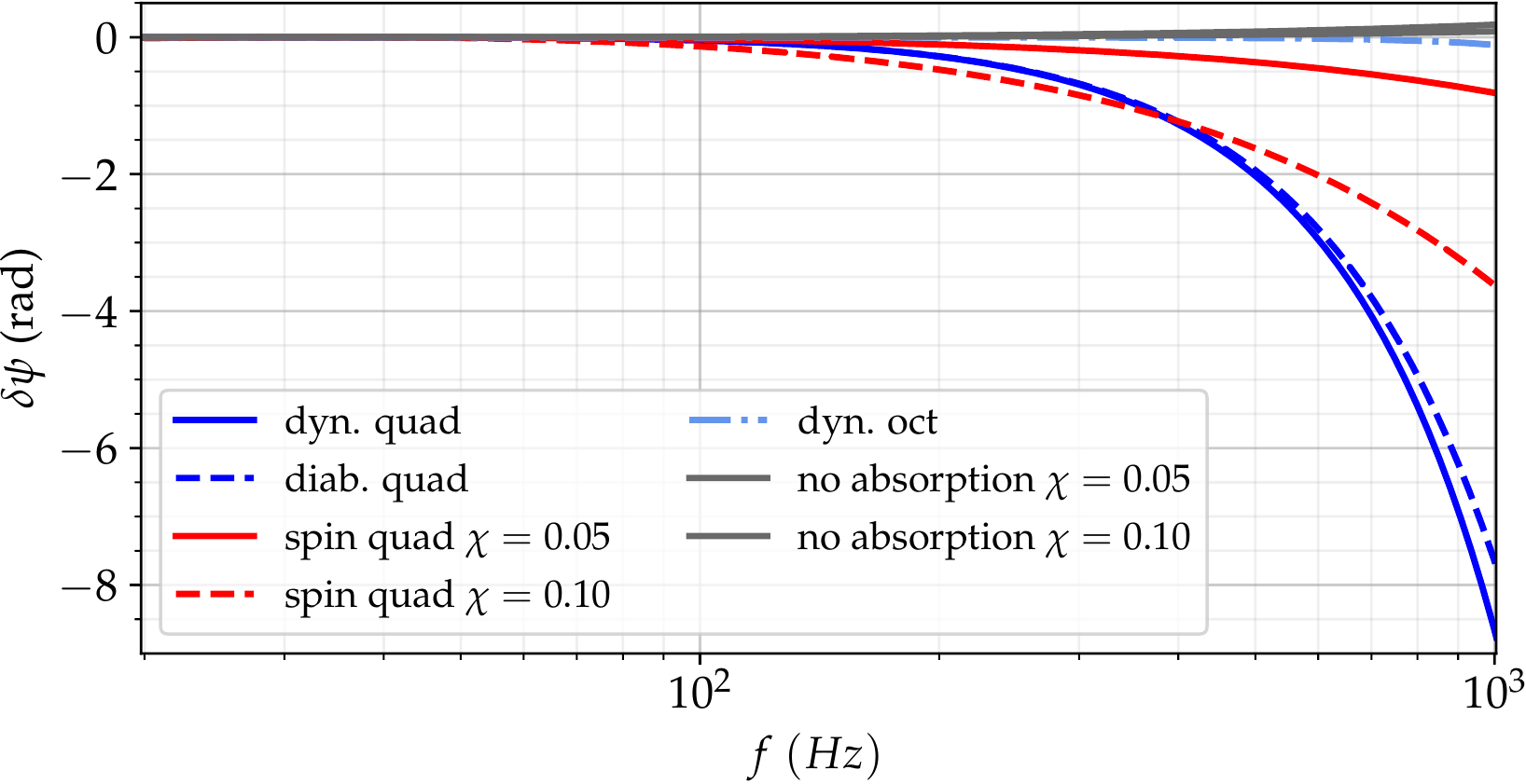}
 \caption{\emph{Example of matter effects in the GW phasing} for an inspiral starting from 20Hz. The waveforms were aligned over a window of 30Hz in the beginning and the parameters correspond to the DD2 EOS. Depending on parameters, the spin-quadrupole effects (red curves) can dominate over tidal effects (blue plus cyan curves), especially at low frequencies.}
 \label{fig:PNphasing}
\end{figure}

Figure ~\ref{fig:PNphasing} illustrates the various contributions to the GW phase discussed above.

From the discussion above, it is apparent that the fractional corrections to the Newtonian point-mass results due to tidal effects scale as a high power of the frequency, $x^5$ and higher, where $x=(\pi M f)^{2/3}$, which corresponds to $O(\alpha^5)$ in the expansion discussed in the beginning of this section. This means that in a loose PN counting that is based on assigning a PN order to each power of $x$, tidal effects first enter effectively as 5PN corrections. As explained above, physically and formally the PN corrections to the point-mass dynamics and GWs and finite size effects are distinct. Nevertheless, the net effect of unknown PN point-mass terms can impact measurements of matter effects, as will be discussed in more detail in Sec.~\ref{sec:PE}. There are two classes of effective or phenomenological models for black hole binaries that effectively include all PN orders for the point-mass dynamics in an approximate way. These are the effective one body (EOB) model~\cite{Buonanno:1998gg,Buonanno:2000ef} and the so-called ``Phenom'' models \cite{Ajith:2007qp,Ajith:2007kx}, both of which aim to combine the available information on the relativistic two-body problem from different regimes into a single framework to generate waveforms for data analysis as we will discuss next. 

\section{Tidal Effective-One-Body Models for binary inspirals}
\label{sec:GWmodels_EOB}

The EOB approach~\cite{Buonanno:1998gg,Buonanno:2000ef} (see also the reviews~\cite{Damour:2016bks,Damour:2011xga,Buonanno:2014aza,Damour:2012mv}) combines results from the PN theory, valid for arbitrary mass ratios, with information about the strong-field effects in the test-particle limit. 
The EOB dynamics involves a nonlinear mapping of the phase space of the binary system to that of an effective particle on a nearly geodesic trajectory in an effective spacetime. The effective particle's mass is the reduced mass while the mass of the spacetime is the total mass of the binary system. This setting reduces to test-mass motion in a BH spacetime in the test-mass limit. For finite mass ratio, corrections are added to the metric potential that depend on the symmetric mass ratio $\nu \in [0,1/4]$ and thus deform the spacetime away from those of a BH. In addition, non-geodesic corrections that depend on higher-than-quadratic powers of momentum appear. These modification are dictated by the requirement to reproduce the known PN results in the weak-field limit. This theoretical structure achieves an extremely concise resummation of PN information into a small number of terms. There is still considerable remaining freedom in the model, for instance on choosing the functional form for the potentials which are generally expressed as non-analytic functions. These choices are constrained by considerations such as a pathology-free physical behavior of quantities such as energetics of the system under variations of parameters. In addition, extra terms that parameterize unknown higher-order information are added to the model that are fixed by comparing to results from NR and gravitational self-force~\cite{Barack:2018yvs,Barack:2009ey,Bini:2016cje,Antonelli:2019fmq,Antonelli:2019ytb,Barack:2010ny}.

The dissipative sector of current EOB models is based on a concise factorized representation of the gravitational waveforms~\cite{Nagar:2019wrt,Damour:2007xr,Damour:2008gu,Messina:2018ghh,Taracchini:2012ig,Taracchini:2013rva,Pan:2013rra,Nagar:2016ayt,Bernuzzi:2010xj,Damour:2008gu,Pan:2010hz} that accurately accounts for the modifications to the GW propagation from the spacetime curvature due to the total mass of the binary. These GW modes are explicit algebraic expressions that depend on the instantaneous EOB dynamics. The evolution of the orbits is computed by imposing energy and angular momentum balance. Calculating a full waveform therefore requires solving the time-domain coupled system of the Hamiltonian equations, the GWs, and the corresponding fluxes that give dissipative contributions to the equations of motion over an inspiral.

For spinning BBHs, there are two families of EOB models, one by A. Buonanno's group and collaborators, e.g.~\cite{Bohe:2016gbl,Taracchini:2012ig,Taracchini:2013rva,Pan:2013rra,Pan:2013tva,Barausse:2009xi,Barausse:2009aa,Babak:2016tgq,Barausse:2011ys,Pan:2011gk,Cotesta:2018fcv}, and one by T. Damour, A. Nagar, and their collaborators, e.g.~\cite{Damour:2014sva,Damour:2014yha,Damour:2012ky,Damour:2009kr,Damour:2008te,Damour:2008gu,Damour:2007vq,Messina:2018ghh,Nagar:2017jdw,Nagar:2016ayt,Damour:2001tu,Damour:2008qf,Nagar:2011fx,Balmelli:2013zna,Nagar:2019wds,Nagar:2017jdw,Damour:2014yha,Bernuzzi:2012ku,Nagar:2011aa,Nagar:2011fx}. 
These BBH EOB models both reproduce the PN and test-mass results in the appropriate limits, use the same energy mapping to the effective phase spaces, and the effective description of a test particle in an effective spacetime. They differ in choices made for the remaining freedom, for instance, the non-analytic functions assumed for the metric potentials and similar aspects that are not fixed by any analytical knowledge, as well as how effects of spins are included, i.e., imposing the exact limit of a spinning particle in Kerr spacetime or not. Both of these models also incorporate information from NR by calibrating parameterized, unknown high-order PN coefficients to NR data to obtain an accurate description of GW signals.

\begin{figure}[t]
 \centering
 \includegraphics[width=\columnwidth]{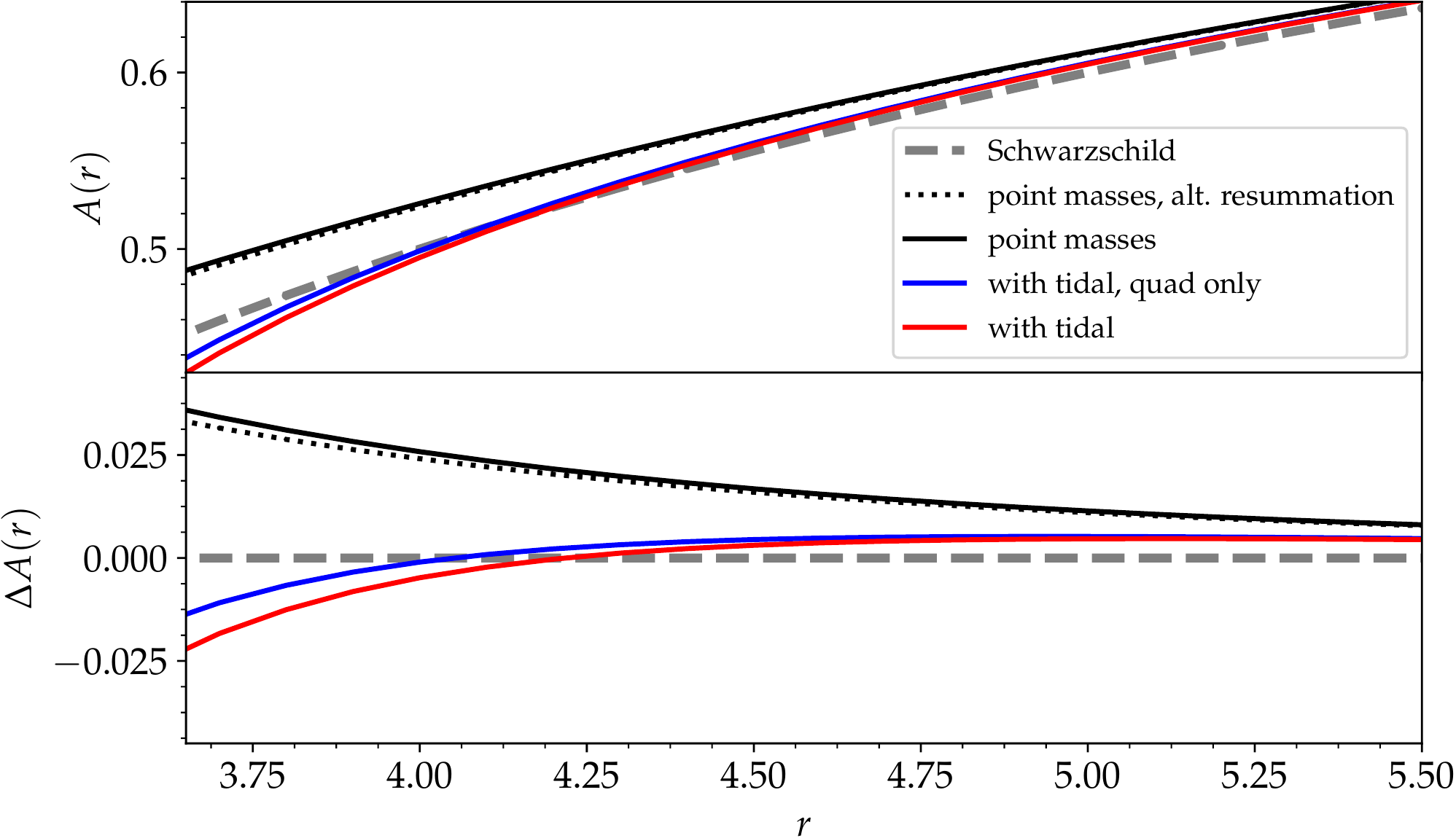}
 \caption{Tidal effects in the EOB potential $A(r)$ which determines the energetics of circular orbits. The grey curve illustrates the test-particle limit of a Schwarzschild BH. Comparing this to the black curves shows the change in the potential due to finite mass ratio, where two different choices of non-analytic functions for resumming the available PN information are shown as the solid and dotted curves. These point-mass results change to those illustrated by the red and blue curves when adiabatic tidal effects are included, where the quadrupole (blue) gives the main contribution and the octupole (red) a small additional imprint. The curves terminate at the light ring of the tidal EOB model. The bottom panels shows the 
 difference in the potential with respect to the Schwarzschild BH.}
 \label{fig:Apotential}
\end{figure}

Tidal effects can be included in EOB models in different ways~\cite{Akcay:2018yyh,Nagar:2018zoe,Bernuzzi:2014owa,Bernuzzi:2012ci,Baiotti:2010xh,Damour:2009wj,Hinderer:2016eia,Steinhoff:2016rfi,Bini:2014zxa,Bini:2012gu}. For simplicity, we will give details for the nonspinning BBH baseline models and note that the inclusion of tidal effects carries over to the spinning baseline models. The effect of the spin-induced multipole moments is also included in these EOB models~\cite{Nagar:2018plt,Lackey:2018zvw}. As noted above, several other matter effects from spins together with tidal effects have not been calculated in detail. The qualitative impact of tidal effects on the EOB dynamics is illustrated in Fig.~\ref{fig:Apotential}, which shows the gravitational potential $A(r)$.

\subsection{Brief summary of EOB models for nonspinning binaries}
The conservative dynamics within the EOB approach are described by the Hamiltonian 
\begin{equation}
H_{\rm EOB}=M\sqrt{1+2\nu\left(H_{\rm eff}-1\right)}-M, 
\end{equation}
where the effective Hamiltonian $H_{\rm eff}$ is that of a particle of mass $\mu$. This energy mapping also emerges in quantum-electrodynamics~\cite{Brezin:1970zr}, scattering theory~\cite{Damour:2016gwp,Vines:2017hyw,Vines:2018gqi,Damour:2017zjx}, and from explicit comparisons of the action variables~\cite{Buonanno:1998gg}.
For nonspinning binaries the the effective Hamiltonian is given by
\begin{equation}
H_{\rm eff}^2=p_{r*}^2+A\left(1+\frac{p_{\phi}^2}{r^2}+2 \left(4-3\nu \right) 
\frac{p_{r*}^4}{\nu r^2}\right). \label{eq:Heff}
\end{equation}
Here, $p_\phi$ is the azimuthal angular momentum 
and $p_{r*}=p_r/\sqrt{D}$ is the rescaled 
radial momentum; all the momenta are per unit reduced mass. 
The EOB potentials $A$ and $D$, describing the effective spacetime,  
can be written as $A=A_{\rm pm}+A_{\rm tidal}$ and $D=D_{\rm pm}$.
The different choices for the 
point-mass potentials  
$A_{\rm pp}$ are either Eqs.~(A1)-(A2h) of Ref.~\cite{Steinhoff:2016rfi} with the 
calibration parameter determined most recently in~\cite{Bohe:2016gbl} or the resummation from~\cite{Akcay:2018yyh,Nagar:2018zoe,Bernuzzi:2014owa}. 
The potential $D$ is either taken from Eq.~(A4) 
of Ref.~\cite{Steinhoff:2016rfi} or~\cite{Nagar:2018zoe}. 

The EOB equations of motion are
\begin{subequations}
\begin{equation}
\begin{aligned}
\frac{d r}{dt}&=\frac{A}{\sqrt{D}} \frac{\partial H_{\rm EOB}}{\partial p_{r*}},  \qquad \qquad
\frac{d p_{r*}}{dt}=-\frac{A}{\sqrt{D}} \frac{\partial H_{\rm EOB}}{\partial r}+{\cal F}_r , \\
\frac{d \phi}{dt}&=\frac{\partial H_{\rm EOB}}{\partial p_\phi},  \qquad \qquad \qquad
\frac{d p_\phi}{dt}= {\cal F}_\phi, 
\end{aligned}
\end{equation}
\end{subequations}
The factors of $A/\sqrt{D}$ appear because $p_{r*}$ and $r$ are not canonically conjugate variables. 
The gravitational radiation reaction forces ${\cal F}_\phi$ and ${\cal F}_r$ are constructed from
\begin{equation}
{\cal F}_\phi=-\frac{1}{\Omega_\phi}\mathcal{\dot E}_{\rm rad}, \ \ \ \ \ {\cal F}_r=\frac{p_{r*}}{p_\phi}{\cal F}_\phi  , \ \ \ \ \ \ \label{eq:EOBforces}
\end{equation}
where the energy flux $\mathcal{\dot E}_{\rm rad}$ is given by 
\begin{equation}
\mathcal{\dot E}_{\rm rad} =\frac{\Omega_\phi^2}{8\pi} \sum_{\ell=2}^8\sum_{m=0}^\ell m^2 |h_{\ell m}^{\rm F}|^2. \label{eq:EOBEdot}
\end{equation}
The sum is only over positive $m$ since $|h_{\ell -m}^{\rm F}|=|h_{\ell m}^{\rm F}|$. The factorized EOB waveforms for point masses have the form~\cite{Damour:2007xr,Damour:2008gu,Damour:2002vi}
\begin{equation}
{h_{\ell m}^{\rm F}}_{\rm pp}=h_{\ell m}^{\rm Newt}S_{\ell m}T_{\ell m}\rho_{\ell m}^\ell N_{\ell m}.  \label{eq:hlmEOBpp}
\end{equation}
Each of these terms depends on the EOB canonical variables and the parameters of the binary, see e.g.~\cite{Bohe:2016gbl,Nagar:2018zoe} for the explicit up to date expressions. 

\subsubsection{Adiabatic tidal effects using PN information}
Adiabatic tidal effects on the energetics of the binary system can be captured in the EOB approach through tidal contributions to the potentials of the form~\cite{Bini:2012gu,Damour:2009wj,Vines:2010ca}
\begin{eqnarray}
A^\text{tidal} &=&-{\cal A}_{\ell=2}^{(A)} \hat{A}^{(A)}  -{\cal A}_{\ell=3}^{(A)} \bigg[ 1 + \left(\frac{15}{2}\frac{m_A}{M}-2\right) r^{-1}  \nonumber\\
  && + \left(\frac{110}{3}\frac{m_A^2}{M^2}-\frac{311}{24}\frac{m_A}{M}+\frac{8}{3}\right)r^{-2} \bigg] +(A \leftrightarrow B). \label{AAT2PN}
\end{eqnarray}
Here ${\cal A}_\ell^{(A)} $ are the Newtonian $\ell-$th multipolar tidal potentials given by
\begin{equation}
{\cal A}_\ell^{(A)}=\frac{2 m_B}{m_A}k_\ell^{(A)}\frac{R_A^{2\ell+1}}{r^{2\ell+2}}. \label{eq:alpha}
\end{equation} 
The 2PN corrections to the tidal potential $\hat{A}^{(A)}$ from Eqs.~(6.6) and (6.18) of Ref.~\cite{Bini:2012gu} are given by
\begin{equation}
\hat{A}_{  {\rm adPN}}^{(A)}= 1 + \frac{5m_A}{2M} r^{-1}  
      + \left(\frac{337}{28}\frac{m_A^2}{M^2}+\frac{1}{8}\frac{m_A}{M}+3\right)r^{-2} \label{AAT}
      \end{equation}
 Tidal effects also influence the dissipative sector since the moving multipole moments of the NS
 contribute to the gravitational radiation. In the EOB model this is accounted 
 for by adding to the waveform modes of Eq.~\eqref{eq:hlmEOBpp} a tidal 
 contribution $h_{\ell m}^{\rm tidal}$ to the GW modes so that
 \begin{equation}
 h_{\ell m}^{\rm F}={h_{\ell m}^{\rm F}}_{\rm pp}+h_{\ell m}^{\rm tidal}. \label{eq:hlmEOB}
 \end{equation}
The explicit results for the adiabatic tidal terms $h_{\ell m}^{\rm tidal}$ are given 
in Eqs.~(A14)-(A17) of Ref.~\cite{Damour:2012yf} for $\ell \leq 3$.  
 
\subsubsection{Dynamical f-mode tidal effects}
The 
effects of dynamic tides from the NS's $f-$modes within the EOB framework are discussed 
in detail in Refs.~\cite{Steinhoff:2016rfi,Hinderer:2016eia}. Here, we consider only the 
approximate model that is used in practical data analysis applications. In this model, the potential from 
Eqs.~\eqref{AAT2PN} and \eqref{AAT} is used but with $k_\ell$ multiplied by a frequency-dependent 
enhancement factor such that 
$k_\ell \to k_\ell \hat{k}^ {\rm dyn}_{\ell}$ with
\begin{equation}
\hat{k}^{\rm dyn}_{\ell}=a_\ell+b_\ell \bigg[ \frac{\omega_{0\ell}^2 }{\omega_{0\ell}^2-(m\Omega^2)}+\frac{\omega_{0\ell}^2}{2\sqrt{\epsilon_m}\hat t \Omega^\prime (m\Omega)^2}+\frac{\sqrt{\pi}\omega_{0\ell}^2}{\sqrt{3}\sqrt{\epsilon_m}(m\Omega)^2} {\cal Q}_{\ell m} \bigg] \ \ \  \label{keffDT}
\end{equation}
with $\Omega^\prime=3/8$, and $\Omega=M^{1/2}r^{-3/2}$. The quantity ${\cal Q}_{\ell m}$ is given by
\begin{equation}\label{QlmDT}
{\cal Q}_{\ell m}=\cos(\Omega^\prime \hat t^2)\left[1+2 {\rm F}_{\rm S}\left(\frac{\sqrt{3}}{\sqrt{4\pi}}\hat t \right)\right]-\sin(\Omega^\prime \hat t^2)\left[1+2 {\rm F}_{\rm C}\left(\frac{\sqrt{3}}{\sqrt{4\pi}}\hat t \right)\right],
\end{equation}
where the functions ${\rm F}_{\rm S}$ and ${\rm F}_{\rm C}$ are 
Fresnel sine and cosine integrals respectively using the conventions 
in Mathematica. The quantities $\hat t$ and $\epsilon_m$ are defined as
\begin{equation}
\hat t=\frac{8}{5\sqrt{\epsilon_m}}\left(1-\frac{r^{5/2}\omega_{0\ell}^{5/3}}{m^{5/3}M^{5/6}}\right),
\qquad 
\epsilon_m=\frac{256 \mu  M^{2/3}\omega_{0\ell}^{5/3}}{5m^{5/3}}.
\end{equation}
In Eq.~\eqref{keffDT} a body label $A,B$ on the quantities $\omega_{0\ell}$, $\hat t$, $\epsilon_m$, and 
${\cal Q}_{\ell m}$ is implied. 
For each $\ell$-multipole only $m=\ell$ contributes in Eq.~\eqref{keffDT} because the effect of modes with
 $m<\ell$ has already been taken into account as adiabatic contributions. 
For the quadrupole and octupole multipole moments the coefficients are given by 
$(a_2, a_3)=(\frac{1}{4},\frac{3}{8})$ and $(b_2,b_3)=(\frac{3}{4}, \frac{5}{8})$.

Similar to the treatment for the conservative 
dynamics, the effect of dynamic $f-$mode tides can be incorporated 
in the dissipative sector in an approximate way by multiplying the occurences of 
$k_\ell$ in $h_{\ell m}^{\rm tidal}$ in Eq.~\eqref{eq:hlmEOB} by an effective function 
$k_\ell \to k_\ell \ \hat{k}_{\ell\ {\rm dyn}}^{\rm diss}$. 
For $\ell=2$ this function is given by
\begin{equation}
\hat{k}_{2\ {\rm dyn}}^{(A)\ {\rm diss}}=\frac{\omega_{02}^2+\hat{k}_{2\ {\rm dyn}}^{(A)}\left(\omega_{02}^2+6 \frac{m_B}{M}\Omega^2\right)}{3\Omega^2\left(1+2 \frac{m_B}{M}\right)}, 
\end{equation}
where $\hat{k}_{2\ {\rm dyn}}^{(A)}$ is the enhancement function 
for body $A$ in the conservative dynamics from Eq.~\eqref{keffDT}.

\subsubsection{Adiabatic tidal effects with strong-field enhancement}
Gravitational self-force calculations have recently computed tidal 
invariants that contain information about strong-field tidal effects in the 
limit of small mass ratios, to linear order in $m_A/M$ \cite{Dolan:2014pja,Bini:2014zxa}. These results have been augmented in Refs.~\cite{Bini:2014zxa,Bernuzzi:2014owa,Akcay:2018yyh,Nagar:2018zoe} by a term 
$\propto m_A^2/M^2$ that would describe currently unknown second-order 
self-force effects. Specifically, in this model Eq.~\eqref{AAT2PN} is employed with 
the potential $\hat{A}^{(A)}$ given by
\begin{equation}
\hat{A}_{ {\rm adGSF^+}}^{(A)}=1+\frac{3r^{-2}}{\left(1-r_{\rm LR}/r\right)} +\frac{m_A}{M}\frac{a_1^{\rm GSF}(r^{-1}) }{  \left(1-r_{\rm LR}/r\right)^{7/2}}+\frac{m_A^2}{M^2} \frac{a_2^{\rm 2GSF}(r^{-1}) }{  \left(1-r_{\rm LR}/r\right)^{p}} +(A\leftrightarrow B)
\label{eq:AGSF}
\end{equation}
In Eq.~\eqref{eq:AGSF}, the coefficient $a_1^{\rm GSF}(r^{-1}) $ is obtained 
from Eqs.~(7.24)--(7.27) of Ref.~\cite{Bini:2014zxa}. 
In the model of Refs.~\cite{Bernuzzi:2014owa,Akcay:2018yyh,Nagar:2018zoe} the choices for the unknown parameters are
$a_2^{\rm 2GSF}=337r^{-2}/28$ and $p=4$. The radius of the light ring 
$r_{\rm LR}$ is obtained from the conservative EOB dynamics by solving
\begin{equation}
\tilde{A}(r_{\rm LR})-\frac{r_{\rm LR}}{2}\tilde{A}^\prime(r_{\rm LR})=0, \label{eq:rLRdet}
\end{equation}
with the potential $\tilde{A}=A_{\rm pp}+A^{\rm tidal}_{\rm adPN}$ 
from the PN model in Eqs.~\eqref{AAT2PN} and \eqref{AAT}. 
These tidal models are included in a slightly different EOB model 
than the dynamical $f$-mode effects. There are a few minor differences in the point-particle sector such as 
the factor $N_{\ell m}$ in the waveform 
modes of Eq.~\eqref{eq:hlmEOBpp}, a different resummation of the potentials $A_{\rm pm}$ and $D$, and the arguments of factors in $h_{\ell m}^{\rm F}$ which involve powers of $v_\phi=(\partial H_{\rm EOB}/\partial p_\phi)^{-2/3}\Omega$ are evaluated for circular orbits instead of $v=\Omega^{1/3}$, 
 Pad{\'e}-resummation of higher-order hereditary terms in $h_{\ell m}^{\rm F}$, and a 
different calibration of the BBH model to NR. 

\section{Phenomenological models}
\label{sec:GWmodels_NRTidal}

Over the last years, there has been an active development of phenomenological BNS waveform models. 
The main idea behind the existing models is the possibility to augment BBH approximants
with a tidal description incorporating finite size effects of the NSs. 
Thus, a fundamental assumption is that the total GW phase can be decomposed into
\begin{equation}\label{eq:phi_omg}
 \phi(M \omega) = \phi_{\rm pm} (M \omega) + \phi_{\rm SO}(M \omega) + \phi_{\rm SS}(M \omega) + 
 \phi_{\rm T}(M \omega) + \cdots \ , 
\end{equation}
where $\phi_{\rm pm}$ denotes the point-mass, $\phi_{\rm SO}$ the spin-orbit, $\phi_{\rm SS}$ the spin-spin, 
and $\phi_{\rm T}$ the tidal phase contribution. We will focus in the following on an accurate representation 
of the tidal phase $\phi_{\rm T}$ that is required for a proper modelling of BNS systems. 
However, in addition and as pointed out in Sec.~\ref{sec:coalescence}, also the spin-spin interaction, $\phi_{\rm SS}$, 
or higher order spin-contributions incorporate information about the EOS due to the presence of the quadrupole moment. Therefore, $\phi_{\rm SS}$
needs to be adjusted to describe the BNS coalescence in comparison to BBH systems. 

As in the time domain, the frequency domain GW phase is typically decomposed as 
\begin{equation}\label{eq:psi_omg}
 \psi(M \omega) = \psi_{\rm pm} (M \omega) + \psi_{\rm SO}(M \omega) + 
 \psi_{\rm SS}(M \omega) + \psi_{\rm T}(M \omega)+ \cdots \ .
\end{equation}

In the following subsections, we will discuss different phenomenological approaches to describe $\phi_{\rm T}$ and $\psi_{\rm T}$.

\subsection{NRTidal-approximants}

We start our discussion with the newest NRTidal version (NRTidalv2) described in Ref.~\cite{Dietrich:2019kaq}. 
Earlier versions use similar principles and assumptions and therefore will not be discussed to avoid repetition, 
but we refer to~\cite{Dietrich:2017aum,Dietrich:2018uni,Dietrich:2018upm} for more details. 

The main idea of the NRTidal~\cite{Dietrich:2017aum,Dietrich:2018uni} approach 
is to provide a closed-form approximation for the tidal phase contribution 
$\phi_{\rm T}$ and $\psi_{\rm T}$. 
In addition NRTidalv2 incorporates spin-spin and cubic-in-spin effects up to 3.5PN, 
these contributions depend on the quadrupole and octupole moment of the individual NSs. 
These effects are included following the PN framework~\cite{Dietrich:2019kaq}.
Furthermore for NRTidalv2, an additional amplitude correction is added. 

As discussed, tidal phase contributions enter 
at the effectively 5th PN order and analytic knowledge exists 
up to the 7.5th PN order (with incomplete knowledge at 7PN);
cf.~Sec.~\ref{sec:GWmodels_PN}. 
Since the current NRTidal models use only leading and next-to-leading 
order mass-ratio effects, we restrict the parameters $c_1^{A,B},c_{3/2}^{A,B},c_2^{A,B},c_{5/2}^{A,B}$ 
in Eqs.~\eqref{eq:PN_coeff} to their equal-mass values. 
Thus, the superscripts $A$ and $B$ are discarded in the following. 

Under this assumption, an effective representation of tidal effects during the inspiral 
is obtained using
\begin{equation}
\phi_T (x) = - \kappa_{\rm eff}^T
\frac{13}{ 8 \nu} x^{5/2} P_{\rm NRTidalv2}(x) \ ,  \label{eq:NRTidal_TD}
\end{equation}
where $P_{\rm NRTidalv2}(x)$ is given by the Pad\'e approximant
\begin{equation}
P_{\rm NRTidalv2}(x) = \frac{1 + n_1 x+ n_{3/2} x^{3/2} + n_2 x^2 + n_{5/2} x^{5/2} + n_3 x^3}
{1+ d_1 x+ d_{3/2} x^{3/2}+ d_2 x^2} \label{eq:Pade_TD}
\end{equation}
and 
\begin{equation}
\kappa_{\rm eff}^T = \frac{3}{16} \tilde{\Lambda}. 
\end{equation}

To enforce consistency with the analytic PN knowledge [Eqs.~\eqref{eq:PhiT_DNV}-\eqref{eq:PN_coeff}],
some of the parameters in Eq.~\eqref{eq:Pade_TD} are fixed: 
\begin{subequations}
\begin{eqnarray}
 n_1     & = & c_1 + d_1, \\
 n_{3/2}     & = & \frac{c_1 c_{3/2}-c_{5/2}-c_{3/2} d_1 + n_{5/2}}{c_1}, \\
 n_{2} & = & c_2 + c_1 d_1 + d_2 , \\
 d_{3/2} & = & - \frac{c_{5/2}+c_{3/2} d_1 - n_{5/2}}{c_1}.
 \end{eqnarray}
  \end{subequations}
with
\begin{subequations}
\begin{align}
  c_{1}   & =  \frac{3115}{624}, \qquad  & c_{3/2} & = -\frac{5 \pi}{2}, \\
  c_2     & =  \frac{28024205}{1100736},  \qquad   & c_{5/2} & =  -\frac{4283 \pi}{312}.
      \end{align}
      \end{subequations}
The remaining parameters are fitted to high-resolution NR data
\begin{subequations}
\begin{align}
  n_{5/2} & =  312.48173, \qquad   & n_3   & = -342.15498, \\
  d_1     & = -20.237200,  \qquad  & d_2  & =  -5.361630;
\end{align}
\end{subequations}
cf.~Sec.~\ref{sec:GWmodels_NR}.

To obtain the tidal corrections in the frequency domain, 
the NRTidal approach uses the stationary phase approximation (SPA); e.g.,~\cite{Damour:2012yf},
\begin{equation}
  \frac{\text{d}^2 \psi_T(\omega)}{\text{d} \omega^2} =
  \frac{1}{\omega} \frac{\text{d} \phi_T(\omega)}{\text{d} \omega} \ . \label{eq:SPA}
\end{equation}
Eq.~\eqref{eq:SPA} is solved numerically and the final result is again approximated 
by a Pad{\'e} approximant
\begin{equation}
\psi_T (x) = - \kappa_{\rm eff}^T
\frac{39}{ 16 \nu} x^{5/2} \tilde{P}_{\rm NRTidalv2}(x) \  \label{eq:NRTidal_FD}
\end{equation}
with
\begin{equation}
\tilde{P}_{\rm NRTidalv2}(x) = \frac{1 + \tilde{n}_1 x+  \tilde{n}_{3/2} x^{3/2} + \tilde{n}_2 x^2+ \tilde{n}_{5/2} x^{5/2} + \tilde{n}_3 x^3}
{1+ \tilde{d}_1 x+ \tilde{d}_{3/2} x^{3/2}+ \tilde{d}_2 x^2} \label{eq:Pade_FD}.
\end{equation}

As for the time domain, some of the Pad{\'e} parameters are constrained to obtain the correct PN limit 
up to 7.5PN (setting unknown terms at 7PN to zero): 
\begin{subequations}\label{eq:Pade_FD_constraints}
\begin{eqnarray}
 \tilde{n}_1     & = & \tilde{c}_1 + \tilde{d}_1, \label{eq:Pade_FD_constraint1} \\
 \tilde{n}_{3/2}     & = & \frac{\tilde{c}_1 \tilde{c}_{3/2}-\tilde{c}_{5/2}-
 \tilde{c}_{3/2} \tilde{d}_1 + \tilde{n}_{5/2}}{\tilde{c}_1},  \label{eq:Pade_FD_constraint2}  \\
 \tilde{n}_{2} & = & \tilde{c}_2 + \tilde{c}_1 \tilde{d}_1 + \tilde{d}_2 ,  \label{eq:Pade_FD_constraint3}  \\
 \tilde{d}_{3/2} & = & - \frac{\tilde{c}_{5/2}+\tilde{c}_{3/2} \tilde{d}_1 - \tilde{n}_{5/2}}{\tilde{c}_1},  \label{eq:Pade_FD_constraint4} 
 \end{eqnarray}
 \end{subequations}
with 
\begin{subequations}\label{eq:Pade_FD_coeffs}
\begin{align}
  \tilde{c}_{1}   & =  \frac{3115}{1248}, \qquad
  & \tilde{c}_{3/2} & = -\pi , \label{eq:Pade_FD_coeff1} \\
  \tilde{c}_2     & =  \frac{28024205}{3302208},  \qquad
  & \tilde{c}_{5/2}  & =  - \frac{4283 \pi}{1092}. \label{eq:Pade_FD_coeff2}
\end{align}
\end{subequations}

The remaining, fitting coefficients are 
\begin{subequations}
\begin{align}
  \tilde{n}_{5/2} & =  90.550822, \qquad   
  & \tilde{n}_3     & = -60.253578, \\        
  \tilde{d}_1     & = -15.111208,  \qquad  
  & \tilde{d}_2     & =  8.0641096. 
\end{align}
\end{subequations}

In addition to the phase correction the NRTidalv2 approach 
also includes tidal amplitude corrections. 
While the GW amplitude is in general less important than a correct phase model, 
one needs a realistic and accurate estimate of the GW amplitude to measure 
the source distance properly, thus, additional tidal amplitude corrections might play a role 
in the measurement of the Hubble constant~\cite{Schutz:1986gp,Abbott:2017xzu} or in determining if a GW source is 
gravitationally lensed~\cite{Pang:2020qow}. 

The amplitude correction is directly derived in the frequency domain and calibrated 
in such a way that the 6PN tidal corrections as well as the phenomenological
corrections of~\cite{Kawaguchi:2018gvj} are recovered: 
\begin{equation}
 \tilde{A}_T^{\rm NRTidalv2} =
 - \sqrt{\frac{5 \pi \nu}{24}} \frac{9 M^2 }{D_L}
 \kappa^T_{\rm eff} x^{13/4} \frac{1 + \frac{449}{108} x + \frac{22672}{9} x^{2.89}}{1+d \ x^4},
 \label{eq:A_FD_NRTP}
\end{equation}
with $d= 13477.8$.

To truncate the waveform after the moment of merger, a Planck taper~\cite{McKechan:2010kp} 
is added at the merger frequency
\begin{equation}
 \omega_{\rm mrg} = \omega_0 \sqrt{\frac{1}{q}}
 \frac{1+ n_1 \kappa_{\rm eff}^T + n_2 (\kappa_{\rm eff}^T)^2}
 {1+d_1 \kappa_{\rm eff}^T + d_2 (\kappa_{\rm eff}^T)^2}, \label{eq:Momega_mrg}
\end{equation}
with $n_1=3.354 \times 10^{-2}, n_2 = 4.315 \times 10^{-5},
d_1 = 7.542 \times 10^{-2}, d_2 = 2.236 \times 10^{-4}$.
The parameter $\omega_0=0.3586$ is chosen so that for the equal-mass cases the correct 
non-spinning BBH limit~\cite{Healy:2017mvh} is recovered.

The tapering ends at $1.2$ or $1.3$ times the merger frequency, $\omega_{\rm mrg}$,  
depending on the exact implementation and NRTidal model version.  
The final amplitude is given by  
\begin{equation}
 \tilde{A} = (\tilde{A}_{\rm BBH} + \tilde{A}_T^{\rm NRTidalv2}) \ \tilde{A}_{\rm Planck};
\end{equation}
we note again that $\tilde{A}_T^{\rm NRTidal}=0$ for the first/original NRTidal version.  
Because of the smooth frequency and amplitude evolution even after the moment of merger,
the Planck taper only introduces negligible errors and does not lead to biases in 
the parameter estimation; cf.~Ref.~\cite{Dudi:2018jzn}. 

\subsection{Phenomenological model of Kawaguchi et al.}

The phenomenological waveform model extension of Kawaguchi et al.~\cite{Kawaguchi:2018gvj}
follows a similar approach as the NRTidal framework, i.e., it uses the assumption that the 
GW phase can be split into individual components, Eq.~\eqref{eq:psi_omg}. 
The main difference between~\cite{Kawaguchi:2018gvj} and the NRTidal model are:
\begin{enumerate}[(i)]
 \item Kawaguchi et al. employ hybrid waveforms combining NR and \texttt{SEOBNRv2T}-waveforms~\cite{Taracchini:2013rva} 
       in the frequency domain and fits directly the frequency domain phase;
 \item Kawaguchi et al. use a model function of the form 
 \begin{equation}
  \psi_T = \frac{3 x^{5/2}}{128 \nu } \left( -\frac{39}{2} \tilde{\Lambda} (1+a \Lambda^{2/3} x^p) \right)
  \left( 1+ \tilde{c}_1 x  + \tilde{c}_{3/2} x^{3/2} + \tilde{c}_2 x^2 + \tilde{c}_{5/2} x^{5/2} \right)
 \end{equation}
 to include non-linear tidal effects instead 
 of higher order PN effects as in Eq.~\eqref{eq:Pade_FD}; 
 \item The model is valid up to a frequency 
 of $1000\rm Hz$ to ensure that no postmerger frequency contributions might effect the 
 inspiral approximant. 
\end{enumerate}

In addition, the model includes a tidal amplitude correction of the form 
\begin{equation}
\begin{split}
 \tilde{A}_T^{\rm Kawaguchi} &=
 \sqrt{\frac{160 \pi \nu}{27}} \frac{M^2}{D_L} \
 {\kappa^T_{\rm eff}} x^{-7/4} \\
&\quad \times  	\left(-\frac{27}{16}x^5-\frac{449}{64}x^6-4251 x^{7.890}\right).
 \label{eq:A_FD_Kawaguchi}
\end{split}
\end{equation}

\subsection{Waveform-model Comparison}

\begin{figure*}[t]
 \centering
 \includegraphics[width=1\textwidth]{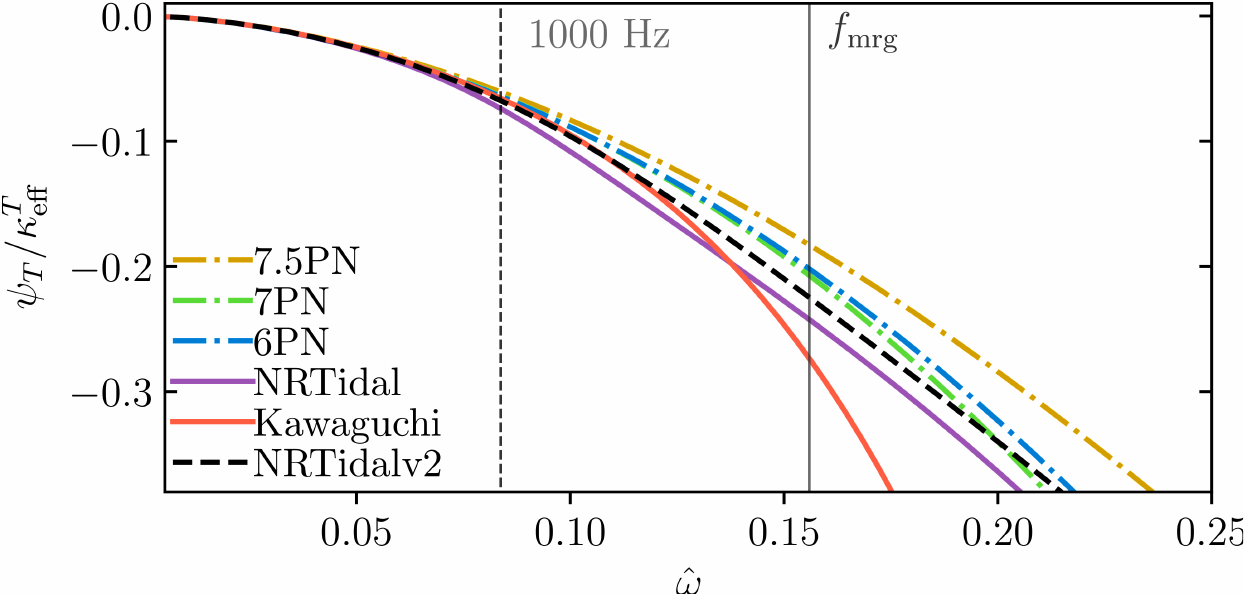}
 \caption{Comparison of the tidal phase of different PN and phenomenological waveform approximant 
 as discussed in the main text; cf.~\cite{Dietrich:2019kaq} for further details.}
 \label{fig:fit_fd}
\end{figure*}

We compare different frequency-domain tidal models in Fig.~\ref{fig:fit_fd}.
The plot shows the 6PN, 7PN, and 7.5PN tidal predictions. 
We find that the 7.5PN approximant is the least attractive
and that all PN predictions are less attractive than 
the phenomenological GW models. 
The original NRTidal approximant reduces to the 6PN curve, 
while NRTidalv2 and the model of Kawaguchi reduce, by construction, to the 
7.5PN approximation. 
It also becomes obvious that up to a frequency of $1000\rm Hz$ or even beyond, the 
NRTidalv2 and Kawaguchi model predict almost identical tidal effects. 
This changes for very large tidal deformabilities, but for configurations similar to GW170817 
as shown in Fig.~\ref{fig:fit_fd} with $\tilde{\Lambda} = 392.1$, 
systematic waveform uncertainties seem to be negligible; see also~\cite{Narikawa:2019xng}.
Extrapolating the model of Kawaguchi et al.~\cite{Kawaguchi:2018gvj} 
beyond its validity region up to the merger leads to an overestimation of tidal effects. 

Considering the difference between NRTidal and NRTidalv2, we find that 
NRTidal predicts larger tidal effects than NRTidalv2, consequently, when the model is employed for 
the analysis of real GW signals, one can expect that it will predict smaller tidal deformabilities to 
compensate for the larger tidal phase contribution. 

\section{Analyzing BNS signals}
\label{sec:PE}

As emphasized in Sec.~\ref{sec:coalescence}, the primary difference of a 
BNS coalescence from a BBH coalescence comes from the presence of finite-size effects. 
From the point of view of estimating source properties, this involves 
parameterizing the matter effects and performing data analysis on a 
larger parameter space. 

\subsection{Estimating source properties}

The first analysis to establish the possibility of using a Bayesian approach to 
estimate BNS source properties was presented in~\cite{DelPozzo:2013ala} using simulations 
of BNS mergers. This initial work was extended in~\cite{Agathos:2015uaa}, 
where higher tidal PN orders and the dominant 
spin-induced quadrupole moment were included. 
These works found that constraints on the binary properties can be improved by combining 
multiple observations. Ref.~\cite{Lackey:2014fwa} put forward a way to parametrize the EOS and combined 
the tidal deformability parameter from multiple events to constrain EOS information. 
All this work however focused on using the PN inspiral-only 
waveform model, \texttt{TaylorF2}. Hybrid waveforms were constructed using NR simulations and 
analytical inspiral waveforms and improvement of measurability of matter effects were
investigated in~\cite{Read:2013zra}. 
The first parameter estimation simulations using different waveform models 
were performed towards a study of systematics connected to GW170817 in~\cite{Abbott:2018wiz}. 
Since the observation of GW170817, more studies on parameter estimation of BNSs have been carried out, 
as discussed below. \\

\noindent Bayesian inference relies on using Bayes' theorem to estimate the 
posterior probability distribution functions. This is done by calculating the likelihood of the observed data 
and folding in prior assumptions about our system, in our case, the parameters characterizing the 
source~\cite{Veitch:2009hd,Veitch:2014wba}. 
The information from individual parameters characterizing the source is encoded 
in posterior probability distributions 
\begin{equation}
 p(\vec{\theta}|\mathcal{H}_s,d) = \frac{p(\vec{\theta}|\mathcal{H}_s)p(d|\vec{\theta},\mathcal{H}_s)}{p(d|\mathcal{H}_s)}.
 \label{eqn:Bayes}
\end{equation}
Here, $\vec{\theta}$ represents the parameter set and $\mathcal{H}_s$ is the hypothesis that a 
GW signal depending on the parameters $\vec{\theta}$ is present in the data $d$. 
The parameter 
set of a BNS source consists of the usual parameters describing a BBH source 
$\{ m_A, m_B, \chi_A, \chi_B, \theta, \phi, \iota, \psi, D_L, t_c, \varphi_c\}$, 
and in addition the tidal deformability components
$\Lambda_A$ and $\Lambda_B$. 
The likelihood of obtaining a signal $h(t)$ in data stream $d(t) = h(t) + n(t)$, 
is proportional to 
\begin{equation}
p(d|\vec{\theta},\mathcal{H}_s) \propto \exp{\left [-\frac{1}{2}(d-h|d-h)\right ]}. 
\label{eqn:lhood}
\end{equation}

\noindent In addition to inferring parameters characterizing a BNS system, Bayesian analysis 
also allows us to perform hypothesis selection. This enables us to do model selection between 
different EOSs. Evidence for each hypothesis is computed by assuming an EOS and estimating 
source properties of a BNS source. By computing evidence for another EOS, the odds in favor of one EOS over 
another is calculated by taking the ratio of evidences. This is also called \emph{Bayes' factor} between two EOSs. 
For any single EOS, the evidence is given by
\begin{equation}
p(d|\mathcal{H}_{\rm EOS_1}, I) = \int p(d|\mathcal{H}_{\rm EOS_1},\vec{\theta},I) p(\vec{\theta}|\mathcal{H}_{\rm EOS_1},I) d\vec{\theta}  
 \label{eqn:evidence}
\end{equation}
By computing $p(d | \mathcal{H}_{\rm EOS_2}, I)$ for another EOS, the ratio between their evidences
\begin{equation}
B^1_2 = \frac{p(d|\mathcal{H}_{\rm EOS_1}, I)}{p(d | \mathcal{H}_{\rm EOS_2}, I)}
 \label{eqn:BF}
\end{equation}

\noindent Ref.~\cite{DelPozzo:2013ala} looked into combining multiple sources to distinguish 
between the EOSs they follow and concluded that $\sim$ 20 sources would be sufficient to 
tell apart a stiff, moderate, or soft EOS. With the additional refinements in their analysis 
and waveform models in~\cite{Agathos:2015uaa}, it was found that imposing a stricter 
prior on NS masses, Gaussian in this case, requires many more signals to make an accurate 
distinction between different EOSs. 
In the future era, the post-merger phase of a GW signal will have a higher 
SNR and therefore be visible in the detector's band, leading to a measurement 
of the frequencies of the post-merger spectrum; 
see e.g.~\cite{Bose:2017jvk,Chatziioannou:2017ixj,Tsang:2019esi,Breschi:2019srl}. 
Besides extracting the source parameters, future detections might also allow us to extract 
EOS-insensitive quasi-universal relations from GW data~\cite{Samajdar:2020xrd}. 

\subsection{Analyses of real signals}

Estimating the source properties from GW170817 provided the first constraints on 
NS-EOS from a GW observation. Within a Bayesian setting, conservative estimates on 
tidal deformability and hence the EOS, were placed using GW170817. 
This was done by using prior distributions independently on the two dimensionless component 
tidal deformabilities~\cite{TheLIGOScientific:2017qsa,Abbott:2018wiz}, i.e., even allowing the
that the two stars have different EOSs. 
The analyses in~\cite{Abbott:2018wiz} used different waveform models, 
combining different BBH baselines and different tidal prescriptions to arrive at 
comparable estimates. The waveform models used were \texttt{TaylorF2}, \texttt{IMRPhenomD\_NRTidal}, 
\texttt{IMRPhenomPv2\_NRTidal}, and \texttt{SEOBNRv4\_ROM\_NRTidal}. 
The models differed in their treatments of the point-mass baselines, the spin-effects 
as well as the tidal information. 
The consistency among results from these models confirmed that 
systematic biases from model inaccuracies were not dominant for GW170817
with the present detector sensitivities. 

\begin{figure*}[t]
 \centering
 \includegraphics[width=\textwidth]{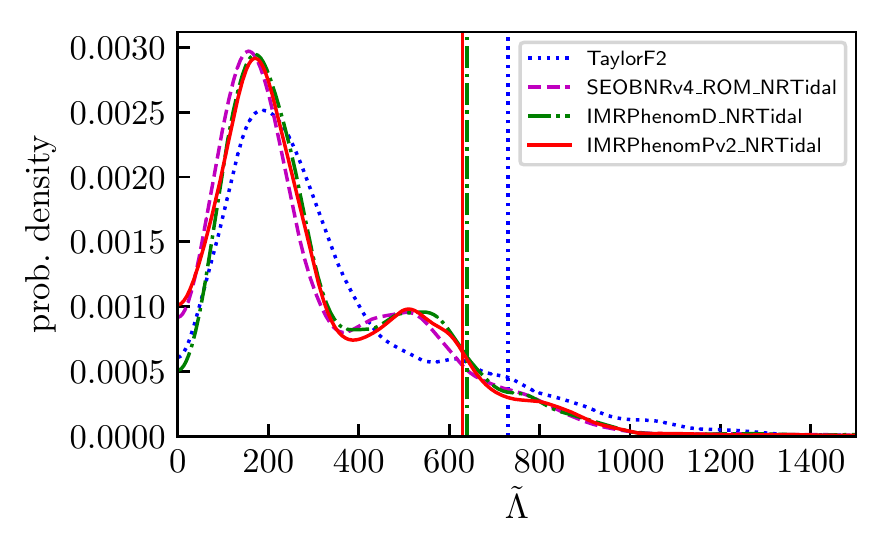}
 \caption{Posterior probability distributions on $\tilde{\Lambda}$ obtained from the waveforms  
          \texttt{TaylorF2} (blue, dotted line), \texttt{SEOBNRv4\_ROM\_NRTidal} (magenta, dashed line), \texttt{IMRPhenomD\_NRTidal} (green, dash-dotted line), 
          and \texttt{IMRPhenomPv2\_NRTidal} (red, solid line) 
          by analyzing the GW170817 signal. 
          The data has been taken from the data release of~\cite{Abbott:2018wiz}. 
          The plot shows the results from imposing the high-spin priors.}
 \label{fig:lamTOfAll}
\end{figure*}

The analyses placed a uniform prior on each of 
$\Lambda_i$ between [0,5000] and estimated $\tilde{\Lambda}$ by effectively 
using a prior uniform on $\tilde{\Lambda}$. 
Two different ranges of uniform spin priors were used; following observations 
of Galactic BNS systems, an upper limit of $0.05$ on the individual components' spin magnitudes 
were placed. In addition, to be non-conservative, upper limits of $0.89$ were used. Estimates for  
intervals of all intrinsic parameters including component masses as well as the tidal deformability $\tilde{\Lambda}$ 
were found to be stronger for the narrower spin priors. With wider spin priors, the component masses 
were estimated to lie between 1.00 $\mathrm{M}_\odot$ and 1.89 $\mathrm{M}_\odot$. With narrower priors on spins, 
the component masses were found to lie between 1.16 $\mathrm{M}_\odot$ and 1.60 $\mathrm{M}_\odot$. The estimate on 
the upper bound of $\tilde{\Lambda}$ was $630$ for wider spin priors, whereas within 90\% credible interval, 
an estimate of $300^{+420}_{-230}$ was made. The posteriors on $\tilde{\Lambda}$ from the high spin priors 
using the four different waveform models mentioned above is shown in Fig.~\ref{fig:lamTOfAll}.

\noindent Constraints on NS radii and EOS were obtained in~\cite{Abbott:2018exr}. 
This analysis was restricted to spins observed in NSs in Galactic binaries and therefore to a 
narrower range. Two methods were used to constrain the NS radii and EOS. Firstly, 
EOS-insensitive relations were used to employ priors on tidal deformabilities 
to enforce that both NS components follow the same EOS. Secondly, a method 
to parameterize directly the EOS $p(\rho)$ was used. 
Radii estimates of $R_A = 10.8^{+2.0}_{-1.7}$km and $R_B=10.7^{+2.1}_{-1.5}$km at 90\% credible intervals 
were obtained for the primary and secondary components by imposing the EOS-insensitive relations. 
Further constraining bounds of $R_A=11.9^{+1.4}_{-1.4}$km and $R_B=11.9^{+1.4}_{-1.4}$km were obtained 
from the method of parametrization of the EOS, in addition to imposing the maximum mass of a NS to 
be $> 1.97 \mathrm{M}_\odot$. 
Ref.~\cite{Carson:2019rjx} updated quasi-universal relations based on priors imposed from 
constraints on presure and density of the EOS from analyzing GW170817. 
A similar analysis was also carried out independently in~\cite{De:2018uhw} although there are differences in 
the waveform models as well as the universal relations used, the results have been overall consistent. 
Preliminary conclusions were drawn in~\cite{TheLIGOScientific:2017qsa,Abbott:2018wiz} 
about EOSs preferred by the GW data. A detailed study of model selection was carried out 
in~\cite{LIGOScientific:2019eut}. The data in this study was analysed with two ranges of priors, 
the narrow prior included uniform priors on spin magnitudes up to $0.05$, as well as a Gaussian prior on 
component masses. The wider prior included uniform spin magnitudes up to $0.7$, and a uniform prior on 
component masses. 
Only the stiffest EOSs could be ruled out completely from GW data. 
Using only the GW data without information from it's EM counterpart, we cannot rule out the BBH hypothesis either. 

\noindent Ref.~\cite{Kastaun:2019bxo} pointed out caveats on quoting lower bounds on quantities 
like the tidal deformability from use of uniform prior in Bayesian inference. However, we note that 
the upper bound on $\tilde{\Lambda}$ is quoted to be constrained from GW data analysis.
In~\cite{Abbott:2020uma}, the second observation of a BNS, GW190425, in the GW spectrum and the heaviest 
chirp-mass of a BNS was reported. Being such a heavy mass event, the constraints on tidal information 
did not improve on the existing ones from GW170817. However the constraints 
extracted were found to be consistent with those obtained from GW170817. 
Due to the high masses, the possibility that this event was gravitationally lensed~\cite{Pang:2020qow} 
or that one or two components have been BHs can not be ruled out, e.g.,~\cite{Han:2020qmn,Kyutoku:2020xka}. 
Incorporating knowledge about the EOS and the fact that no EM counterpart has been observed can be used to further 
constrain the properties of GW190425~\cite{Coughlin:2019zqi,Foley:2020kus,Most:2020kyx}

\subsection{Uncertainties from waveform modeling}

\begin{figure*}[t!]
   \includegraphics[keepaspectratio,width=\textwidth]{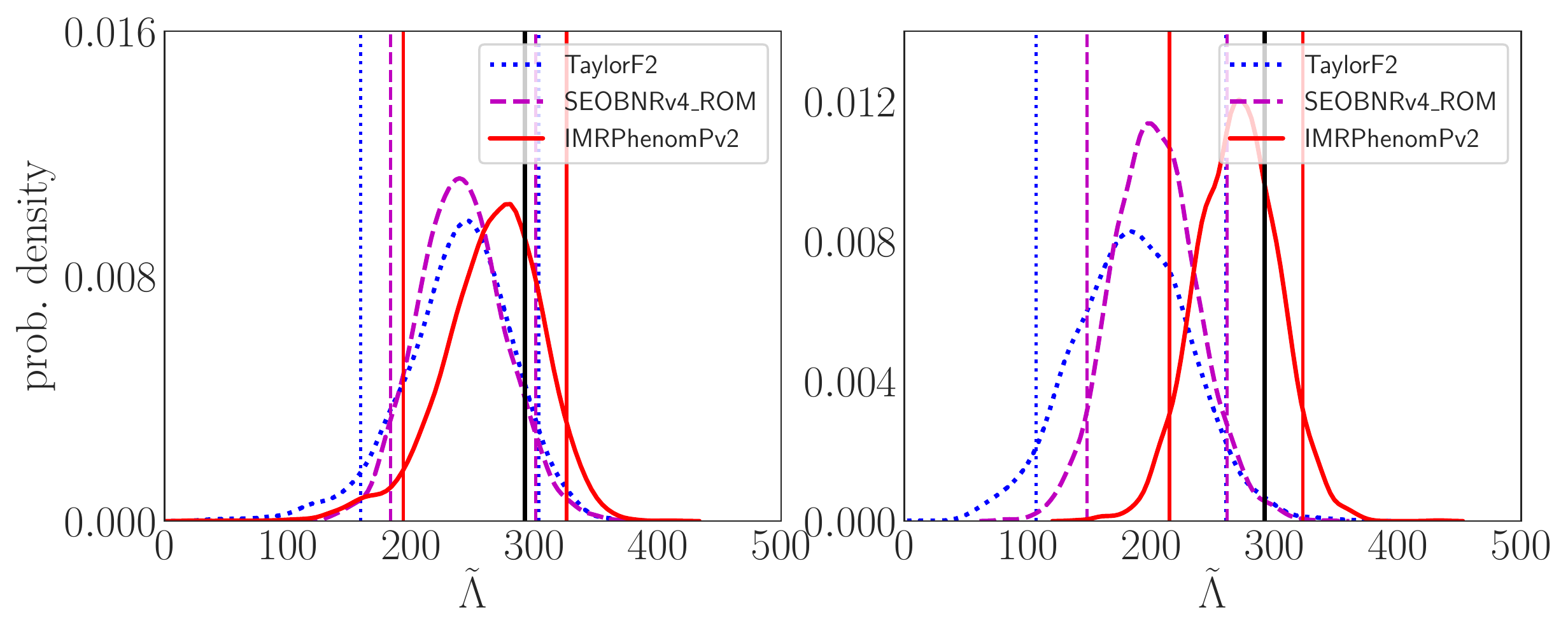}
  \caption{Posterior probability distributions on $\tilde{\Lambda}$ when differing BBH baselines, the tidal description is always NRTidal. The injected value is shown by the vertical black line and the 90\% credible intervals are marked for each model. The injected sources are equal-mass (left panel) with $m_1=m_2=1.375 \ \mathrm{M}_\odot$ and unequal-mass (right panel) with $m_1 = 1.68 \ \mathrm{M}_\odot$ and $m_2 = 1.13 \ \mathrm{M}_\odot$. The injected waveform model for both sources is with \texttt{IMRPhenomPv2\_NRTidal}.}
  \label{fig:bias1}
\end{figure*}

\begin{figure*}[t!]
   \includegraphics[keepaspectratio,width=\textwidth]{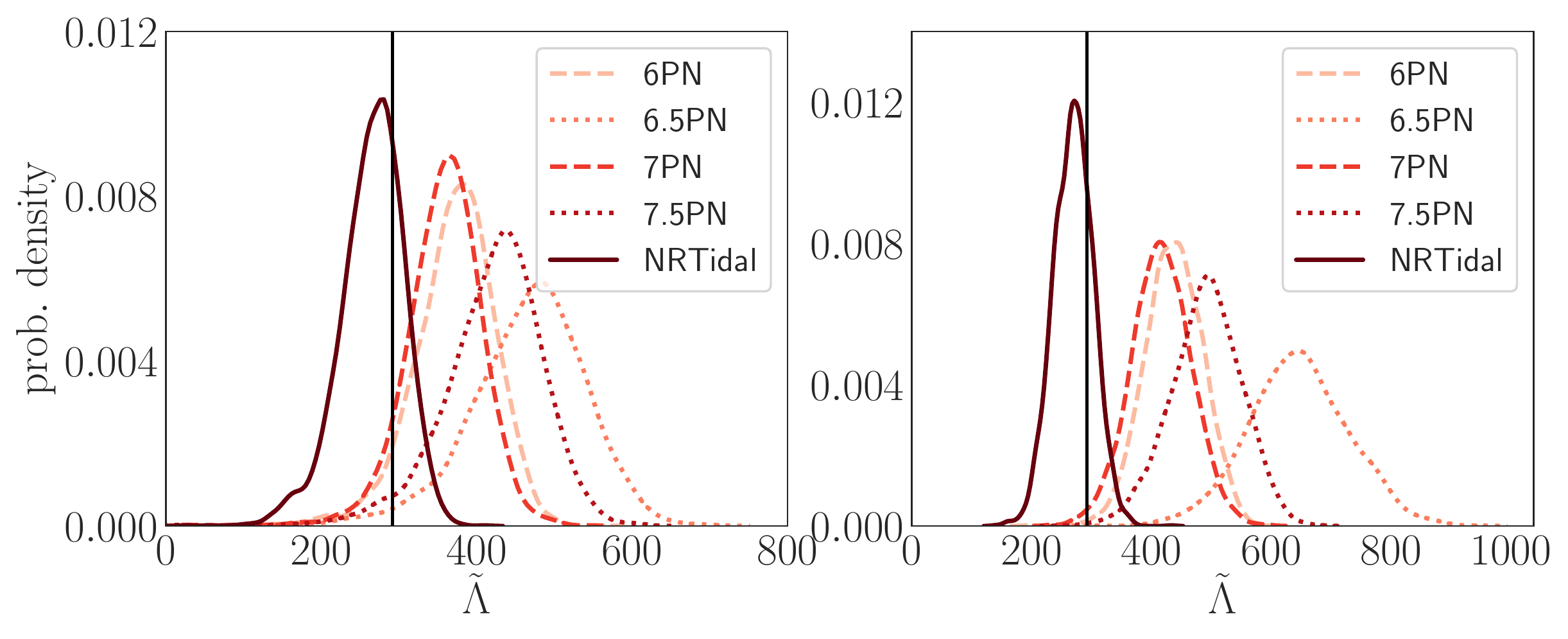}
  \caption{Posterior probability distributions on $\tilde{\Lambda}$ for different tidal descriptions keeping the same 
           baseline, \texttt{IMRPhenomPv2}. The injected value is shown by the vertical black line and the 90\% credible intervals are marked for 
           each waveform with dashed lines. The injected sources are equal-mass (left panel) with $m_1=m_2=1.375 \ \mathrm{M}_\odot$  
           and unequal-mass (right panel) with $m_1 = 1.68 \ \mathrm{M}_\odot$ and $m_2 = 1.13 \ \mathrm{M}_\odot$. The injected 
           waveform model for both sources is with \texttt{IMRPhenomPv2\_NRTidal}.}
  \label{fig:bias2}
\end{figure*}

Although tidal estimates between different waveform models were found to be in agreement 
for GW170817, with improved detector sensitivities, the systematic uncertainties 
will start becoming dominant~\cite{Abbott:2018exr}. 
Ref.~\cite{Samajdar:2018dcx} simulated non-spinning 
GW170817-like BNS sources in the era of Advanced LIGO and Advanced Virgo's design sensitivity 
to look for biases in the estimates of the tidal deformability parameter. This analysis was performed 
by using different BBH baselines combined with two different tidal prescriptions, the one derived from 
PN, and the NRTidal framework. The SNRs of these simulations were $\sim$ 90 and the analysis 
focused on BBH baselines derived from PN, EOB, and phenomenological models. 
The simulated sources were an equal-mass and an unequal-mass binary, both following the APR4 EOS. 
To estimate the effect of changing point-mass BBH baselines, the NRTidal framework was used 
by varying all four underlying models (\texttt{IMRPhenomPv2}, \texttt{IMRPhenomD}, \texttt{TaylorF2}, and 
\texttt{SEOBNRv4\_ROM}). For the equal-mass source, the injected tidal deformability was 
recovered within 90\% credible interval whereas in case of the unequal-mass source, only 
the two phenomenological models recovered the injected value within a 90\% credible interval; 
note that the phenomenological model \texttt{IMRPhenomPv2\_NRTidal} was used for injection. 
Other non-tidal intrinsic parameters like total mass, mass ratio, and the effective spins (in this case, 0) were found to be 
recovered reliably and consistently across all models and for both the sources. 
In addition, systematics arising from using 
the two different tidal prescriptions by keeping the point-mass baseline the same, 
were also investigated. It was found that increasing the PN order did not make the estimates more 
accurate, instead, the half-integer PN orders result in making the binary's interaction potential repulsive and 
therefore lead to an overestimate of the tidal deformabilities; cf.~Fig.~\ref{fig:fit_fd}.  
Even for this case, one finds that non-tidal intrinsic parameter estimates of total mass, mass ratio, 
and the effective spin are reliably and consistently recovered across all tidal descriptions. Out of the two sources, 
it was shown that unequal-mass sources 
showed greater biases in tidal deformability estimates. 
The biases in tidal deformability for both the equal-mass and unequal-mass source for the Phenom, EOB, and 
TaylorF2 baselines when the tidal information is kept the same, NRTidal, in all cases, are shown in Fig.~\ref{fig:bias1}. 
Fig.~\ref{fig:bias2} shows the bias in tidal deformability when the baseline is the same, \texttt{IMRPhenomPv2}, and 
the tidal descriptions are changed.
As an extension to this work, Ref.~\cite{Samajdar:2019ulq} investigated the biases on tidal parameter estimates 
from highly spinning BNSs, also in the era of design sensitivity, and the importance of spin-induced quadrupole moments. 
Particularly, it was found that neglecting EOS-dependent spin-induced quadrupole moments leads to a bias in 
parameter estimation once the dimensionless spin magnitude of a component NS is $\chi > 0.2$. In addition, \cite{Samajdar:2019ulq} studied different spin magnitudes and orientations as well as the advantage of having envisaged EM counterparts like a short GRB or a kilonova or both were studied. 
For aligned-spin configurations, in-plane spins, and spin vectors oriented at $45^\circ$ to the orbital angular momentum vector, the waveform models including the EOS-dependent spin-induced quadrupole moment performed 
better in estimating the tidal deformability parameter as spin magnitudes increased. 
In addition, having EM counterparts produced improved slightly the estimates 
of tidal deformability, however, no noticeable improvement in estimates of non-tidal parameters were observed. 

While the above works focused on tidal estimates from the inspiral only, Ref.~\cite{Dudi:2018jzn} 
used hybrid simulations to include the inspiral as well as post-merger regimes in a BNS coalescence. 
It was shown that non-tidal parameters were estimated reliably for SNRs as low as $25$ and for soft EOSs, even when 
analyzing the signal with a point-particle waveform model without tidal effects. For 
stiffer EOSs or larger SNRs, significant biases in estimates of non-tidal parameters were observed 
when analyzing a BNS signal with a point-particle waveform model. 
Additionally, it was found that neglecting the post-merger effects in the recovery of a BNS signal 
would not lead to significant biases in parameter estimation in the second generation era of detectors. 
In~\cite{Messina:2019uby}, the authors 
calculated a higher order PN term at 5.5PN and included it for parameter estimation of tidal waveforms on a non-spinning injection. The estimates using this waveform model which includes the higher order PN term is more consistent with the signal and the estimates obtained from the full \texttt{IMRPhenomD\_NRTidal} model than lower order PN approximants.

\subsection{A GW data-analysis outlook}

The envisaged era of third generation (3G) detectors will lead to 
increase in sensitivities at frequencies as low as $f \sim 5$ Hz. 
This will result in GW signals spending several cycles in the detector's 
band and much higher SNRs, BNS signals may spend up to hours in the 
bandwidth of the detector. As the signal enters the band much earlier, this 
will give the opportunity of early-warning alerts for possible EM 
counterparts~\cite{Akcay:2018aqh}. Within the large timescales of the signal 
lasting in band, the detectors' power spectrum may not be stationary and 
the Earth may also involve further subtleties in GW data analysis with such long 
signals. Ref.~\cite{Regimbau:2012ir} investigated overlapping signals in the third generation era, 
the work however focused on the problem of GW detections only, finding that overlap between two or more signals 
did not affect the performance of search algorithms unless the lower frequency cutoff was reduced. 
For long signals, the possibility of BNS signals 
being overlapped with other GW signals from BNSs or other GW sources is much higher. As no parameter estimation studies have been 
done on such overlapping signals, it remains to be seen whether retrieving source properties will be affected. 
The above work also identified critical issues for matched filtering for a signal lasting in band 
for up to a few days, making our current availability of computational resources inadequate. 
While techniques like faster sampling algorithms and waveform models will improve the ability to perform data analysis, 
we need better approaches to be able to extract information from the signals 
in the 3rd generation detector era.

\section{Summary}

In the last years, GW astronomy has been revolutionized by the direct detection of 
GWs from BH and NS binary systems. In particular the possibility to detect GW and EM signals 
in combination is a driving force in the field of multi-messenger astronomy and 
highlights the importance of the detection of BNS configurations.
In this review, we have focused on the measurement of tidal information from GW signals 
in the current GW detector era, i.e., our discussion focused primarily on the inspiral and on matter effects that are incorporated in state-of-the-art models.
We reviewed key aspects in the field of 
NR (Sec.~\ref{sec:GWmodels_NR}), 
PN theory (Sec.~\ref{sec:GWmodels_PN}),
the EOB formalism (Sec.~\ref{sec:GWmodels_EOB}), 
and phenomenological waveform approximants (\ref{sec:GWmodels_NRTidal})
to model GW signals for BNSs. 
We finalized our discussion by a short review about the detected GW signals 
GW170817 and GW190425, and by briefly describing GW data analysis 
results focusing on the imprint and extraction of 
tidal effects from GW signals. In conclusion, the planned improvements in experimental capabilities in the next years to decades will open unprecedented opportunities for fundamental physics and astrophysics with BNSs. However, capitalizing on the science potential will require significant further effort on addressing the challenges in numerical and analytical modeling as well as data analysis outlined above.

\begin{acknowledgement}
T.~Dietrich acknowledges support by the
European Union's Horizon 2020 research and innovation
program under grant agreement No 749145, BNSmergers. 
T.~Hinderer acknowledges support from the DeltaITP and NWO Projectruimte grant GW-EM NS, and thanks the YKIS 2019.  
A.~Samajdar is supported by the research
programme of the Netherlands Organisation for Scientific Research (NWO). 
We thank Kyohei Kawaguchi, Nathan Johnson-McDaniel, Luciano Rezzolla, and Jan Steinhoff for feedback on the manuscript. 
\end{acknowledgement}

\bibliographystyle{spmpsci}
\bibliography{paper20200406.bbl}

\end{document}